\documentclass[fleqn,usenatbib]{mnras}
\usepackage{graphicx}	
\usepackage{amsmath}	
\usepackage{newtxtext,newtxmath}

\usepackage[T1]{fontenc}
\usepackage{ae,aecompl}

\usepackage{lscape}

\title[FAST GPPS Survey: III. Timing of 30 pulsars]{
The FAST Galactic Plane Pulsar Snapshot Survey -- III. Timing results of 30 newly discovered pulsars}

\author[Su et al.]{W.~Q. Su,$^{1,2}$
J.~L. Han,$^{1,2,3}$\thanks{E-mail: hjl@nao.cas.cn }
P.~F. Wang,$^{1,2,3}$\thanks{E-mail: pfwang@nao.cas.cn }
J.~P. Yuan,$^{4,3}$
Chen Wang,$^{1,2,3}$
D.~J. Zhou,$^{1,2}$
Tao Wang,$^{1,2}$ \and
Yi Yan,$^{1,2}$ 
W.~C. Jing,$^{1,2}$
Z.~L. Yang,$^{1,2}$  
N.~N. Cai,$^{1,2}$ 
Xue Chen,$^{1,2}$ 
Jun Xu,$^{1,2,3}$
Lang Xie,$^{1,2,3}$   \and
H.~G. Wang,$^{5,6}$   
R.~X. Xu,$^{7}$   and   X.~P. You$^{8}$
\\
$^{1}$National Astronomical Observatories, Chinese Academy of Sciences, Jia-20 Datun Road, ChaoYang District, Beijing
   100012, China\\ 
$^{2}$School of Astronomy and Space Science, University of Chinese Academy of Sciences, Beijing 100049, China\\ 
$^{3}$Key Laboratory of Radio Astronomy and Technology,  Chinese Academy of Sciences, Beijing 100101, China \\   
$^{4}$Xinjiang Astronomical Observatory, Chinese Academy of Sciences, 150 Science 1-street, Urumqi 830011, China \\
$^{5}$Department of Astronomy, School of Physics and Materials Science, Guangzhou University, Guangzhou 510006, Guangdong Province, China \\ 
$^{6}$National Astronomical Data Center, Great Bay Area, Guangzhou 510006, Guangdong Province, China \\ 
$^{7}$Department of Astronomy, Peking University, Beijing 100871, China \\ %
$^{8}$School of Physical Science and Technology, Southwest University, Chongqing 400715, China 
}

\date{Accepted XXX. Received YYY; in original form ZZZ}

\pubyear{2023}

\begin{document}
\label{firstpage}
\pagerange{\pageref{firstpage}--\pageref{lastpage}}
\maketitle

\begin{abstract}
Timing observations are crucial for determining the basic parameters of newly discovered pulsars. Using the Five-hundred-meter Aperture Spherical radio Telescope (FAST) with the L-band 19-beam receiver covering the frequency range of 1.0–1.5 GHz, the FAST Galactic Plane Pulsar Snapshot (GPPS) Survey has discovered more than 600 faint pulsars with flux densities of only a few or a few tens of $\mu$Jy at 1.25 GHz. To obtain accurate position, spin parameters and dispersion measure of a pulsar, and to calculate derived parameters such as the characteristic age and surface magnetic field, we collect available FAST pulsar data obtained either through targeted follow-up observations or through coincidental survey observations with one of the 19 beams of the receiver. From these data we obtain time of arrival (TOA) measurements for 30 newly discovered pulsars as well as for 13 known pulsars. We demonstrate that the TOA measurements acquired by the FAST from any beams of the receiver in any observation mode (e.g. the tracking mode or the snapshot mode) can be combined to get timing solutions. We update the ephemerides of 13 previously known pulsars and obtain the first phase-coherent timing results for 30 isolated pulsars discovered in the FAST GPPS Survey. Notably, PSR J1904+0853 is an isolated millisecond pulsar, PSR J1906+0757 is a disrupted recycled pulsar, and PSR J1856+0211 has a long period of 9.89 s that can constrain pulsar death lines. Based on these timing solutions, all available FAST data have been added together to obtain the best pulse profiles for these pulsars.
\end{abstract}


%
%
\section{Introduction}

Pulsars are highly magnetized neutron stars. Their fast and stable rotations make them highly precise astronomical clocks and powerful tools for studying physics and astrophysics. Pulsar timing is the most important step in determining the basic parameters. By measuring and modelling the times of arrival (TOAs) of pulses, one can determine the spin period, period derivative, position, and dispersion measure (DM), and then derive the characteristic age, surface magnetic field strength, and spin-down luminosity. Pulsar timing can show if a pulsar is in a binary system, and then be used to obtain the binary parameters. The timing of pulsars in compact binary systems can be used to test theories of gravity through the measurement of changes in their orbital parameters caused by strong gravitational effects \citep{1989ApJ...345..434T,2021PhRvX..11d1050K}. The timing can show the additional Shapiro delay, which can be used to determine the masses of the pulsars \citep{dpr+10,afw+13}, which in turn constrains the equation of state of neutron stars. The long-term timings of some pulsars show irregularities of TOAs. The sudden increases in the rotation frequency of some young pulsars are known as ‘glitches’ \citep[e.g.][]{2013MNRAS.429..688Y,bsa+22}, and can be used as a probe of the internal structure of neutron stars. The monitoring of a series of millisecond pulsars (MSPs) can reveal the stochastic gravitational wave background \citep[e.g.][]{2010CQGra..27h4013H,pdd+19,fbb+23,2023RAA....23g5024X}. Furthermore, exact measurements of pulsar TOAs can be used as a tool for time-keeping \citep[e.g.][]{2020MNRAS.491.5951H}.

To date, there are over 3300 pulsars in the Australia Telescope National Facility (ATNF) Pulsar Catalogue \citep{2005AJ....129.1993M}.\footnote{\url{http://www.atnf.csiro.au/research/pulsar/psrcat/}} 
Most of them were discovered in pulsar surveys carried out with large radio telescopes, and precise basic parameters were determined mostly by subsequent monitoring and timing \citep[e.g., ][]{2001MNRAS.328...17M,2006ApJ...637..446C,2014ApJ...791...67S,2019A&A...626A.104S,tkt22}. In the $P-\dot{P}$ diagram, different types of pulsars, such as MSPs, magnetars, normal pulsars and young pulsars, occupy distinct regions. 

The Five-hundred-meter Aperture Spherical radio Telescope \citep[FAST, ][]{Nan2006,2011IJMPD..20..989N} is the most sensitive single-dish radio telescope in the world. There are three ongoing pulsar surveys using FAST: the FAST Galactic Plane Pulsar Snapshot (GPPS) survey \citep[][i.e. Paper I]{2021RAA....21..107H}, which has discovered more than 600 pulsars\footnote{\url{http://zmtt.bao.ac.cn/GPPS/}}, including more than 76 transient pulsars \citep[][i.e. Paper II]{zhx+23} and five fast radio bursts \citep[][i.e. Paper IV]{zhj+23}; the Commensal Radio Astronomy FAST Survey \citep[CRAFTS,][]{2018IMMag..19..112L} which has discovered more than 172 pulsars \citep{mzl+23}; the FAST Globular Cluster Pulsar Survey \citep{pqm+21}, which has discovered more than 40 pulsars. The FAST GPPS Survey covers the Galactic plane in the Galactic latitude range of $\pm10\degr$ visible to the FAST, by using a specially designed observation mode, dubbed the ‘snapshot mode’, through the quick switches of four pointings using the L-band 19-beam receiver of the FAST \citep[see details in][]{2021RAA....21..107H}. Pulsars discovered in the FAST GPPS Survey are very faint, with some of them having an extremely low flux density of a few $\mu$Jy \citep{2021RAA....21..107H,zhx+23}.

This paper is the third paper of the FAST GPPS Survey, reporting the timing results of 30 isolated pulsars discovered in the FAST GPPS Survey. It is hard to detect such faint pulsars discovered by the survey using other radio telescopes in a limited observation time. In order to obtain the basic parameters of these newly discovered pulsars, we collect available FAST data from the FAST GPPS Survey observations or the follow-up tracking observations, either through targeted observations or through coincidental observations with one of the beams of the L-band 19-beam receiver. We demonstrate that TOA measurements acquired by FAST from any beams of the receiver can be combined for pulsar timing. We obtain coherent timing solutions of 13 known pulsars and 30 newly discovered pulsars. In Section 2, we briefly describe the observations of the newly discovered pulsars and verify the precision of observations with various beams. In Section 3, we present the timing results and discuss the pulsar properties. In Section 4, we summarize our work.

\section{FAST observation data and processing}

Our data have been taken mostly from the FAST GPPS Survey and other FAST observation projects, including PT2020\_0071, PT2021\_0037, PT2021\_0126, PT2021\_0132, PT2022\_0047, PT2022\_0174 and PT2022\_0186. In addition, we obtained one TOA value from the FAST archival open data.\footnote{\url{https://fast.bao.ac.cn/cms/article/125/}}

\subsection{FAST observations and data collection}

In observations for the FAST GPPS Survey and follow-up verification observations of pulsar candidates, very often a pulsar, either a known pulsar or a newly discovered pulsar, is detected from one beam or even from several beams of the L-band 19-beam receiver. Data from 4096, 2048 or 1024 frequency channels covering the band of 1.0–1.5~GHz are recorded in the search mode with a sampling time of 49.152 $\mu$s. For each frequency channel, data for two polarization channels ($XX$ and $YY$, mostly in the GPPS Survey observations) or four polarization channels ($XX$, $X^*Y$, $XY^*$ and $YY$, mostly in follow-up tracking observations) of each frequency channel are recorded. The GPPS Survey snapshot observations last for 5 min for each pointing, and the verification tracking observations often last for 15 min, and all data from the 19 beams are recorded. The 19-beam receiver has a system temperature of about 22 K. Often, at the beginning or the end of each observation session, periodic calibration noise signals are injected for 2 min (for the 15-min sessions) or 40 s (for the 5-min sessions), and the data of four polarization channels in this duration are recorded for the system calibration for the bandpass and polarization characteristics.

The specific FAST timing observations in some follow-up projects were made with the tracking mode or the SwiftCalibration mode or the SnapShotCal mode. In the tracking and SwiftCalibration observations, each pulsar was tracked for 5–15 min, often using only the central beam of the L-band 19-beam receiver. The SnapShotCal mode observations were conducted for some pulsars to collect data from 76 beams so that the data of two or three other pulsars could be collected by using some of the 4$\times$19 beams in one cover without costing any valuable FAST time for additional slewing.

When the data of all FAST beams in any observation mode are searched, known pulsars and newly discovered pulsars are detected and catalogued. It is important to verify if all these data can be used together for pulsar timing.

\subsection{Data preparing and pulsar timing}

To carry out timing analysis, the initial parameters and TOAs of all observations have to be obtained first. Using the accumulated TOA data over more than one year, one can in principle obtain the precise spin parameters and position of a pulsar through timing analysis.

By using the initial parameters from the ATNF pulsar catalogue (for a known pulsar) or the searching output (for a newly discovered pulsar), including the approximate position, the initial period and DM, we fold the survey data or follow-up FAST observation data in the searching mode by using \textsc{dspsr}\footnote{\url{http://dspsr.sourceforge.net/}} \citep{2011PASA...28....1V}. This creates fits files with the initial parameters properly set in the head information, which are then used to find the optimal barycentric period and DM and obtain an integrated pulse profile with the highest signal-to-noise ratio (S/N) using the \textsc{pdmp} tool from \textsc{psrchive}\footnote{\url{http://psrchive.sourceforge.net}} \citep{2004PASA...21..302H}. 

The basic timing model for each pulsar includes its position, spin frequency (F0), derivative of the spin frequency (F1), and DM. We first use the beam position for a pulsar obtained from the survey (for a newly discovered pulsar or a known pulsar with a poor position in the ATNF catalogue), and F0 and DM obtained from the best period  obtained by \textsc{pdmp} as the initial parameters for timing analysis, while F1 is initially assumed to be zero. 

These initial parameters are important for obtaining more TOAs from observations. The observation data are folded to form PSRFITS archive files with a subintegration of every 10–20 s and with 256–2048 phase bins per period with these initial parameters. After removing radio-frequency interference (RFI) using \textsc{paz} and \textsc{psrzap}, we integrate all subintegrations and all channels using \textsc{pam} in the \textsc{psrchive} tool. A noise-free standard profile template is created using \textsc{paas}, mainly through the Gaussian fitting to such an integrated pulse profile with the highest S/N typically a few tens. Finally, by using \textsc{pat}, we obtain TOAs from cross-correlating the template with all observed pulse profiles, typically with an uncertainty of much less than 1 ms for normal pulsars, depending on the pulsar period and observation length.

\begin{figure}
\includegraphics[width=0.98\columnwidth]{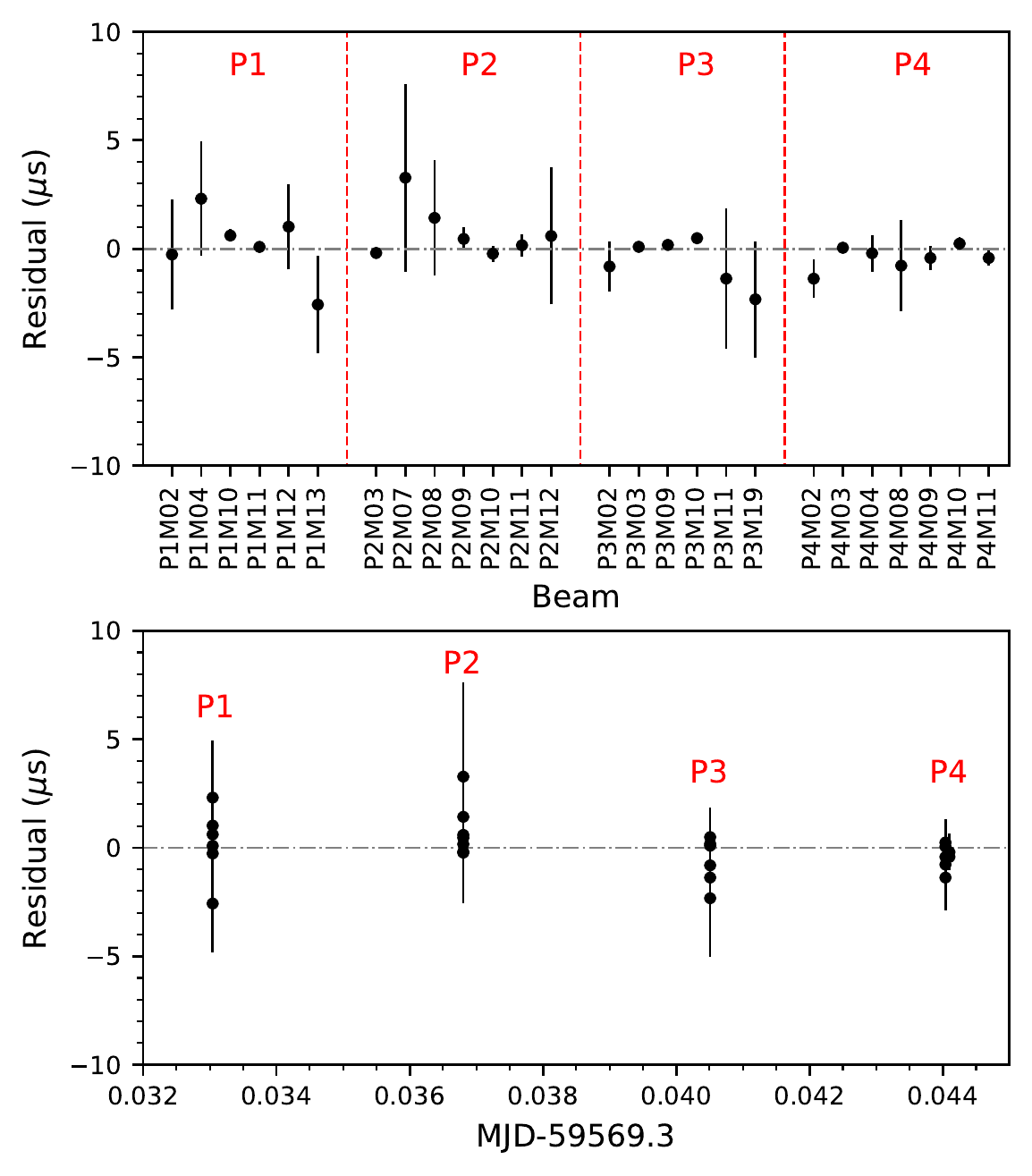}
\caption{Direct measurements of TOA residuals for PSR B1937+21 detected by many beams (M$k$, $k=$ 01 -- 19, in the {\it upper panel}, most of them via side-lobes) of the FAST L-band 19-beam receiver in the four pointings (P$n$, $n =$1, 2, 3 and 4, in the {\it lower panel}) conducted in one snapshot observation of the FAST GPPS Survey, G57.49$-$0.17\_20211221. All TOAs are very consistent with each other within $0.4~\mu s$ for this bright high-precision MSP, without any further calibration on the different delays between the receivers for 19 beams, indicating that data of different beams and different pointings can be used together for timing analysis. The large error bars of some data are caused by a low signal-to-noise ratio.}
\label{beams}
\end{figure}

\begin{figure}
\includegraphics[width=0.46\textwidth]{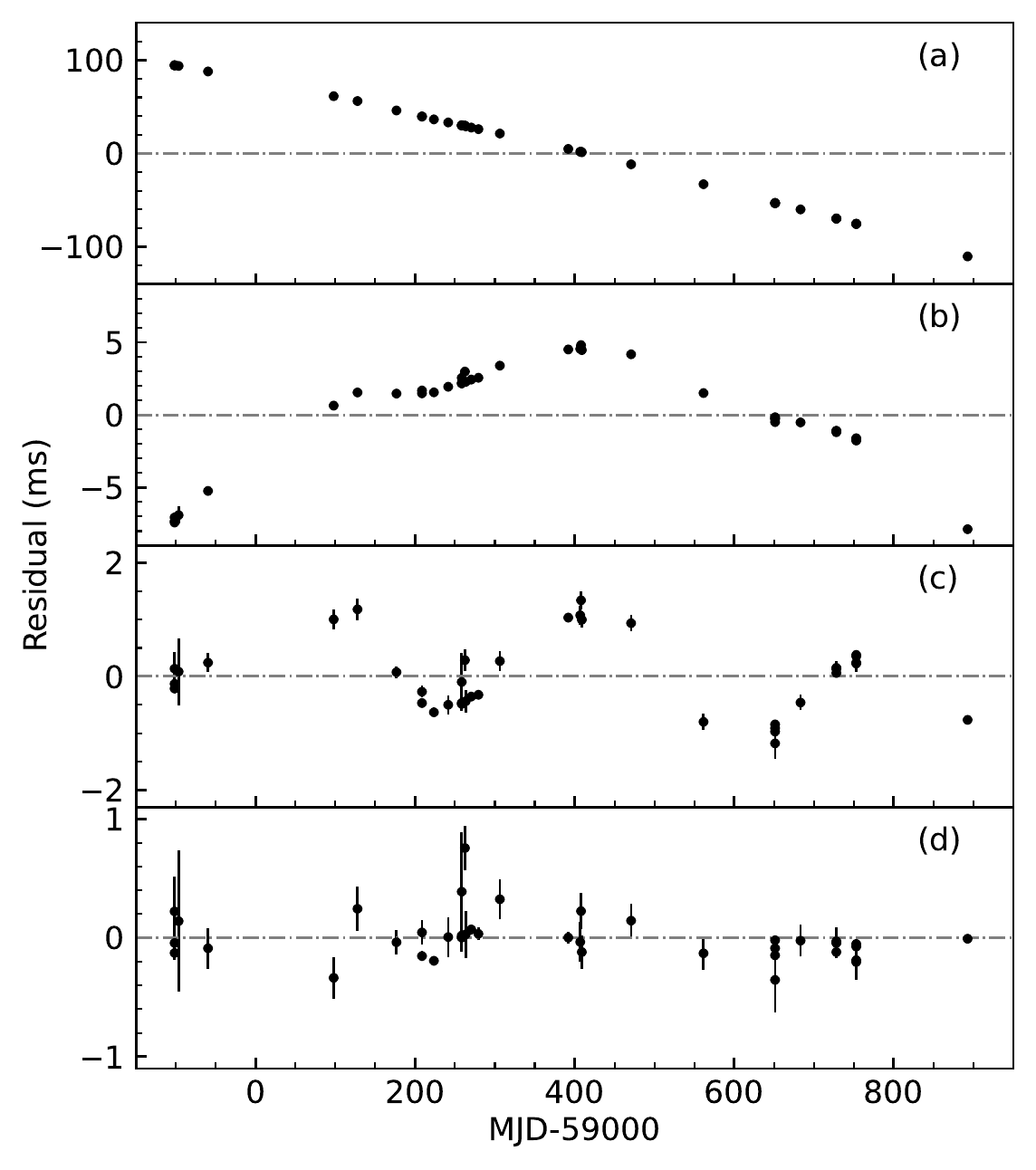}
\caption{An example of pulsar timing analysis for a known pulsar, PSR J1905+0600. The panels show the residuals obtained by using (a) the published ephemeris in \citet{2004MNRAS.352.1439H}, (b) a new fitting for F0, (c) a new fitting for F0 and F1, (d) a new fitting for F0 and F1, and also the position.}
\label{J1905+0600}
\end{figure}

\begin{table*}
\begin{center}
\caption[]{Basic parameters for 13 previously known pulsars, including right ascension, declination, barycentric period, period derivative, dispersion measure, with the $1\sigma$ uncertainty given in brackets, together with the epoch for the period, data span, and weighted rms residuals. For each pulsar, previously published parameters in the literature are given in the first line, and our new measurements are given in the second line for every pulsar. For comparison with previously published parameters, our timing solutions are presented in TDB units.
}
\label{knowntab1}
 \tabcolsep 4pt 
 \footnotesize
 \begin{tabular}{llllllrlcl}
  \hline
Name          & Ref.      & RA                & Dec.               & $P$                  & $\dot{P}$    & DM                & Epoch & Span  & $W_{rms}$ \\
              &           & (hh:mm:ss)        & (dd:mm:ss)        & (s)                & ($10^{-15}~\rm s~s^{-1}$)     & ($\rm cm^{-3}$pc) & (MJD) & (yr)  & (ms)  \\
\hline\noalign{\smallskip}
J1852$-$0000  & [1]        &  18:52:40.167(9) &  $-$00:00:25.5(3) &  1.92066632921(2) &  251.9666(3)  & 593(1)  &    58205.0 &5.7\\    
              & This work  &  18:52:40.15(1)  &  $-$00:00:26.3(7)  &  1.92068580196(3)  &  251.933(1)  &  593.1(3)  &  59403.0  &  3.74  &  0.523 \\
J1852$+$0008$^*$  & [2]        & 18:52:42.78(3)    & $+$00:08:09.6(8)  & 0.467894113075(20) & 5.679(7)   &  254.9(18)  & 52584.0    & 1.50  &    \\
              & This work  &  18:52:42.802(2)  &  $+$00:08:09.6(1)  &  0.4678973753676(9)  &  5.68435(6)  &  255.40(6)  &  59228.0  &  2.78  &  0.174 \\
J1853$+$0029$^*$  & [1]        & 18:53:17.745(17)  & $+$00:29:23.8(7)  &  1.8767576226(3)  &  2.431(2)   &  232(4)     &57826.0&3.5 & \\
              & This work  & 18:53:17.783(4)  &  $+$00:29:22.1(4)  &  1.87675787185(1)  &  2.428(2)  &  227.39(8)  &  59013.0  &  1.60  &  0.208 \\
J1858$+$0241$^*$  & [2]        & 18:58:53.81(14) &   $+$02:41:38(6)   &  4.6932329333(12)   &  24.32(9)  &  336(15)    &  52111.0 & 2.31   \\
              & This work  &  18:58:53.81(2)  &  $+$02:41:37.5(5)  &  4.69324828044(5)  &  24.274(6)  &  325.9(5)  &  59425.0  &  2.88  &  0.794 \\
J1901$+$0320$^*$  & [2]        &  19:01:03.01(9)  & $+$03:20:18(4)     & 0.63658447822(8) &    0.52(3)   &  393(7) & 52503.0 &1.22 & \\
              & This work  &  19:01:03.082(6)  &  $+$03:20:17.5(2)  &  0.63658476124(1)  &  0.5096(4)  &  394.8(3)  &  59428.0  &  2.87  &  0.296 \\
J1901$+$0510$^*$  & [2]        &  19:01:57.85(11) & $+$05:10:34(4)     &  0.61475669408(12) &  31.10(4)  &  429(7)    &   52618.0 & 1.15 \\
              & This work  &  19:01:57.90(5)  &  $+$05:10:36.5(7)  &  0.61477450398(2)  &  31.113(2)  &  435.5(2)  &  59243.0  &  1.91  &  0.618 \\
J1905$+$0600$^*$  & [2]        & 19:05:04.35(5)    & $+$06:00:59.9(14) & 0.441209731966(18) & 1.1123(10) &  730.1(19)  & 52048.0   & 3.17  &    \\
              & This work  &  19:05:04.3352(8)  &  $+$06:01:00.65(2)  &  0.4412104290545(3)  &  1.10719(2)  &  728.44(3)  &  59395.0  &  2.72  &  0.074 \\
J1905$+$0902$^*$  & [3],[4]    & 19:05:19.535(2)   & $+$09:02:32.49(8) & 0.2182529126846(9) & 3.49853(8) &  433.4(1)   & 54570.0  &   &    \\
              & This work  &  19:05:19.5414(8)  &  $+$09:02:32.39(2)  &  0.2182543240059(3)  &  3.499036(9)  &  433.497(7)  &  59355.0  &  3.35  &  0.042 \\
J1906$+$0649$^*$  & [2]        & 19:06:11.97(3)    & $+$06:49:48.1(10) & 1.28656437956(10)  & 0.152(5)   &  249(4)     & 52317.0   & 2.80  &    \\
              & This work  &  19:06:12.003(2)  &  $+$06:49:48.5(2)  &  1.286564468927(6)  &  0.1498(4)  &  250.16(9)  &  59200.0  &  2.47  &  0.292 \\
J1927$+$1852$^*$  & [5],[6],[7]& 19:27:10.422(8)   & $+$18:52:08.5(2)  & 0.482766273821(7)  & 0.116(1)   &  264.5(4)   & 51600.0   &   &    \\
              & This work  &  19:27:10.414(5)  &  $+$18:52:08.62(9)  &  0.482766349761(2)  &  0.11606(6)  &  264.66(6)  &  59310.0  &  4.05  &  0.190 \\
J1930$+$1408  & [1],[8]    & 19:30:18.9526(18) & $+$14:08:55.39(5) & 0.425720327378(5)  & 0.00190(1) &  210.87(13) & 56885.0   & 8.6      &    \\  
              & This work  & 19:30:18.949(5)  &  $+$14:08:55.3(1)  &  0.425720327769(1)  &  0.00194(9)  &  212.11(9)  &  59388.0  &  3.17  &  0.227 \\
J1936$+$2042  & [1],[8]    & 19:36:27.42(2)    & $+$20:42:04.5(4)  & 1.39072342303(15)  & 49.3744(14)&  197.4(5)   &  56065.0  & 5.8  &    \\
              & This work  &  19:36:28.702(1)  &  $+$20:41:26.62(3)  &  1.390726665137(2)  &  5.2926(1)  &  195.11(3)  &  59441.0  &  2.96  &  0.080 \\
J1954$+$2923  & [9],[10]   & 19:54:22.554(2)  &  $+$29:23:17.29(4)  &  0.4266767865302(4)  & 0.001711(3) & 7.932(7)  & 48719.0  & 21.4 &      \\
              & This work  &  19:54:22.5107(4)  &  $+$29:23:16.15(1)  &  0.4266767880541(2)  &  0.001663(6)  &  7.932(7)  &  59336.0  &  4.25  &  0.040 \\
  \noalign{\smallskip}\hline
\end{tabular}
\end{center}
{$^*$ Pulsars with improved parameters.}
References: 
[1] \citet{2022ApJ...924..135P}; [2] \citet{2004MNRAS.352.1439H}; [3] \citet{2006ApJ...637..446C}; [4] \citet{2013ApJ...772...50N}; [5] \citet{1975ApJ...201L..55H}; [6] \citet{2002AJ....123.1750L}; [7] \citet{2021RAA....21..107H}; [8] \citet{2015ApJ...812...81L}; [9] \citet{1970Natur.227.1123D}; [10] \citet{2004MNRAS.353.1311H}.
\end{table*}

We conduct a timing analysis using the \textsc{tempo2}\footnote{\url{http://bitbucket.org/psrsoft/tempo2}} package \citep{2006MNRAS.369..655H}, and the final timing solutions are obtained after a few iterations. We first fit F0 using TOAs obtained from several observations in nearby epochs, then fit the other parameters in the timing model as needed, and obtain RA, DEC, F0 and F1 as the first iteration. The JPL DE440 planetary ephemeris is used in our analysis. In the analysis of GPPS pulsars, the Barycentric Coordinate Time (TCB) unit is utilized, while the Barycentric Dynamical Time (TDB) unit is employed for previously known pulsars for comparison with parameters in the literature. After these basic parameters of the first iteration are obtained for the phase-connected solutions, FAST-observed data are folded again with these improved parameters, and then the new template is formed from the integrated profile obtained by combining all data using \textsc{psradd}, typically with a much improved S/N, depending on the pulsar flux density and the sharpness of the profile. All TOAs are obtained again from every FAST observation session, and these TOAs in general have a better precision than those from profiles with a low S/N. These TOAs are fitted again to obtain an improved timing solution and residuals via \textsc{tempo2} as the second iteration. To improve the DM precision, we integrate all available archive files to one subintegration and four subbands, and then compute the TOAs of different frequency subbands, and then all parameters except DM are fixed, and we use \textsc{tempo2} to fit the final DM. With the basic parameters together with the newly derived DM, the FAST data are folded again, and the TOAs are obtained for the final timing analysis, and the solutions are given as final results.

To obtain the integrated pulse profiles with the best S/N, the data are refolded with the final ephemeris. Following the polarization calibration procedure described in \citet{whj+22}, we obtain the total power pulse profile or polarization profiles of pulsars for which data with four polarization channels are recorded. The new rotation measure (RM) is obtained via \textsc{rmfit}. The Faraday rotation of linear polarization is corrected for the final polarization profiles, and data from several sessions are then summed with a weight of S/N. If the polarization data of a pulsar are not recorded for many sessions, the  total intensity pulse profile is obtained from the sum of all observations with a weight of S/N. Using the method described in \citet{2021RAA....21..107H}, we estimate the flux densities of these newly discovered pulsars from the integrated total intensity profiles.

\begin{figure*}
\includegraphics[width=0.9\textwidth]{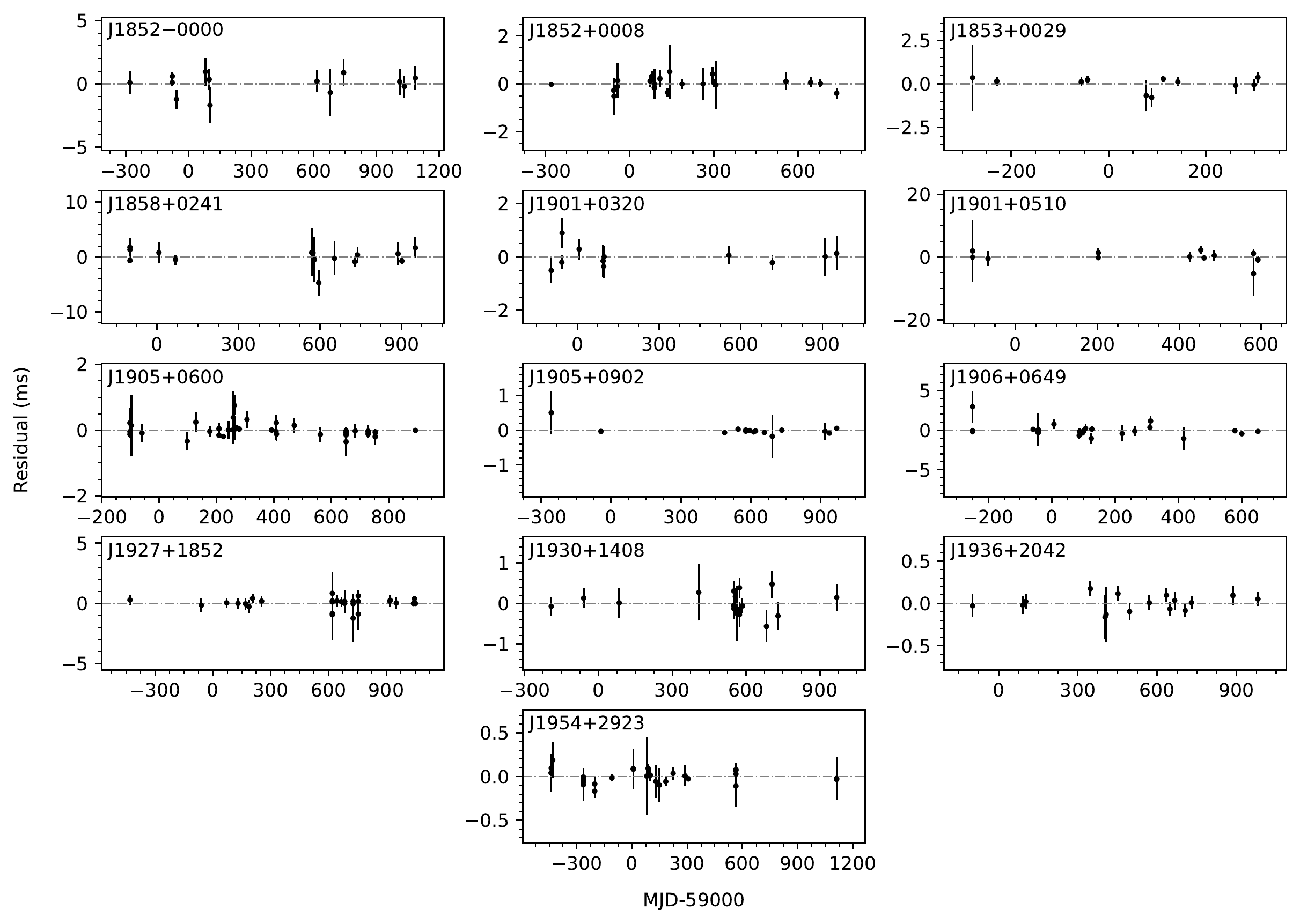}
\caption{Post-fit timing residuals for 13 known pulsars measured by FAST after re-fitting a new timing model given in Table~\ref{knowntab1}. To make the the reduced $\chi^2 \sim 1$, TOA uncertainties have been scaled by a factor termed EFAC.}
\label{knownPSRpostfit}
\end{figure*}

\subsection{Verification of timing using data from different beams of FAST 19-beam receiver}

Here we verify if data collected by different beams of the FAST 19-beam receiver can be used together for pulsar timing analysis. For this subtle work, we study the TOA data of PSR B1937+21, which is a bright MSP with a period of 1.56 ms. Previously, the high-precision timing solution of PSR B1937+21 \citep{2021MNRAS.507.2137R} was obtained from long-term timing observations through the International Pulsar Timing Array (including observations by the European Pulsar Timing Array, the North American Nanohertz Observatory for Gravitational Waves, and the Parkes Pulsar Timing Array). PSR B1937+21 was observed by chance in a FAST snapshot survey in 2021 December in the cover of G57.49-0.17\_20211221. Among four pointings of the 19 beams, the strong pulsar was detected from 31 of the 76 beams in the sky, mostly via the side-lobes of the receiver beams. We folded the data into 10-s subintegrations using the PPTA ephemeris and removed RFI using \textsc{paz}, then integrated the data in both time and frequency dimensions. The template is formed from data collected by the beam of P4M03, which has the highest S/N. We obtained TOAs using \textsc{pat} and calculated pre-fit residuals using \textsc{tempo2}. 

Clearly, the uncertainties of TOAs depend on the S/Ns. We have 26 TOA measurements with an uncertainty smaller than 5~$\mu$s, as shown in Fig.~\ref{beams}. For all beams with pulsar signal detection, TOA residuals are consistent with each other with $\sigma\sim 0.4~\mu$s. This demonstrates that pulsar signals detected from any beams of the FAST L-band 19-beam receivers from any pointings can be used for pulsar timing directly. The differences for the signal delay in different receivers in the L-band 19-beam receivers are almost negligible.

\section{FAST timing results}

Based on FAST observations, we obtain timing solutions for 30 pulsars newly discovered by the GPPS Survey and 13 previously known pulsars. First, we present the results for the 13 pulsars.

\subsection{Timing results for 13 previously known pulsars}

We have collected a good number of FAST observations for 13 known pulsars from the GPPS Survey observations and the follow-up observations. Through timing analysis of these data, we obtain new timing solutions as presented in Table~\ref{knowntab1}, including precise coordinates in right ascension and declination, barycentric periods, period derivatives, and DMs. Also listed in Table~\ref{knowntab1} are the epoch for the period, the data span, and the weighted rms residual. 

For example, PSR J1905+0600 is a 0.468-s pulsar, and its discovery and timing solutions were reported by \cite{2004MNRAS.352.1439H}. During the GPPS Survey and follow-up observations, this pulsar was detected by chance many times in a span of 2.72 yr. We finally obtained 40 TOAs. By using the official ephemeris given in \cite{2004MNRAS.352.1439H}, we obtained the residuals shown in Fig.~\ref{J1905+0600}(a), which indicates that further improvements on the ephemeris are desired. We then fitted F0, F1, and the position in turn, and finally obtained reasonable residuals, as shown in Fig.~\ref{J1905+0600}(d). The plots for post-fit residuals of these 13 known pulsars are shown in Fig.~\ref{knownPSRpostfit}. 

Among these 13 known pulsars, the ephemerides in Table~\ref{knowntab1} have been improved for nine pulsars, PSRs J1852+0008, J1853+0029, J1858+0241, J1901+0320, J1901+0510, J1905+0600, J1905+0902, J1906+0649, and J1927+1852, compared with those in the literature. The ephemerides of three pulsars, PSRs J1852$-$0000, J1930+1408 and J1954+2923, are consistent with previous ones in the references. Owing to the proper motion in J1954+2923, the position we obtained is slightly different from that in the reference.

The most remarkable difference is found for the ephemeris of PSR J1936+2042. This pulsar was discovered by \citet{2015ApJ...812...81L} in the Arecibo survey, and its timing solution was recently published by \citet{2022ApJ...924..135P}. Using our FAST data, we obtained the best-fitting phase-coherent timing solution (see Table~\ref{knowntab1} and Fig.~\ref{knownPSRpostfit}), which is significantly different for the period derivative and also the declination. In principle, such a large difference of the period derivative may be caused by glitches, but it is hard to have such a large-magnitude difference (49.4$\times 10^{-15}$~s/s to 5.3$\times 10^{-15}$~s/s). However, note that timing data for this pulsar in fig. 3 of \citet{2022ApJ...924..135P} are observed mostly in the second half of 2014, with very few data outside this date span, and the position is not well constrained by their data; therefore, it is understandable that an offset of 38 arcsec is obtained from our FAST determination. More timing data are needed to confirm our new ephemeris.

\subsection{Timing results for 30 GPPS pulsars}

\begin{figure*}
\includegraphics[width=0.90\textwidth]{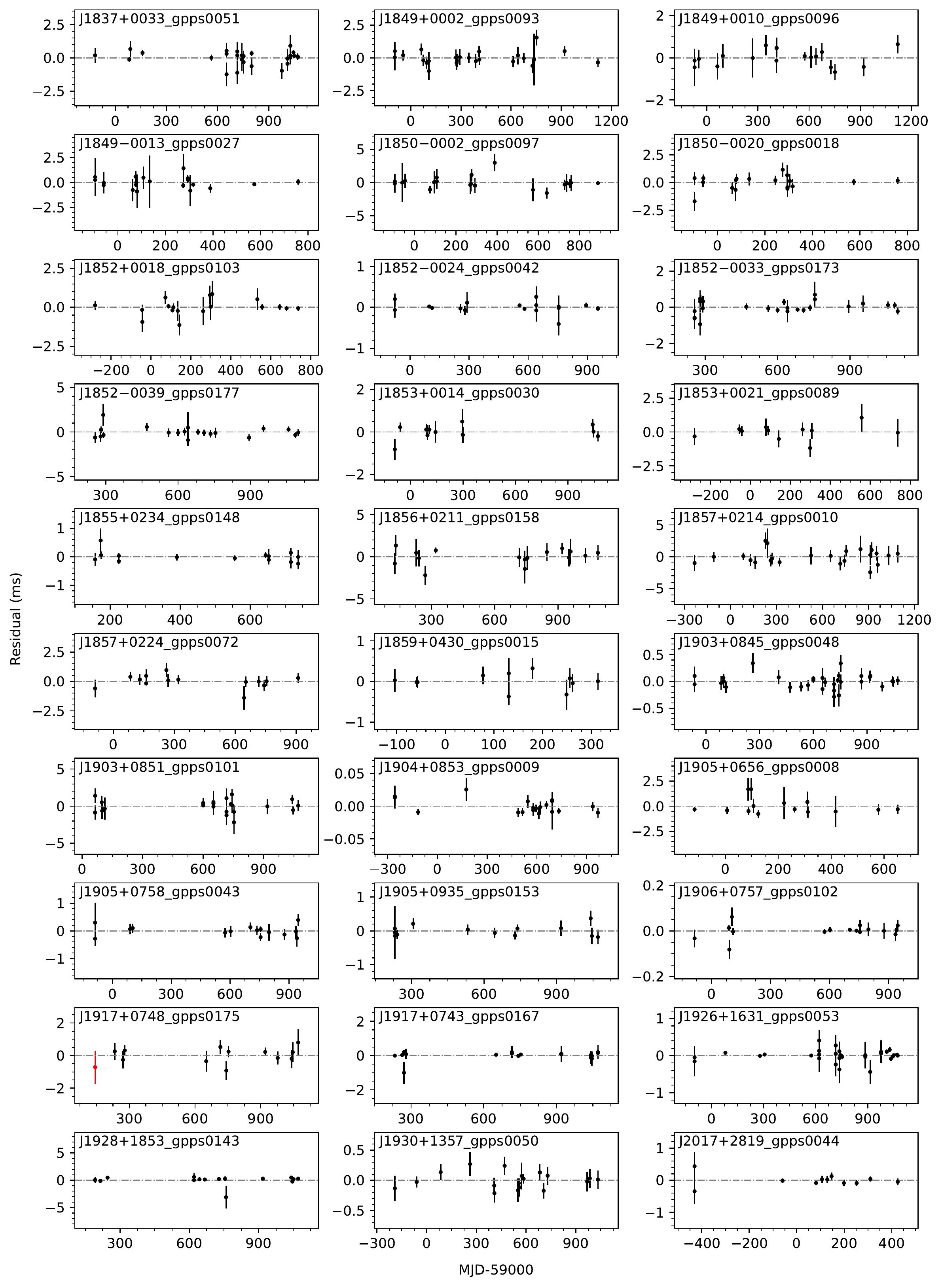}
\caption{Post-fit timing residuals  with phase-connected timing solutions in Table~\ref{GPPStab1} for 30 pulsars newly discovered in the FAST GPPS Survey. The TOA uncertainties have been scaled by EFAC so that the reduced $\chi^2 \sim 1$. The red dot indicates one TOA obtained from FAST open archival data: the list of FAST open archival data can be found at the website \url{https://fast.bao.ac.cn/cms/article/125/}.}
\label{gppsPSRpostfit}
\end{figure*}

Since the discovery of these 30 GPPS pulsars, we have observed them either in targeted follow-up observations or in other survey observations by chance. We obtained TOAs from some beams of these observations and obtained the timing solutions as presented in Table~\ref{GPPStab1}. Their positions are determined to be accurate with an uncertainty of less than 1 arcsec (or a few milliarcseconds for a MSP), verifying that the coarse positions previously obtained by FAST pointings are good enough for the follow-up studies. Based on the newly measured pulsar period and period derivatives, we derive the characteristic age, surface magnetic field strength, and spin-down luminosity of these pulsars. Post-fit timing residuals after the model fittings for these 30 newly discovered pulsars are shown in Fig.~\ref{gppsPSRpostfit}. The residuals for most of these isolated pulsars are white-noise-like, indicating a good fit with the basic timing parameters. No glitch or timing noise was detected owing to the small observational time span and the fact that these pulsars are not young. Some TOAs exhibit large uncertainties owing to the low S/N, caused by a short observation duration for a faint pulsar or by a large offset of the pulsar in a FAST beam.

\begin{figure}
\includegraphics[width=0.46\textwidth]{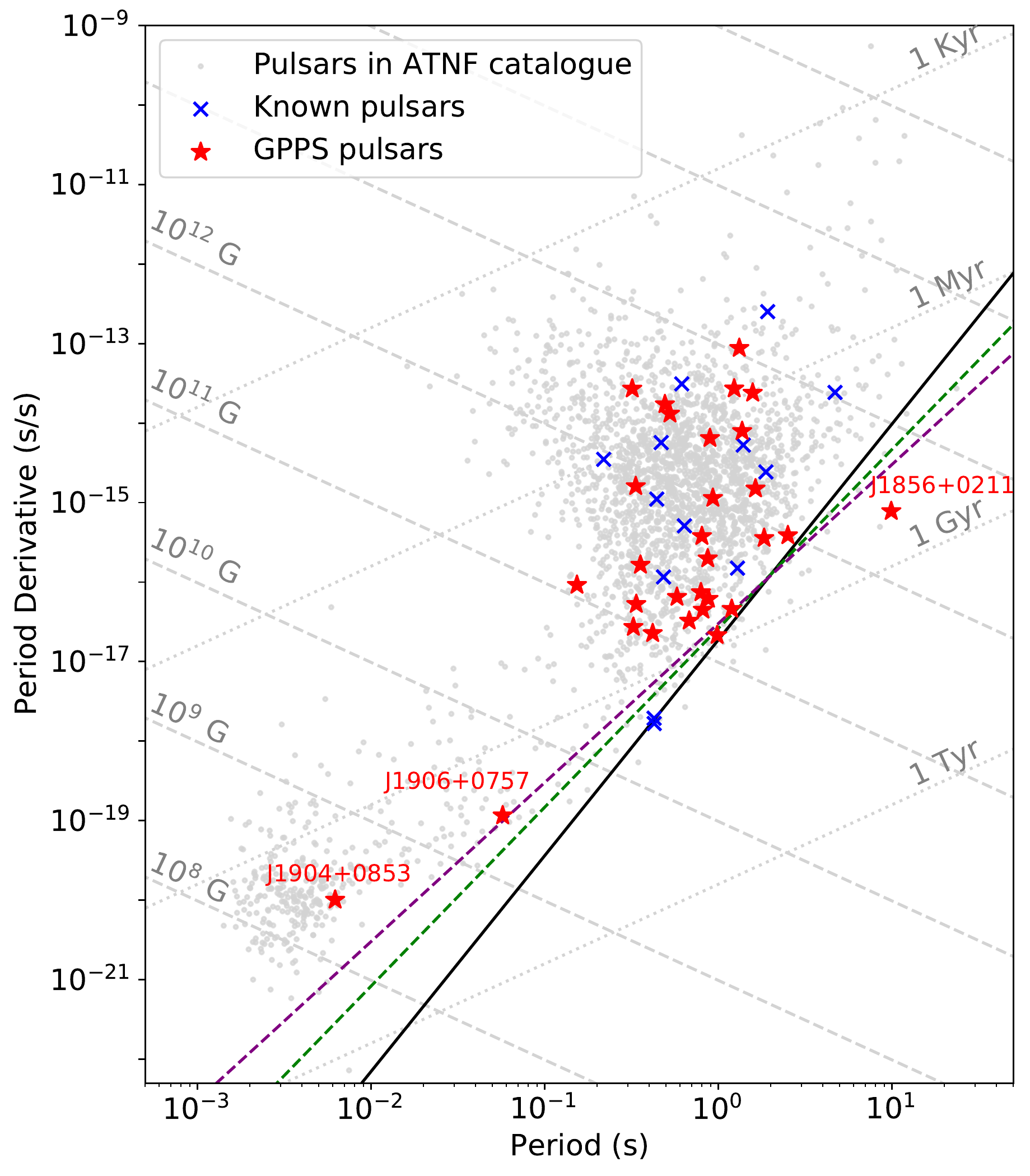}
\caption{
$P-\dot{P}$ diagram for pulsars based on data from the ATNF catalogue (grey points) and on new data in this paper: blue crosses denote the 13 previously known pulsars, and red stars denote pulsars newly discovered in the GPPS Survey. The dashed and dotted light grey lines in the background indicate constant surface magnetic field strengths and characteristic ages, respectively. The solid black line represents the death line from \citet{1993ApJ...402..264C}, and the dashed green and purple lines represent the death lines from equations (4) and (9) of \citet{2000ApJ...531L.135Z}, , corresponding to curvature radiation from the vacuum gap model and the space-charge-limited flow model.
}
\label{PPdot}
\end{figure}

Fig.~\ref{PPdot} illustrates the positions of these 30 pulsars in the $P-\dot{P}$ diagram, together with 13 previously known pulsars with new timing results. Notably, PSR J1904+0853 is a MSP, and PSR J1906+0757 is found to be an aged pulsar near the death line. A long-period pulsar, PSR J1856+0211, with a spin period of approximately 10 s, is not a magnetar but a normal pulsar, even though it is below the death lines in Fig.~\ref{PPdot}. 

\subsubsection{An isolated millisecond pulsar: PSR J1904+0853}
PSR J1904+0853 is a MSP with a period of 6.2 ms and a period derivative of $1.009 \times 10^{-20}~\rm s~s^{-1}$. It has a relatively broad pulse profile, and the duration of each observation is short, resulting in a 7-$\mu$s rms residual. Its TOAs are well modelled by the basic timing model, and the residuals are noise-like, indicating that PSR J1904+0853 is an isolated MSP.

\begin{landscape}
\begin{table}
\begin{center}
\caption[]{Best-fitting parameters of the timing models of 30 isolated pulsars newly discovered in the GPPS Survey and derived parameters. Columns are set for the pulsar name, the previous temporary name in \citet{2021RAA....21..107H}, together with the GPPS discovery number, flux density at 1.25 GHz, right ascension (2000), declination (2000), barycentric period, period derivative, dispersion measure, epoch for the period, data span, weighted rms residuals, and derived parameters for the characteristic age, surface magnetic field strength, and spin-down luminosity. The 1$\sigma$ uncertainty is given in brackets, and the results are presented in TCB units.}  
\label{GPPStab1}
 \tabcolsep 3pt  
 \footnotesize
 \begin{tabular}{llcllllrcccccc}
   \hline\noalign{\smallskip}
PSR name & Previous name: GPPS no.  &  $S_{\rm 1.25GHz}$&  RA (2000)         & Dec. (2000)         & $P$                    &$\dot{P}$& DM               & Epoch    & Span  & $W_{rms}$ &log[$\rm \tau$/yr]       & log[B/G]            &  log[$\rm \dot{E}$/erg $\rm s^{-1}$]\\
              &                     &  ($\mu$Jy)        & (hh:mm:ss)       & (dd:mm:ss)        & (s)                  &($10^{-15}~\rm s~s^{-1}$) & ($\rm cm^{-3}$pc)& (MJD)    & (yr)  & (ms) \\
  \hline\noalign{\smallskip}
J1837$+$0033  &J1837$+$0033g: gpps0051 & 33.1  &  18:37:41.100(4)  &  $+$00:33:46.1(1)  &  0.418131244964(1)  &  0.02273(6)  &  187.4(1)  &  59477.0  &  3.26  &  0.248  & 8.46  &  10.99  &  31.09 \\
J1849$+$0002  &J1849$+$0001g: gpps0093 & 16.3  &  18:49:22.557(5)  &  $+$00:02:17.0(2)  &  0.525650459766(2)  &  13.1274(1)  &  188.2(3)  &  59511.0  &  3.33  &  0.444  & 5.80  &  12.42  &  33.55 \\
J1849$+$0010  &J1849$+$0009g: gpps0096 & 17.8  &  18:49:42.670(6)  &  $+$00:10:00.4(1)  &  1.318430596396(5)  &  87.8234(3)  &  505.1(2)  &  59524.0  &  3.26  &  0.365  & 5.38  &  13.04  &  33.18 \\
J1849$-$0013  &J1849$-$0014g: gpps0027 & 149.8 &  18:49:26.996(3)  &  $-$00:13:46.8(1)  &  0.491754129798(2)  &  17.2224(1)  &  345.9(1)  &  59331.0  &  2.34  &  0.243  & 5.66  &  12.47  &  33.76 \\
J1850$-$0002  &J1850$-$0002g: gpps0097 & 60.7  &  18:50:02.340(9)  &  $-$00:02:09.7(3)  &  0.893355410120(6)  &  6.4624(7)  &  541.7(4)  &  59396.0  &  2.70  &  0.636  & 6.34  &  12.39  &  32.55 \\
J1850$-$0020  &J1850$-$0020g: gpps0018 & 66.3  &  18:50:01.658(5)  &  $-$00:20:37.2(2)  &  1.574737324031(7)  &  24.0214(8)  &  602.8(2)  &  59331.0  &  2.34  &  0.380  & 6.02  &  12.79  &  32.39 \\
J1852$+$0018  &J1852$+$0018g: gpps0103 & 45.1  &  18:52:26.742(3)  &  $+$00:18:42.4(1)  &  0.318762753335(1)  &  27.09983(8)  &  454.8(1)  &  59228.0  &  2.78  &  0.218  & 5.27  &  12.47  &  34.52 \\
J1852$-$0024  &J1852$-$0024g: gpps0042 & 58.2  &  18:52:14.1296(8)  &  $-$00:24:17.67(3)  &  0.3554290457076(3)  &  0.16460(2)  &  292.31(8)  &  59438.0  &  2.82  &  0.049  & 7.53  &  11.39  &  32.16 \\
J1852$-$0033  &J1852$-$0033g: gpps0173 & 22.7  &  18:52:26.526(3)  &  $-$00:33:39.71(9)  &  1.369012367933(4)  &  7.9154(3)  &  321.6(1)  &  59676.0  &  2.30  &  0.216  & 6.44  &  12.52  &  32.09 \\
J1852$-$0039  &J1852$-$0040g: gpps0177 & 79.7  &  18:52:13.195(5)  &  $-$00:39:44.8(2)  &  0.802905417851(4)  &  0.3783(3)  &  354.0(3)  &  59676.0  &  2.30  &  0.368  & 7.53  &  11.75  &  31.46 \\
J1853$+$0014  &J1853$+$0013g: gpps0030 & 35.2  &  18:53:52.502(3)  &  $+$00:14:12.8(5)  &  0.928595784010(3)  &  1.1422(3)  &  309.6(2)  &  59490.0  &  3.18  &  0.217  & 7.11  &  12.02  &  31.75 \\
J1853$+$0021  &J1853$+$0023g: gpps0089 & 19.4  &  18:53:07.748(5)  &  $+$00:21:31.2(4)  &  0.576939240923(6)  &  0.0650(4)  &  205.6(3)  &  59228.0  &  2.78  &  0.378  & 8.15  &  11.29  &  31.13 \\
J1855$+$0234  &J1855$+$0235g: gpps0148 & 14.9  &  18:55:52.374(2)  &  $+$02:34:55.58(6)  &  0.983029834436(3)  &  0.0215(4)  &  103.71(6)  &  59447.0  &  1.59  &  0.109  & 8.86  &  11.17  &  29.95 \\
J1856$+$0211  &J1856$+$0211g: gpps0158 & 32.4  &  18:56:59.42(1)  &  $+$02:11:10.9(4)  &  9.89009157558(8)  &  0.78(1)  &  124.5(5)  &  59607.0  &  2.64  &  0.719  & 8.30  &  12.45  &  28.50 \\
J1857$+$0214  &J1857$+$0214g: gpps0010 & 59.6  &  18:57:07.406(8)  &  $+$02:14:37.2(3)  &  0.333917722223(2)  &  1.6059(2)  &  982.0(7)  &  59427.0  &  3.63  &  0.899  & 6.52  &  11.87  &  33.23 \\
J1857$+$0224  &J1857$+$0224g: gpps0072 & 35.2  &  18:57:10.706(5)  &  $+$02:24:43.2(2)  &  0.875874011838(6)  &  0.0615(5)  &  398.0(5)  &  59410.0  &  2.74  &  0.319  & 8.35  &  11.37  &  30.56 \\
J1859$+$0430  &J1859$+$0430g: gpps0015 & 18.5  &  18:59:08.750(5)  &  $+$04:30:18.7(1)  &  0.336324930220(2)  &  0.053(1)  &  782.9(4)  &  59104.0  &  1.14  &  0.154  & 8.00  &  11.13  &  31.74 \\
J1903$+$0845  &J1903$+$0845g: gpps0048 & 30.4  &  19:03:39.6082(9)  &  $+$08:45:58.00(2)  &  0.1531502678628(1)  &  0.091961(7)  &  129.56(6)  &  59503.0  &  3.10  &  0.085  & 7.42  &  11.08  &  33.00 \\
J1903$+$0851  &J1903$+$0851g: gpps0101 & 52.9  &  19:03:04.53(2)  &  $+$08:51:28.4(1)  &  1.23188198312(1)  &  27.1523(5)  &  83.0(1)  &  59567.0  &  2.75  &  0.554  & 5.86  &  12.77  &  32.76 \\
J1904$+$0853  &J1904$+$0852g: gpps0009 & 56.8  &  19:04:55.29234(9)  &  $+$08:53:09.823(2)  &  0.0061974296377438(4)  &  0.00001009(3)  &  195.159(3)  &  59355.0  &  3.35  &  0.007  & 9.99  &   8.40  &  33.22 \\
J1905$+$0656  &J1905$+$0656g: gpps0008 & 60.7  &  19:05:44.343(6)  &  $+$06:56:35.0(2)  &  2.51176254691(2)  &  0.387(2)  &  24.2(2)  &  59267.0  &  2.10  &  0.381  & 8.01  &  12.00  &  29.98 \\
J1905$+$0758  &J1905$+$0758g: gpps0043 & 35.2  &  19:05:50.964(3)  &  $+$07:58:52.62(5)  &  1.192648616445(2)  &  0.0459(2)  &  197.1(1)  &  59429.0  &  2.83  &  0.149  & 8.61  &  11.37  &  30.03 \\
J1905$+$0935  &J1905$+$0936g: gpps0153 & 60.8  &  19:05:26.088(4)  &  $+$09:35:59.41(9)  &  1.634497451794(4)  &  1.5008(4)  &  417.96(4)  &  59650.0  &  2.30  &  0.141  & 7.24  &  12.20  &  31.13 \\
J1906$+$0757  &J1906$+$0757g: gpps0102 & 28.9  &  19:06:41.7449(3)  &  $+$07:57:01.842(4)  &  0.05718905678761(1)  &  0.000116(1)  &  79.104(6)  &  59430.0  &  2.83  &  0.011  & 9.89  &   9.42  &  31.39 \\
J1917$+$0748  &J1916$+$0748g: gpps0175 & 11.9  &  19:17:00.899(8)  &  $+$07:48:43.9(2)  &  0.867880485703(5)  &  0.1981(6)  &  153.6(2)  &  59607.0  &  2.54  &  0.353  & 7.84  &  11.62  &  31.08 \\
J1917$+$0743  &J1917$+$0743g: gpps0167 & 32.8  &  19:17:25.912(2)  &  $+$07:43:16.69(4)  &  0.8134532288005(8)  &  0.04485(8)  &  198.39(4)  &  59652.0  &  2.29  &  0.072  & 8.46  &  11.29  &  30.52 \\
J1926$+$1631  &J1926$+$1631g: gpps0053 & 78.6  &  19:26:23.9083(6)  &  $+$16:31:27.78(1)  &  0.6783666636602(3)  &  0.03267(2)  &  196.73(3)  &  59487.0  &  3.21  &  0.050  & 8.52  &  11.18  &  30.62 \\
J1928$+$1853  &J1928$+$1852g: gpps0143 & 27.9  &  19:28:03.448(6)  &  $+$18:53:04.9(1)  &  0.792806958052(3)  &  0.0743(2)  &  290.0(3)  &  59631.0  &  2.40  &  0.250  & 8.23  &  11.39  &  30.77 \\
J1930$+$1357  &J1930$+$1357g: gpps0050 & 16.0  &  19:30:11.991(2)  &  $+$13:57:06.43(3)  &  0.3235617566079(4)  &  0.02723(3)  &  184.8(1)  &  59420.0  &  3.35  &  0.110  & 8.28  &  10.98  &  31.50 \\
J2017$+$2819  &J2017$+$2819g: gpps0044 & 61.0  &  20:17:23.297(2)  &  $+$28:19:14.26(3)  &  1.832558917299(6)  &  0.3580(4)  &  66.0(2)  &  58997.0  &  2.34  &  0.074  & 7.91  &  11.91  &  30.36 \\
  \noalign{\smallskip}\hline
\end{tabular}
\end{center}
\end{table}
\end{landscape}

One possible formation scenario for such an isolated MSP not in a globular cluster is that the companion was ablated by the wind from the MSP, as is happening in PSR B1957+20 \citep{fst+1988}. However, some recent studies show that the evaporation time-scale is too long to produce these isolated MSPs within the Hubble time \citep{Chen2013,lzw+2017}, and many novel mechanisms have been proposed \citep{pvv+2011,nzg+2019,jwc+2020}. The formation mechanism of isolated MSPs remains an unsolved problem.

\subsubsection{A disrupted recycled pulsar: PSR J1906+0757}

PSR J1906+0757 has a period of 57 ms, and could be a young pulsar or an old pulsar. With many new  TOAs for this pulsar, we obtain the period derivative of $1.16 \times 10^{-19}~\rm s~s^{-1}$, indicating that it is an old pulsar, with a characteristic age of $7.8 \times 10^{9}$~yr and surface magnetic field of $2.6 \times 10^{9}$~G. 

Such a weakly magnetized isolated pulsar seems to be an example of a disrupted recycled pulsar \citep{2004MNRAS.347L..21L}. In high-mass X-ray binary systems, after the neutron star has accreted matter from the companion, its spin rate increases, and the magnetic fields are buried out. Thereafter, the binary system is disrupted in the following supernova explosion, and the recycled pulsar becomes isolated. Typically, such isolated pulsars  \citep{2010MNRAS.407.1245B} have a spin period in the range 20–100 ms and a magnetic field weaker than $3 \times 10^{10}$~G. The parameters of PSR J1906+0757 are consistent with such expectations. However, some neutron stars are born with weak magnetic fields as anti-magnetars \citep{hg+2010,gha+2013}. Therefore it is hard to figure out the origin of such an isolated pulsar.

\subsubsection{A long-period pulsar: PSR J1856+0211}

PSR J1856+0211 has a spin period of 9.89 s and it could be a magnetar if the period derivative is large. With FAST observations, we obtained the period derivative of $7.8 \times 10^{-14}~\rm s~s^{-1}$, which implies a characteristic age of 201~Myr, a surface magnetic field strength of $2.8\times 10^{12}$~G, and a spin-down luminosity of $3.2\times10^{28}\ {\rm  erg\ s}^{-1}$. Obviously, it is not a magnetar but a very old pulsar that is located below the conventional death line, which poses a significant challenge to death-line models.  

\begin{figure*}
\includegraphics[width=0.24\textwidth]{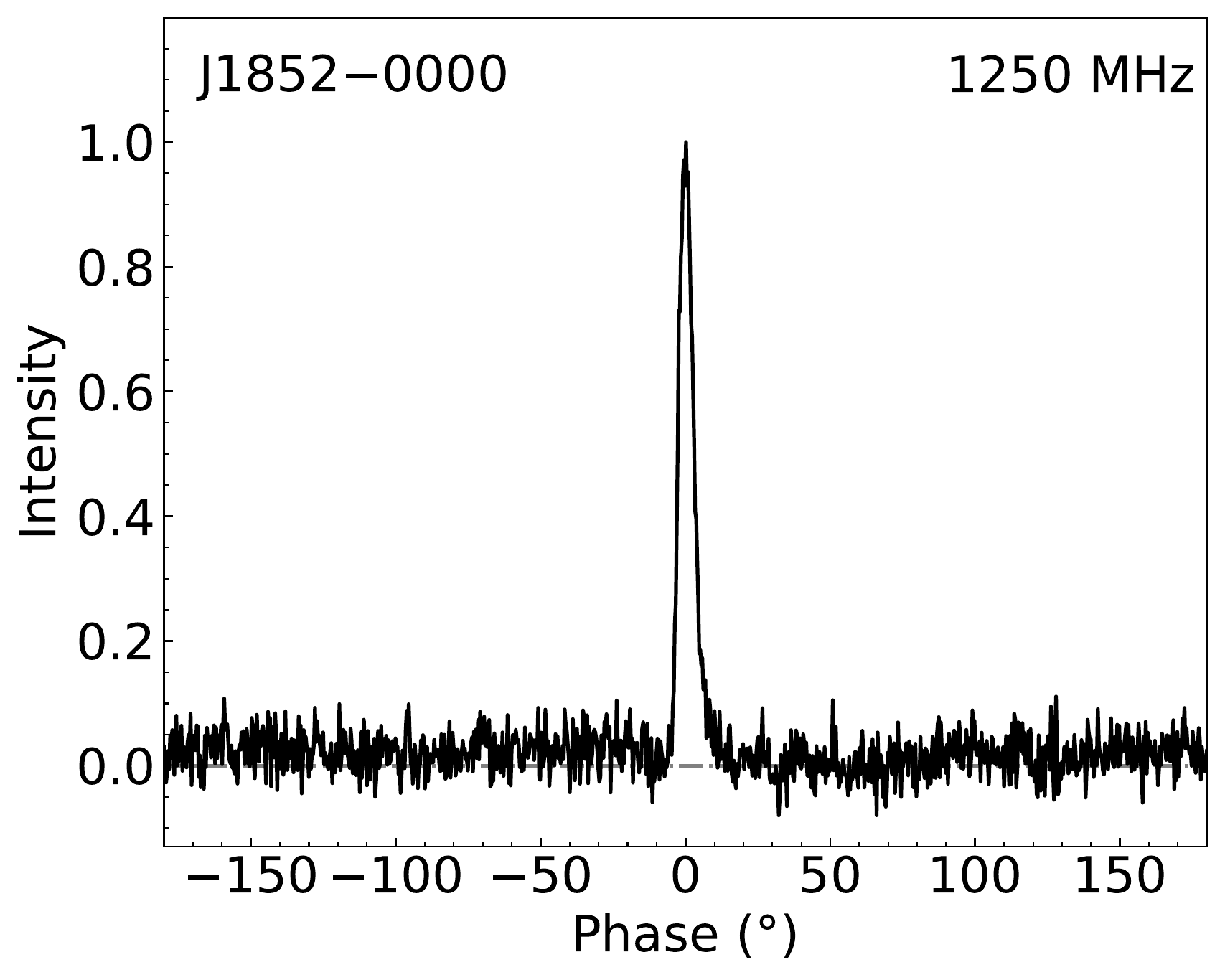}
\includegraphics[width=0.24\textwidth]{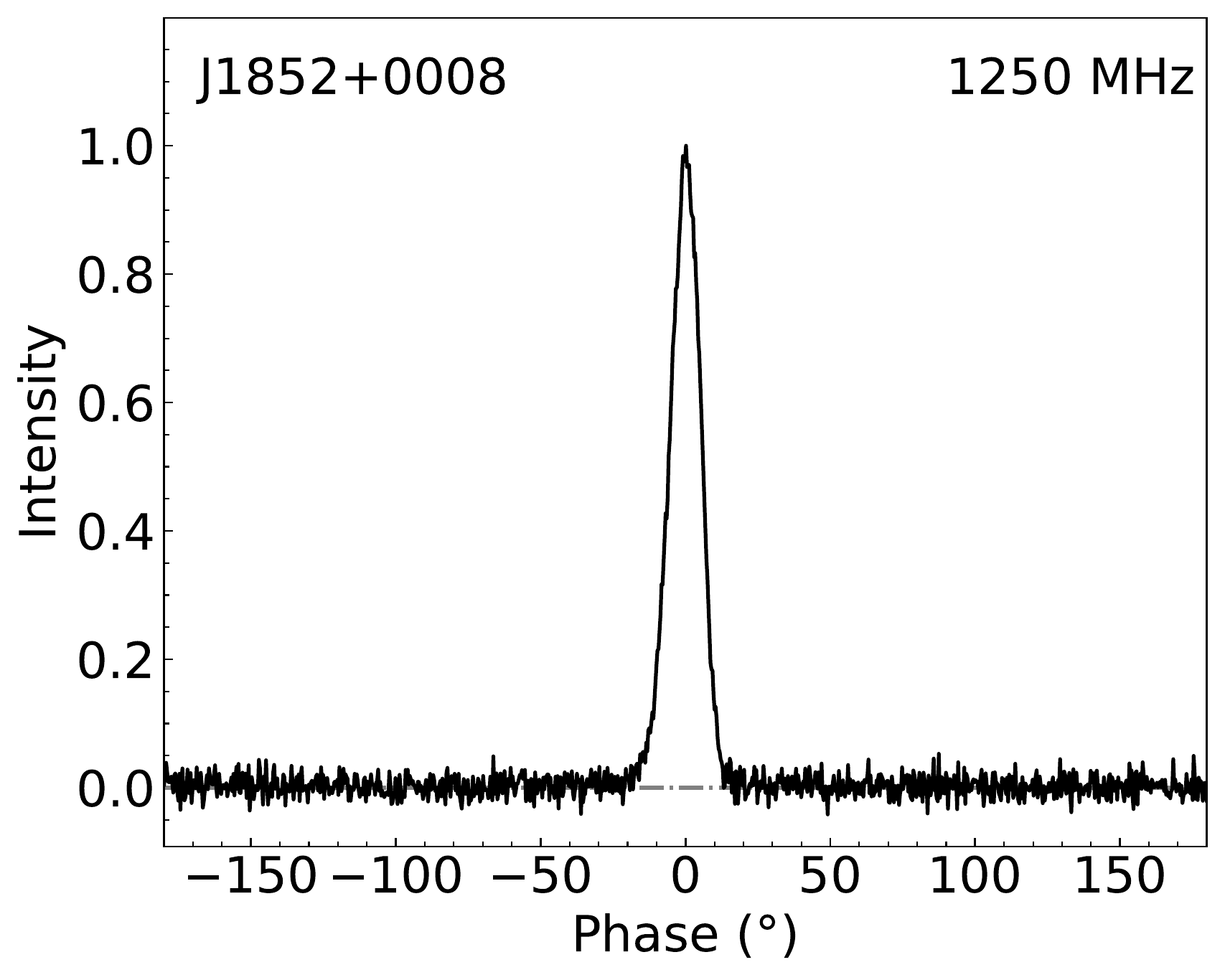}
\includegraphics[width=0.24\textwidth]{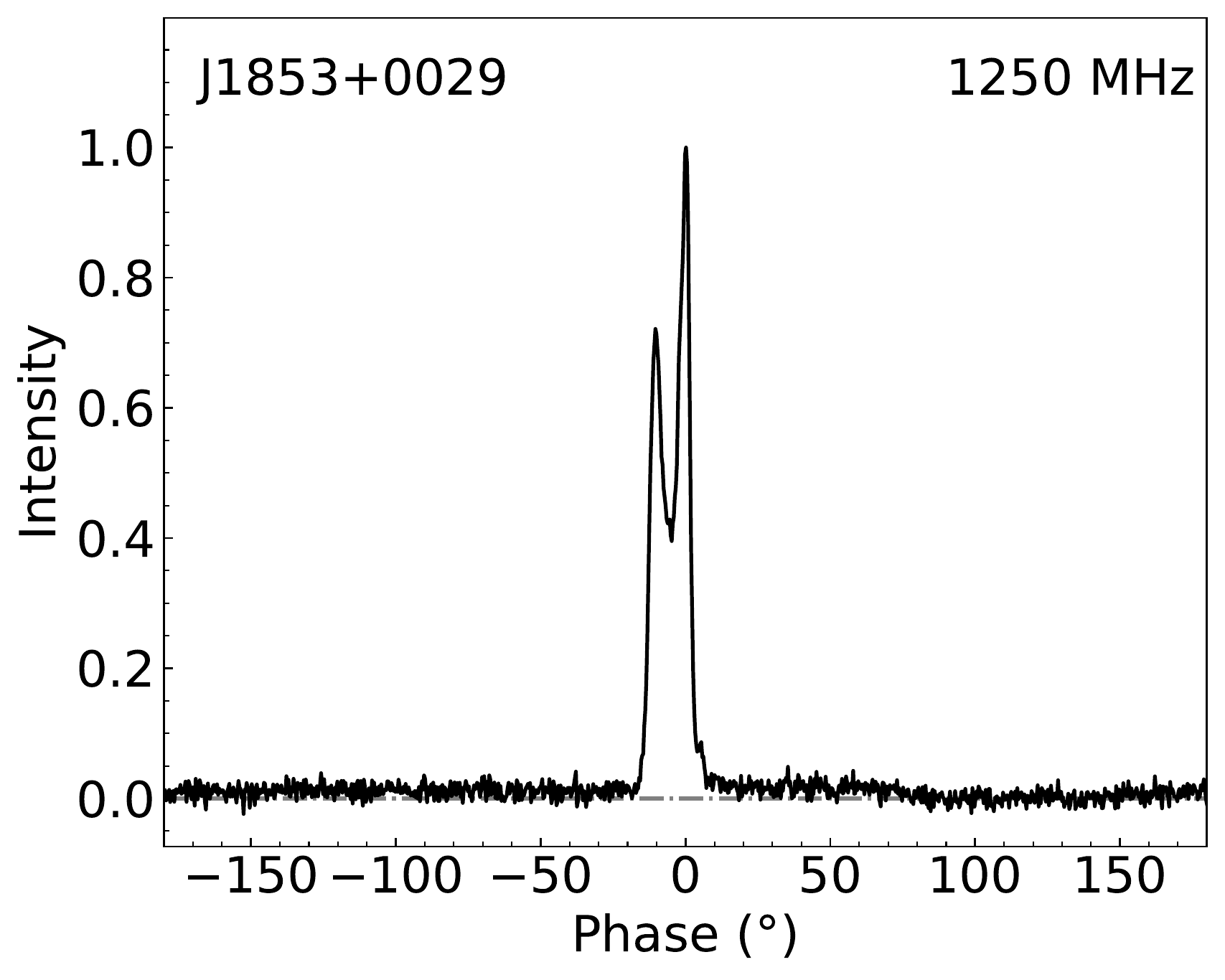}
\includegraphics[width=0.24\textwidth]{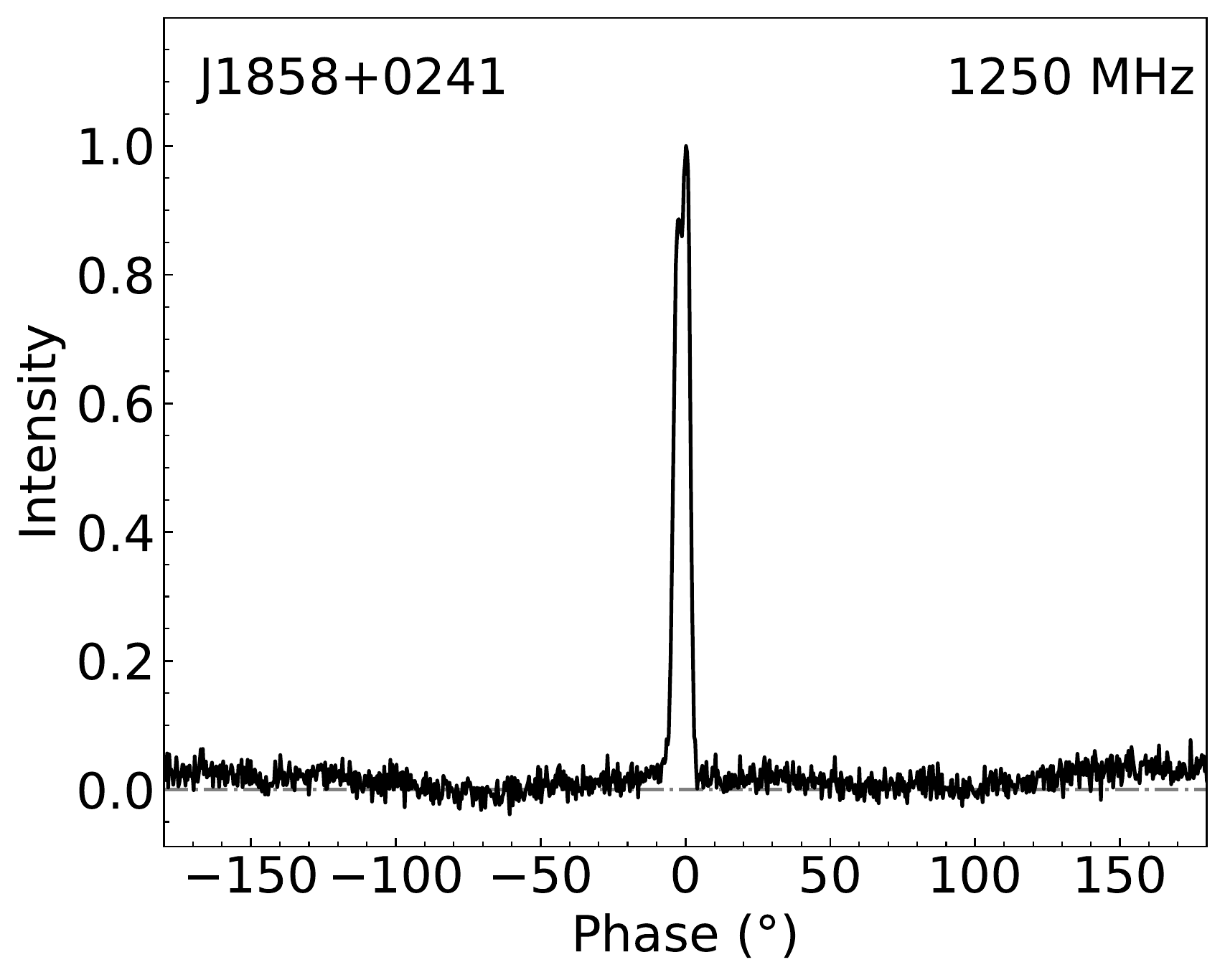}
\includegraphics[width=0.24\textwidth]{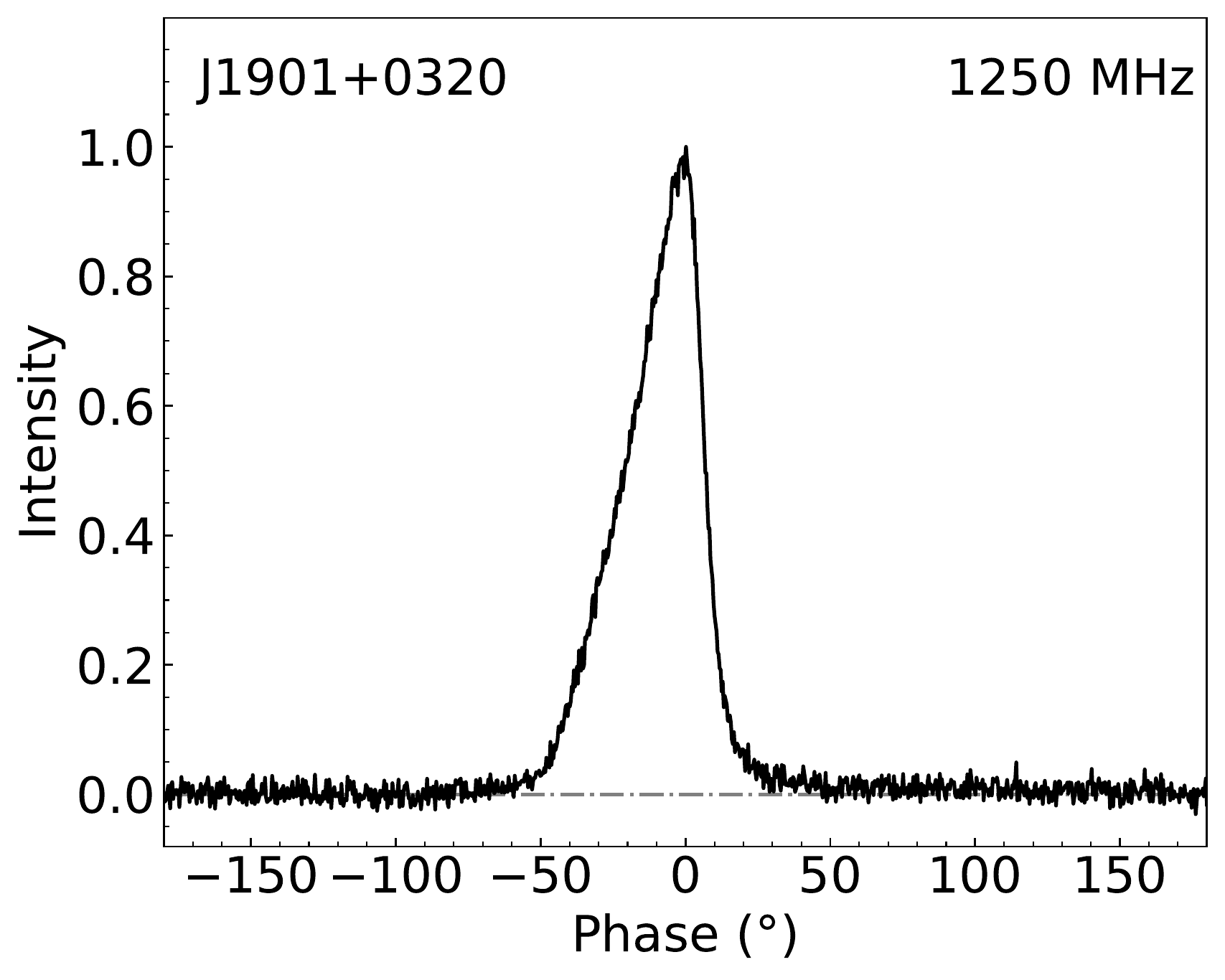}
\includegraphics[width=0.24\textwidth]{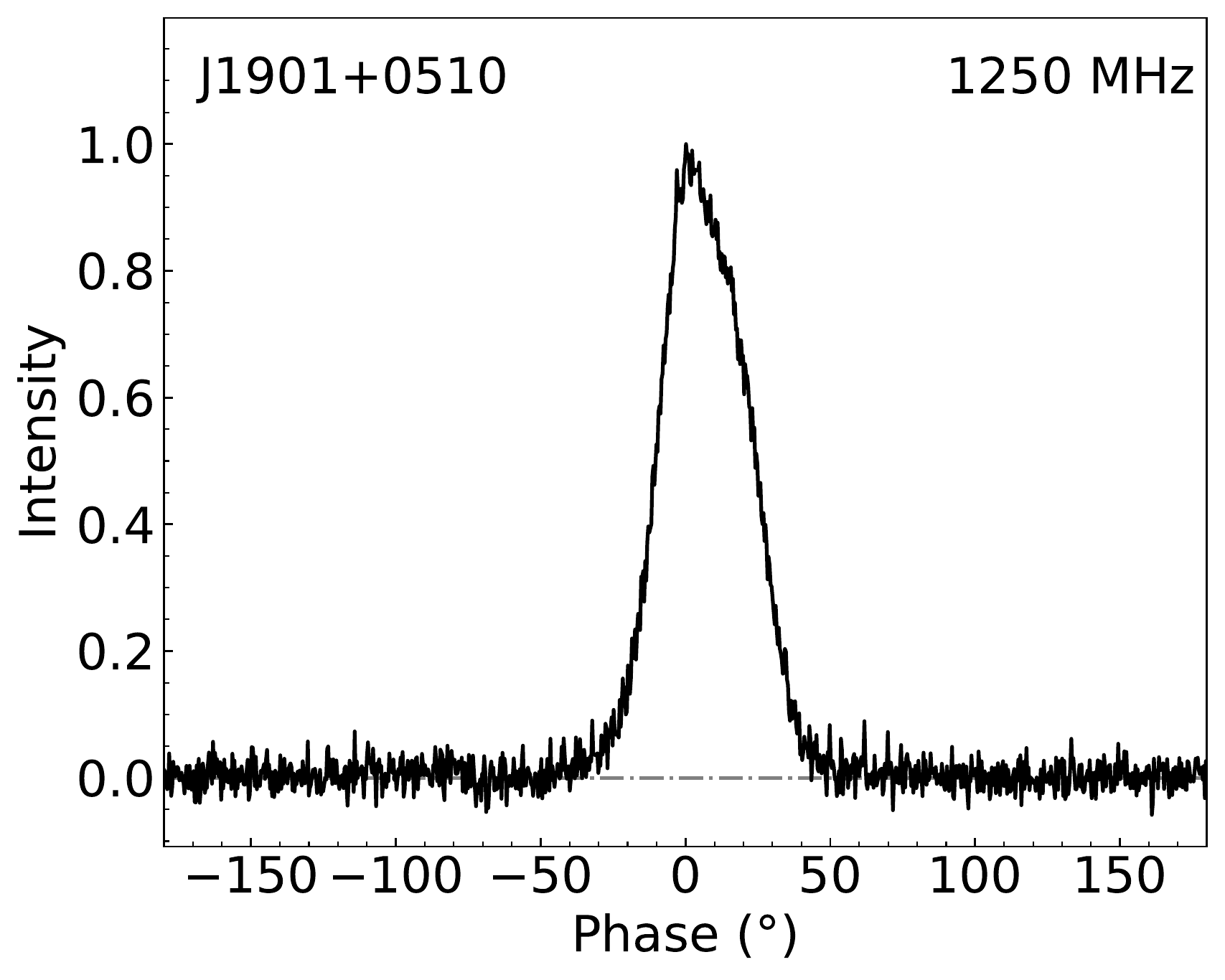}
\includegraphics[width=0.24\textwidth]{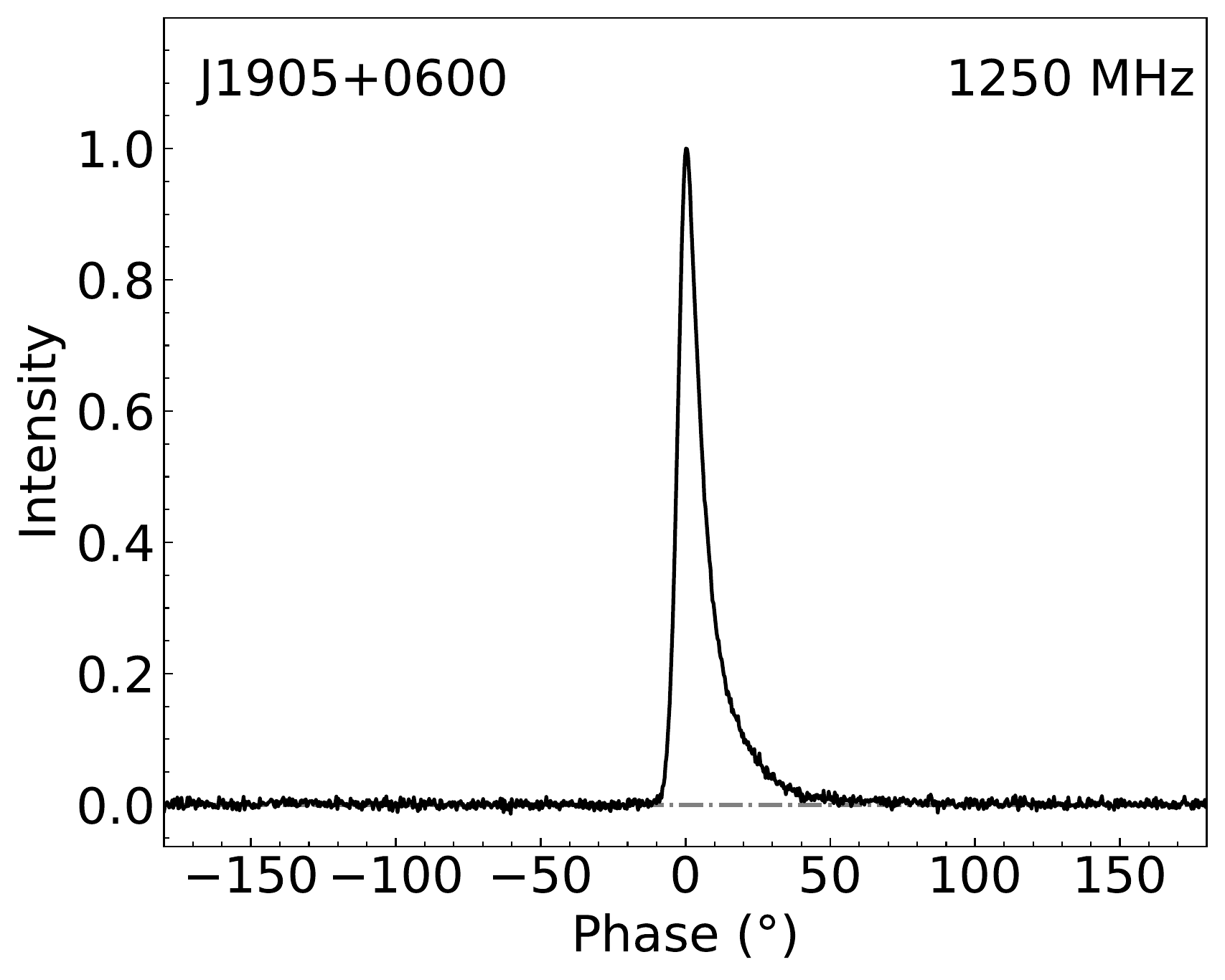}
\includegraphics[width=0.24\textwidth]{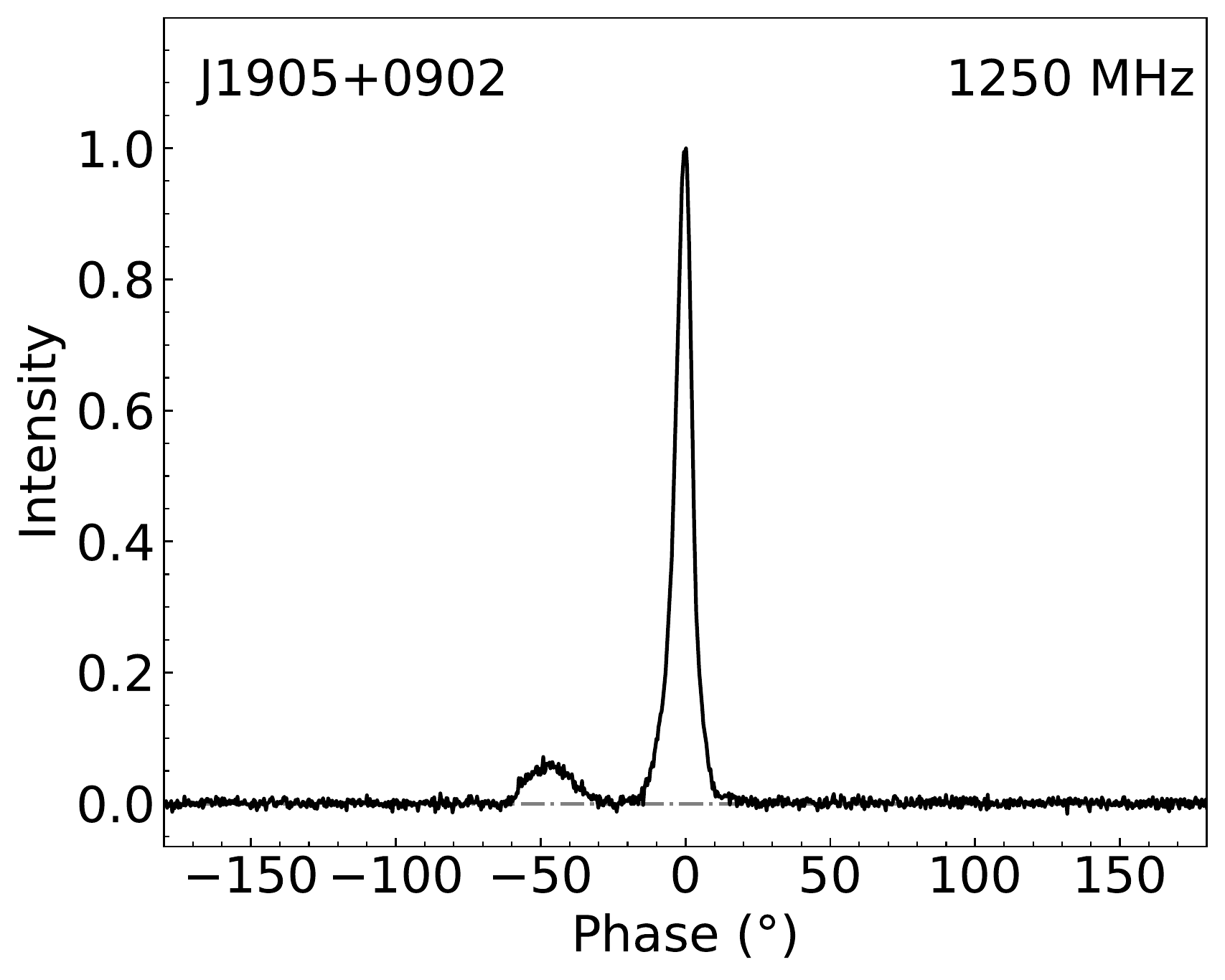}
\includegraphics[width=0.24\textwidth]{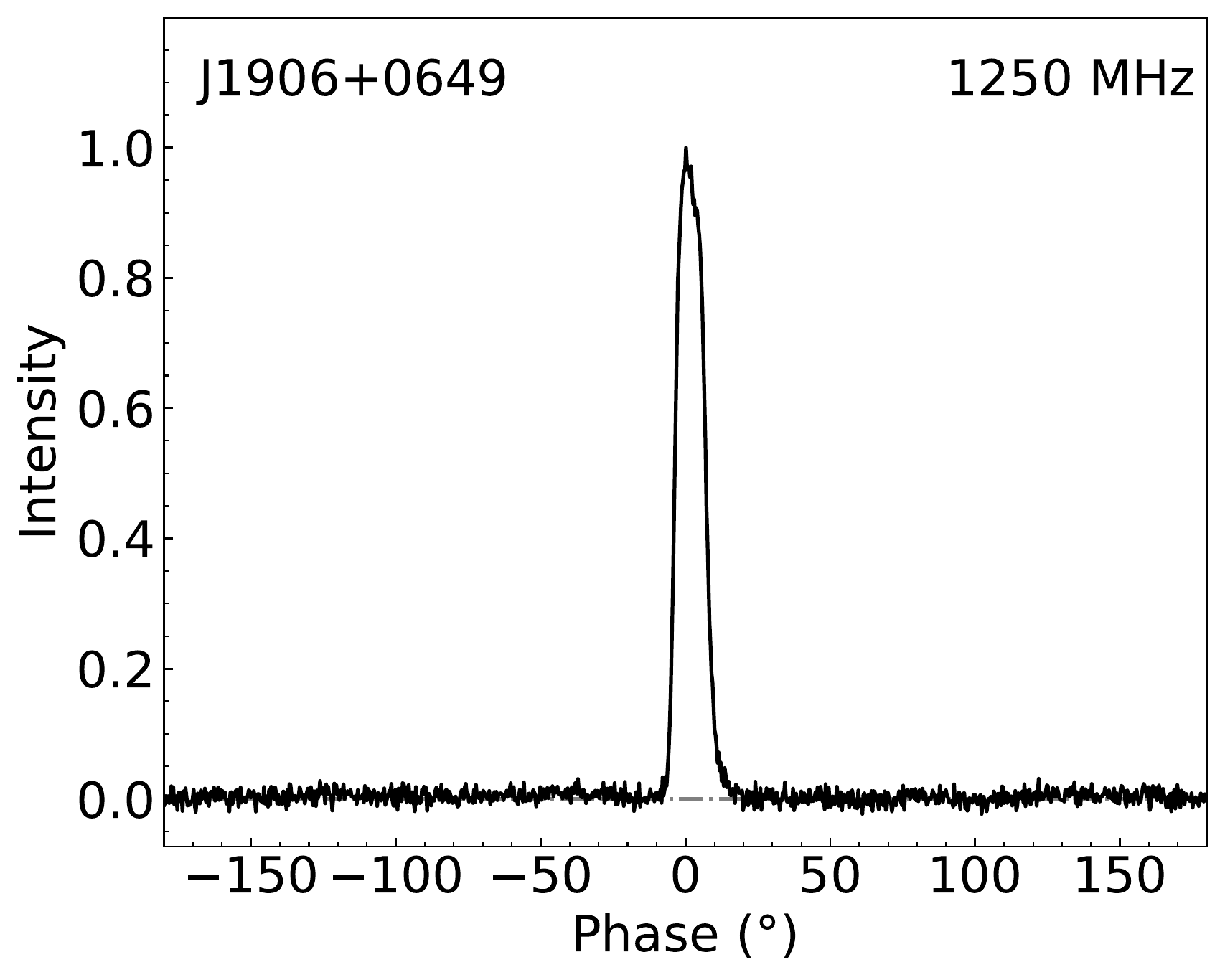}
\includegraphics[width=0.24\textwidth]{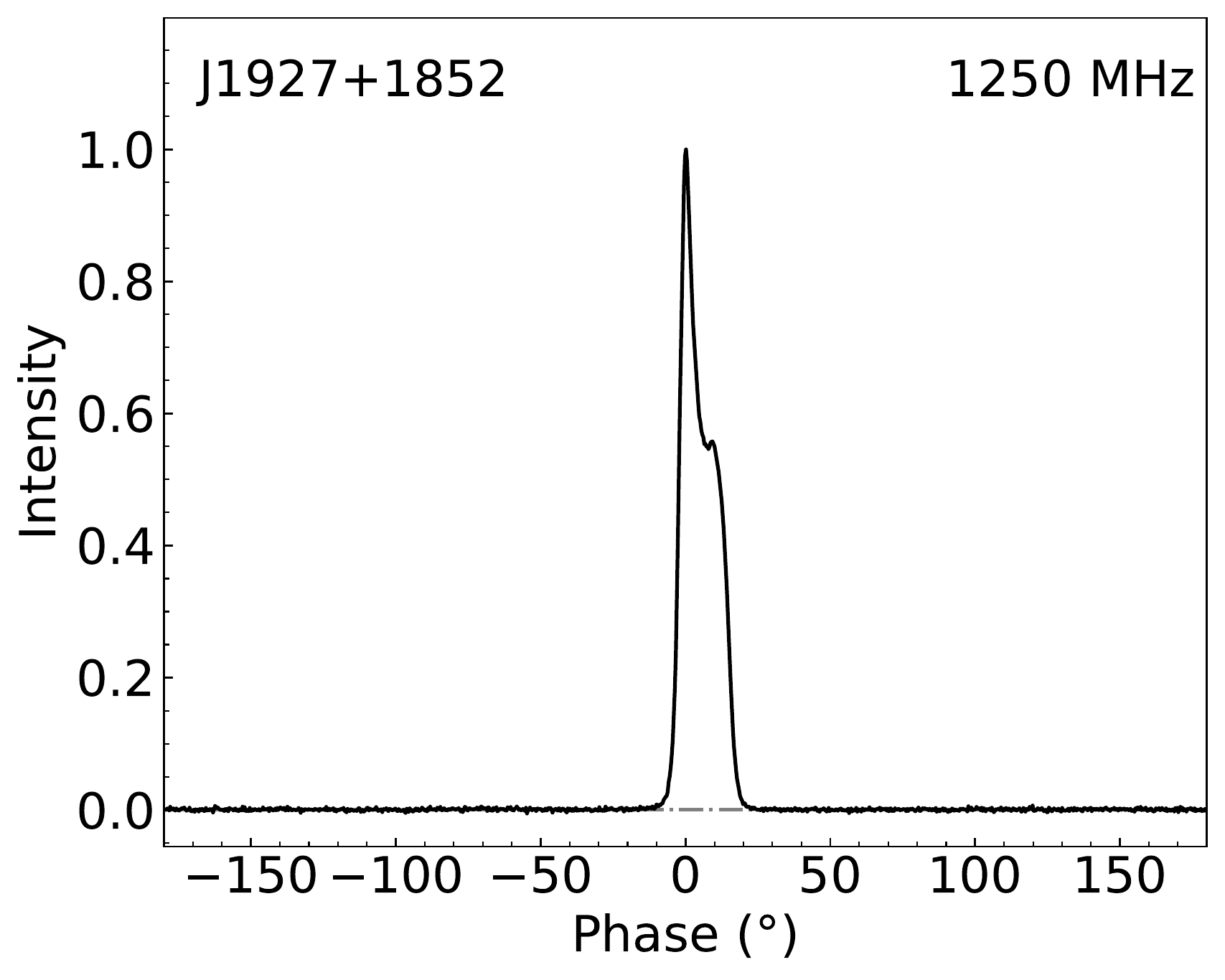}
\includegraphics[width=0.24\textwidth]{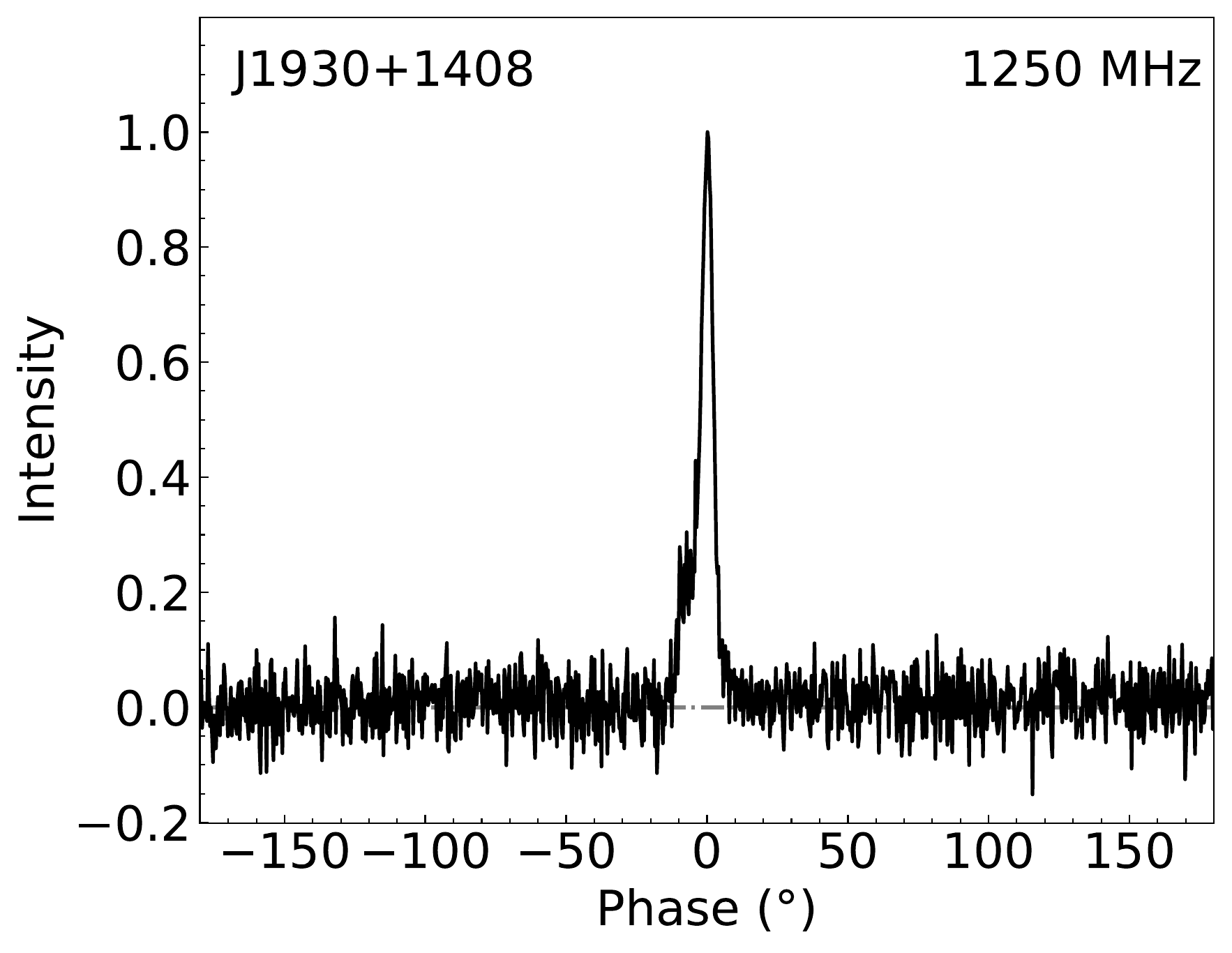}
\includegraphics[width=0.24\textwidth]{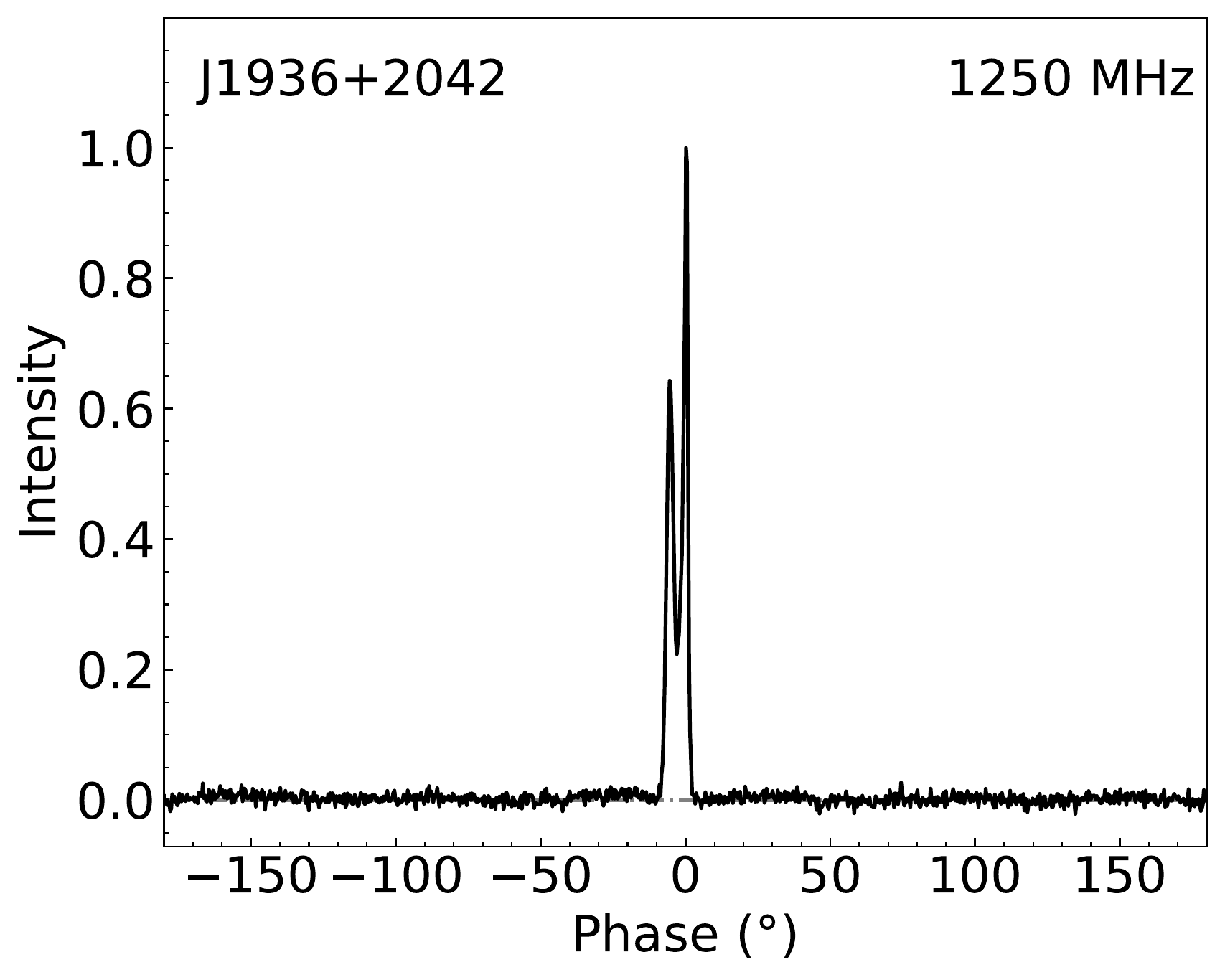}
\includegraphics[width=0.24\textwidth]{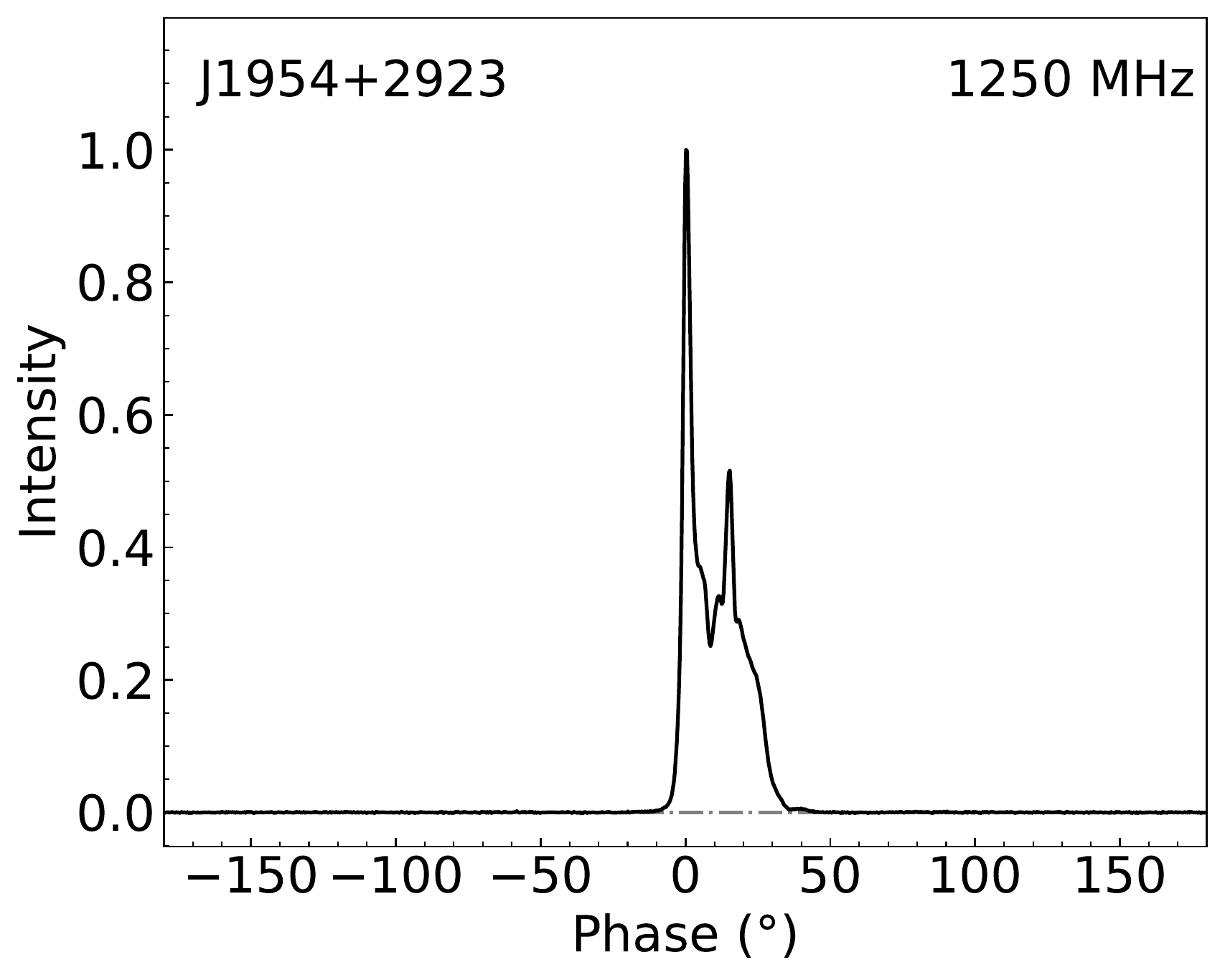}
\includegraphics[width=0.24\textwidth]{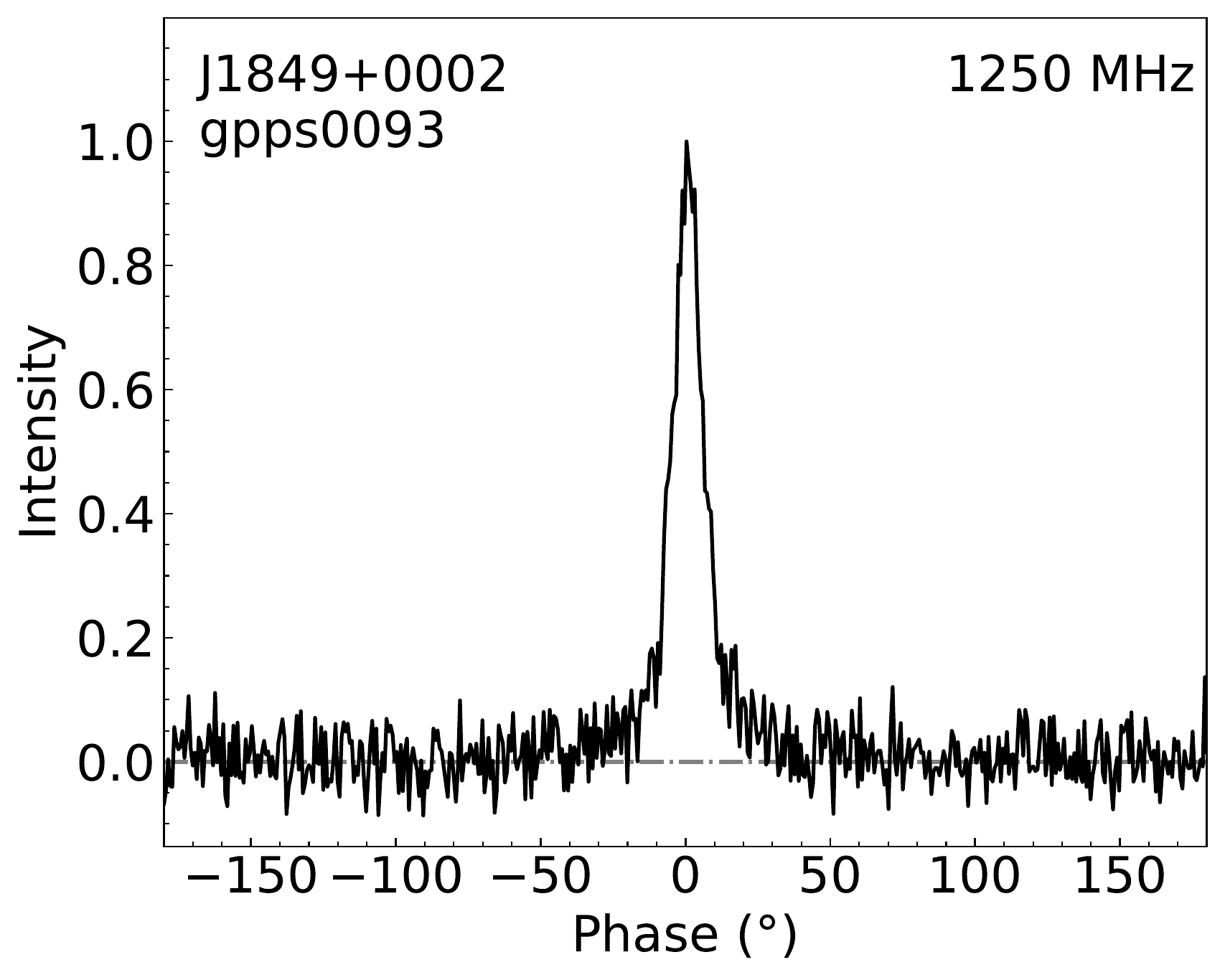}
\includegraphics[width=0.24\textwidth]{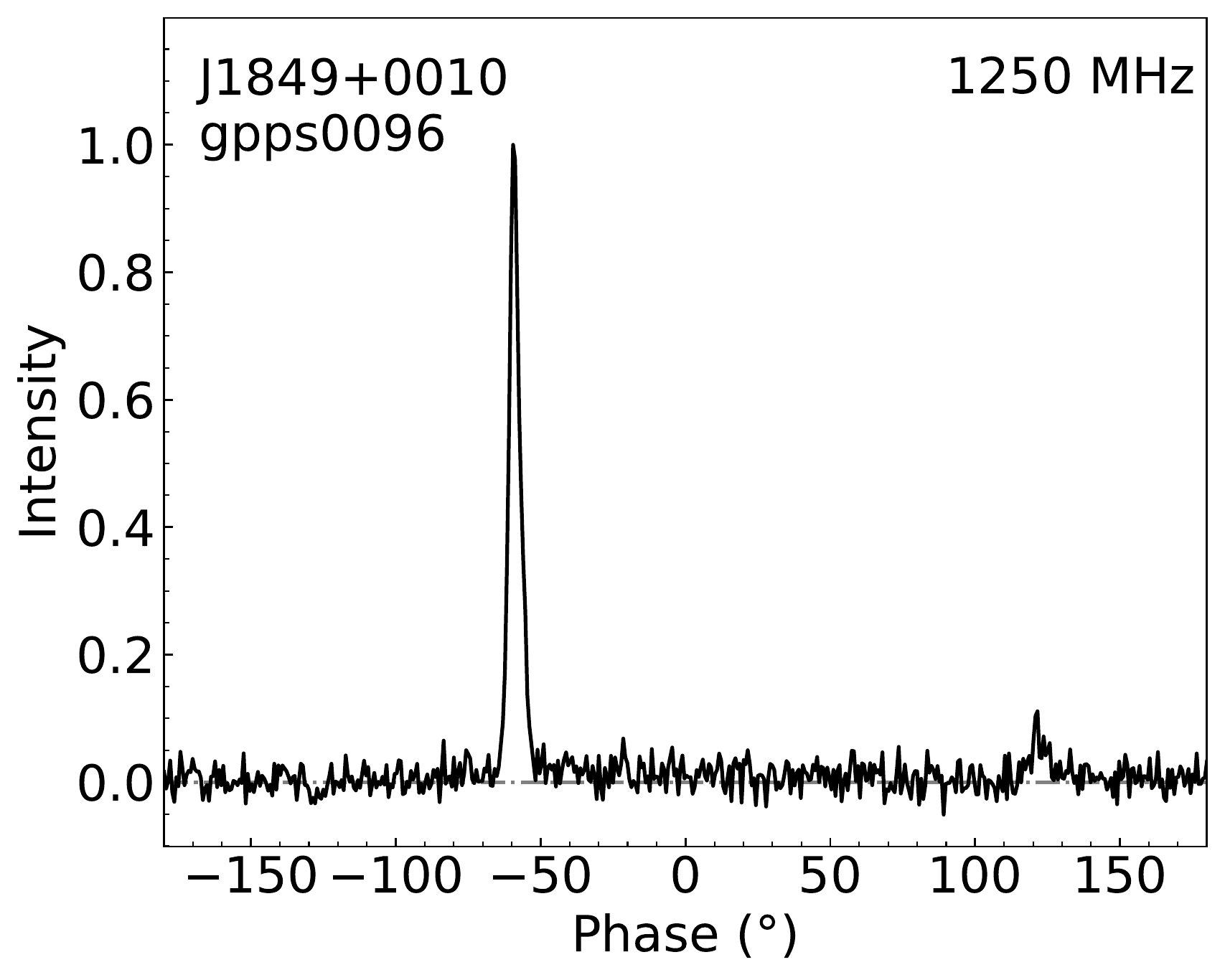}
\includegraphics[width=0.24\textwidth]{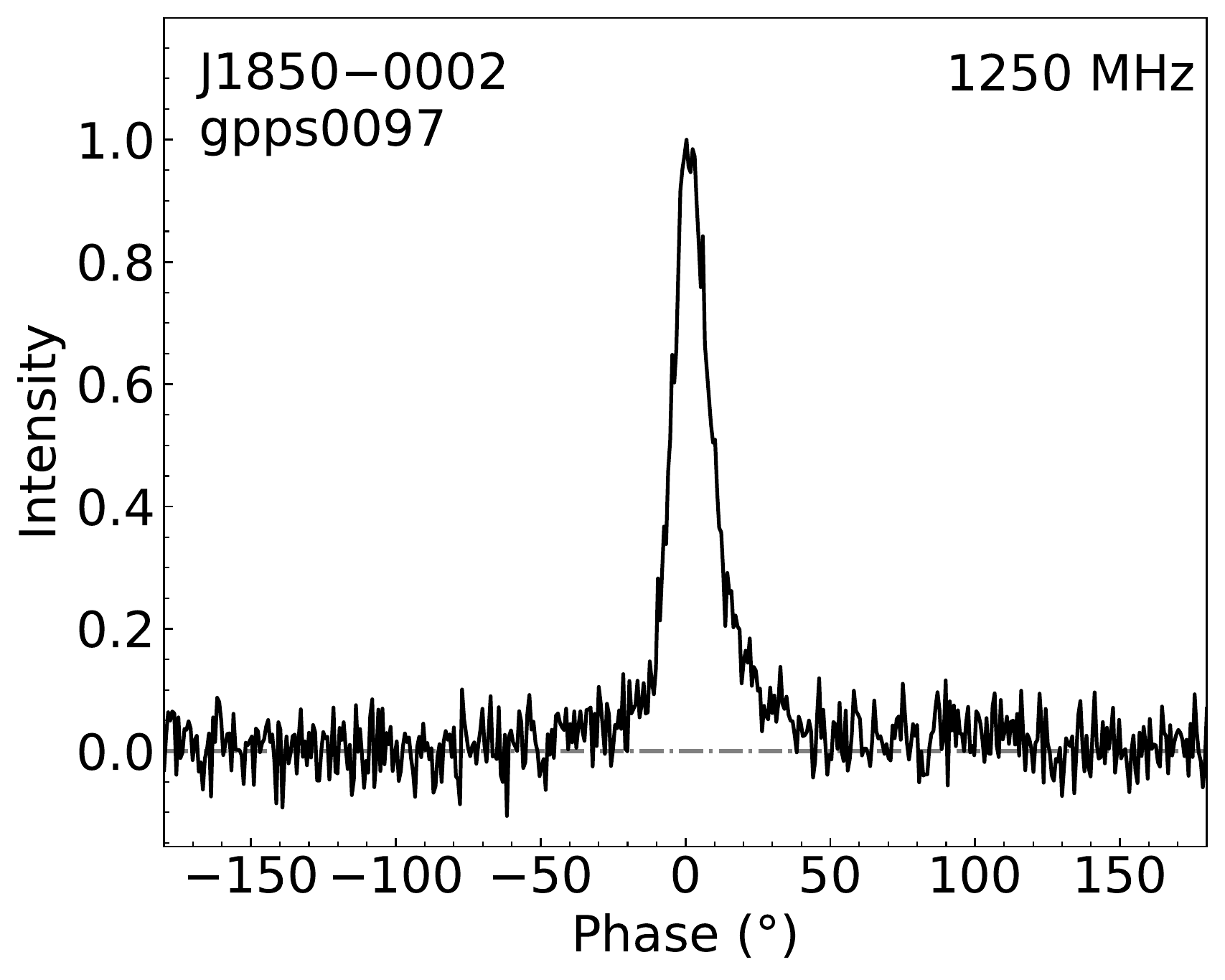}
\includegraphics[width=0.24\textwidth]{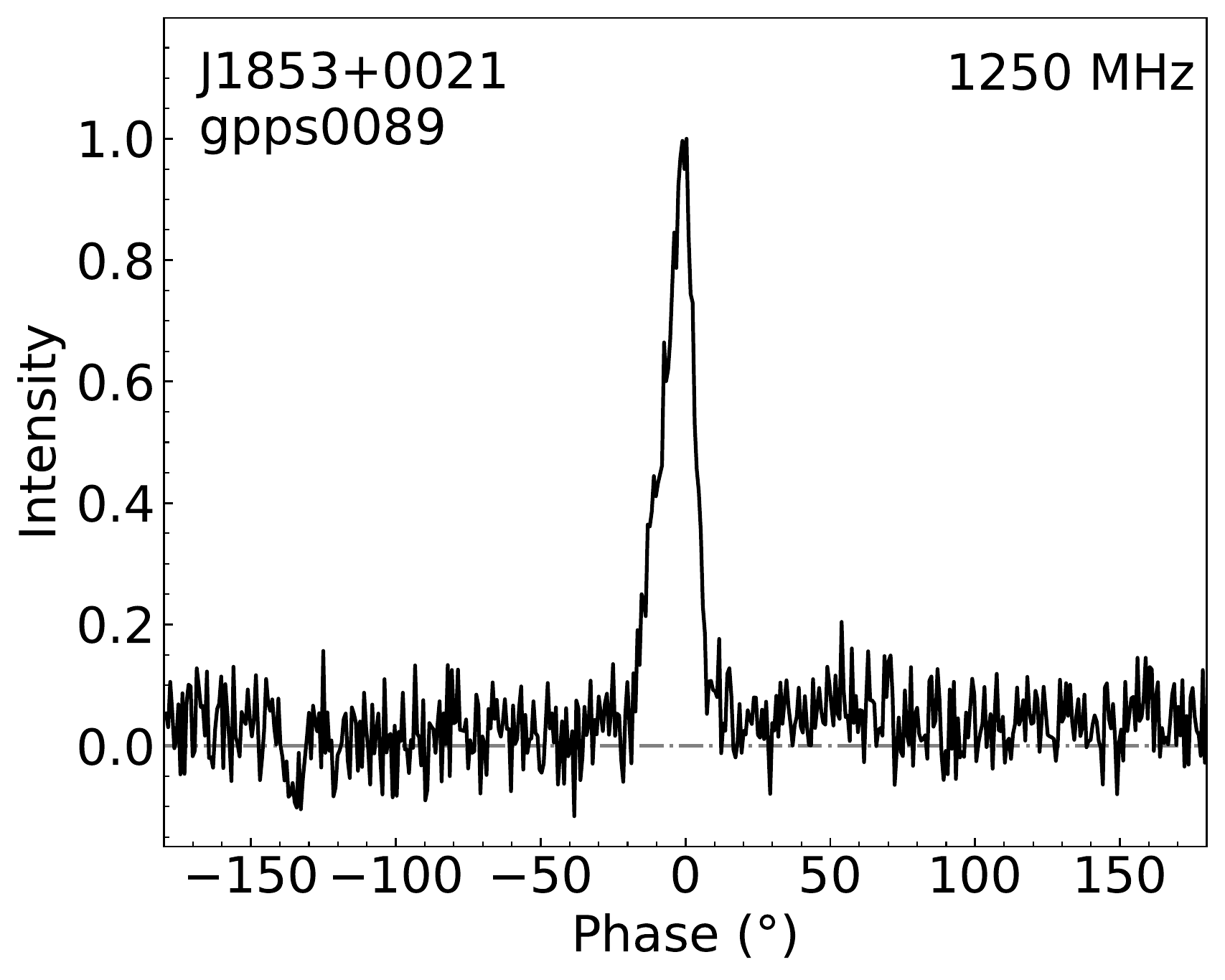}
\includegraphics[width=0.24\textwidth]{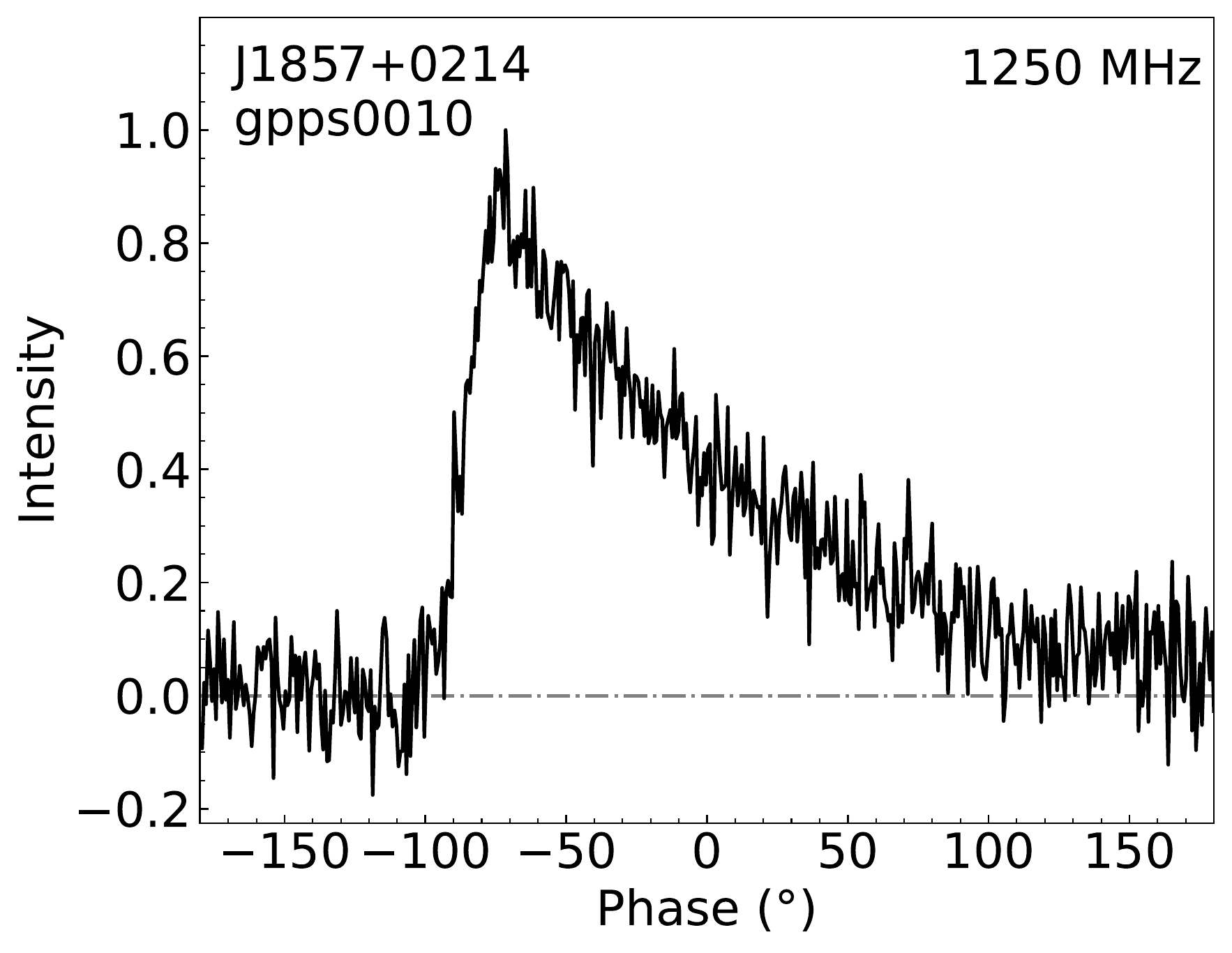}
\includegraphics[width=0.24\textwidth]{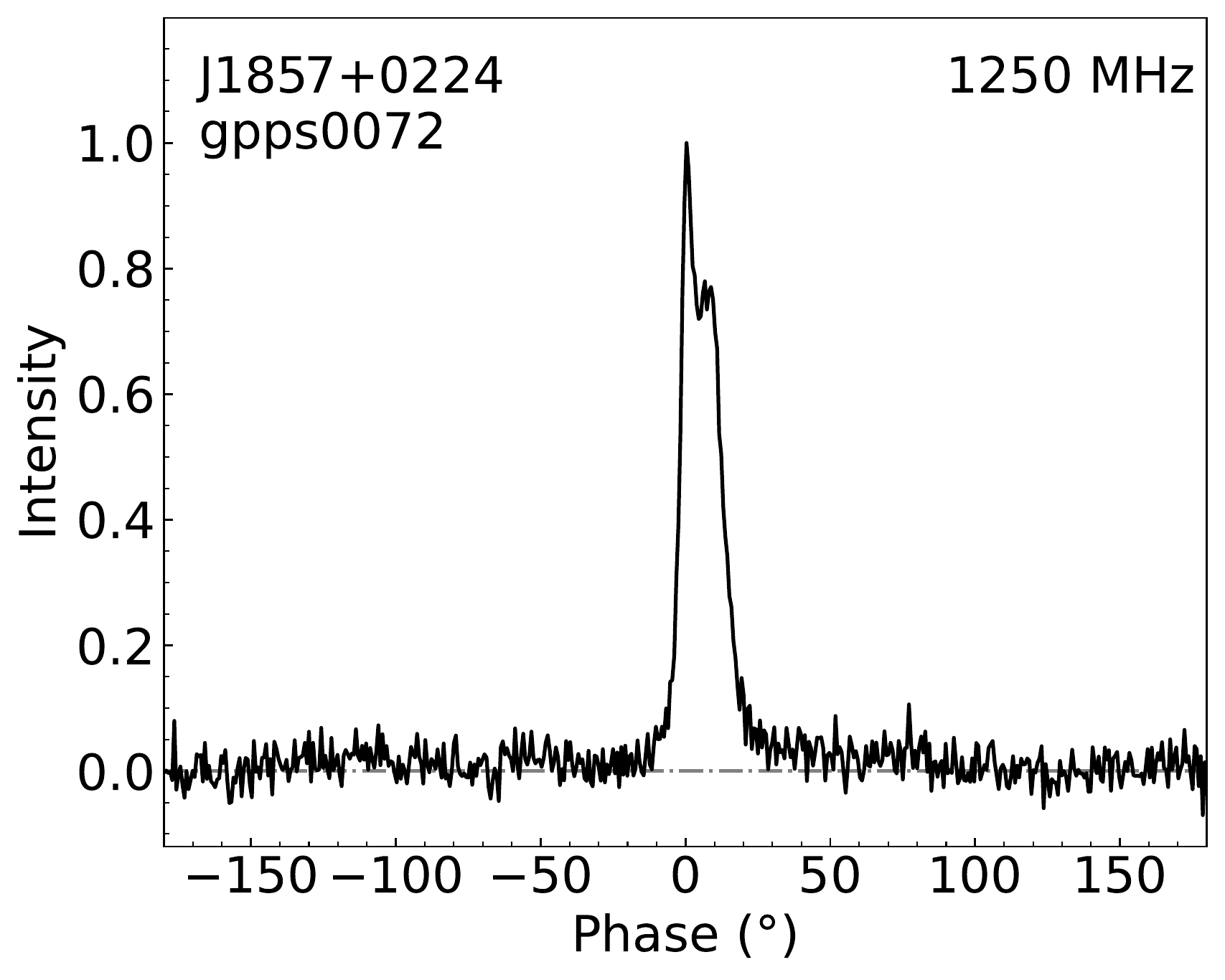}
\includegraphics[width=0.24\textwidth]{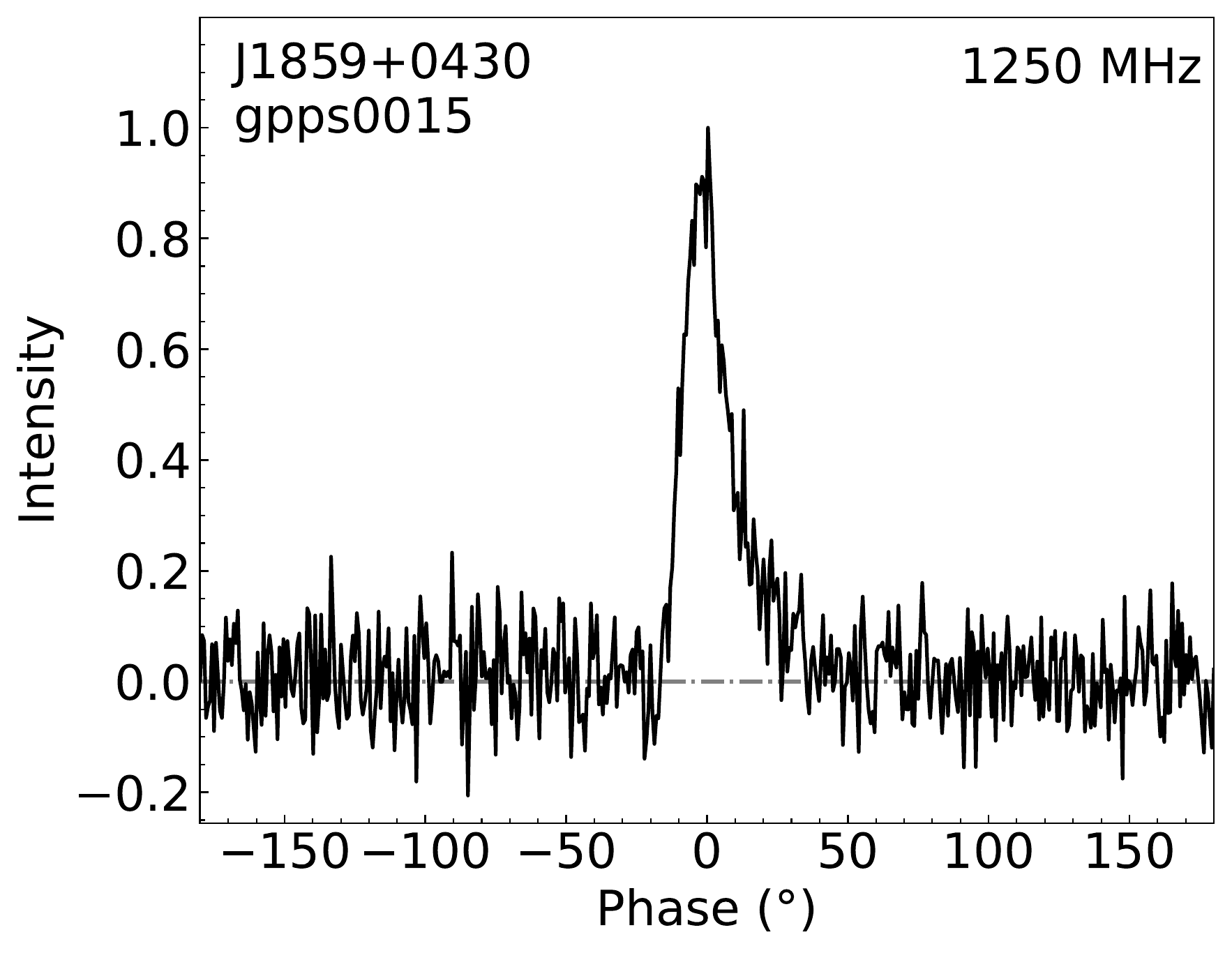}
\includegraphics[width=0.24\textwidth]{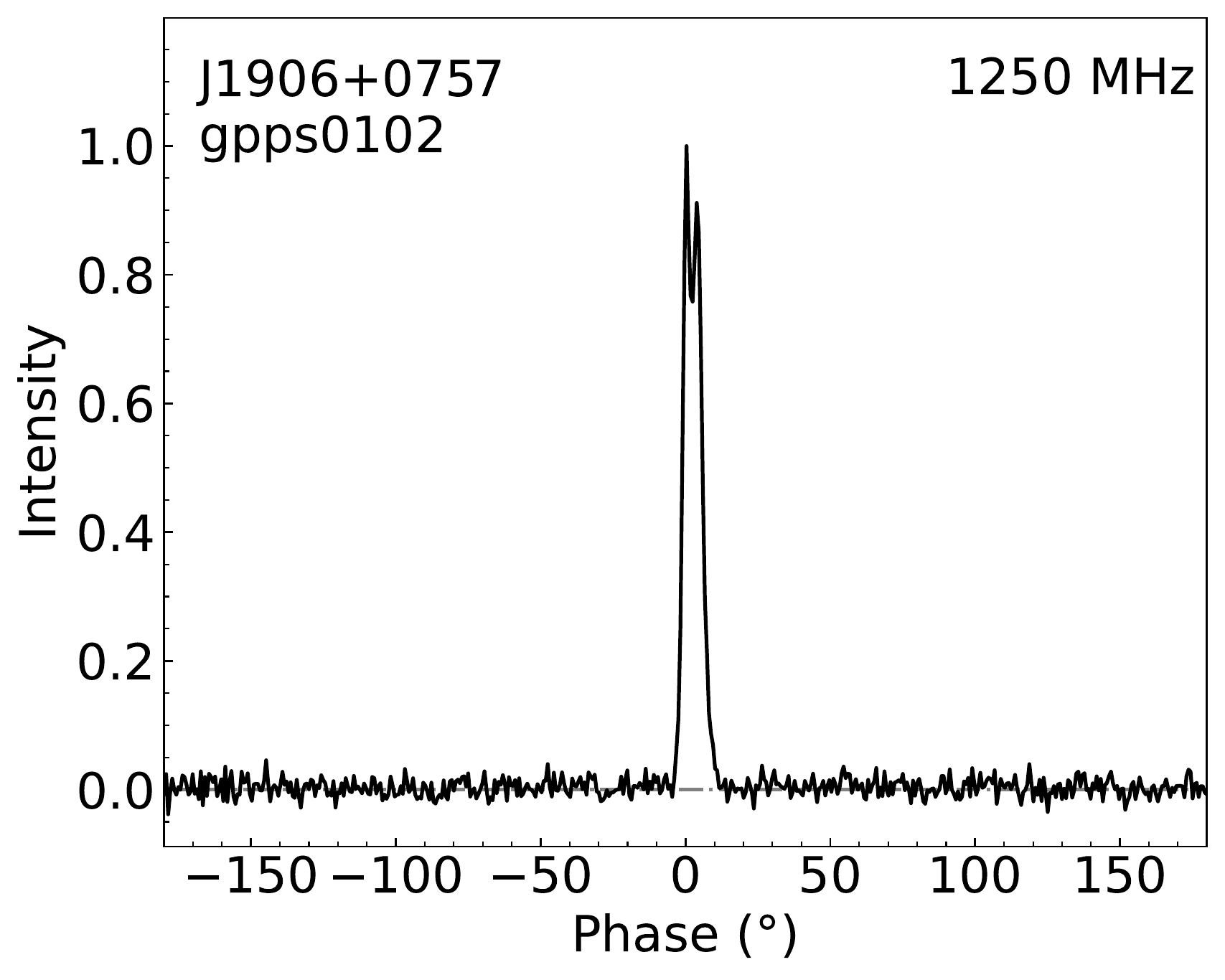}
\includegraphics[width=0.24\textwidth]{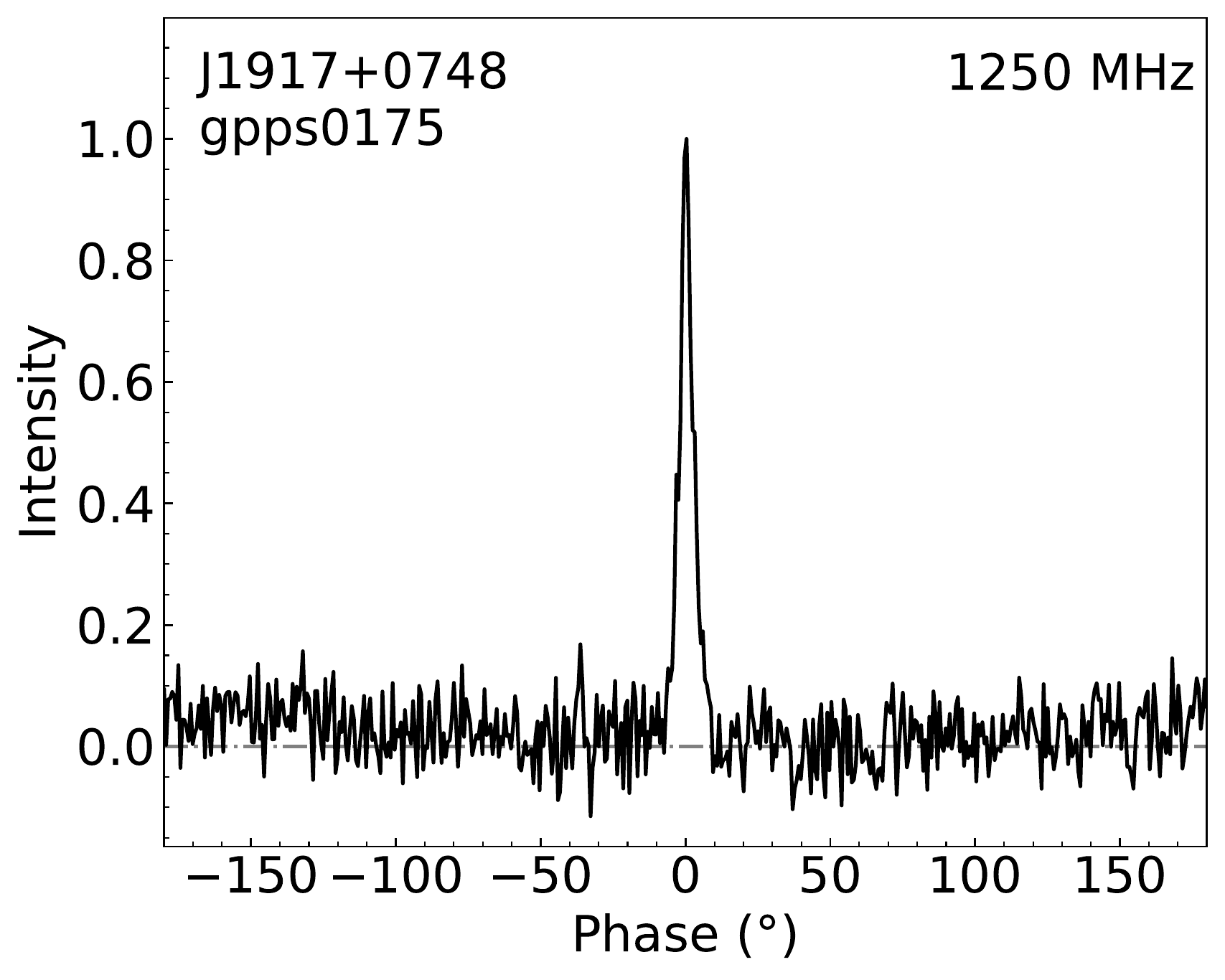}
\includegraphics[width=0.24\textwidth]{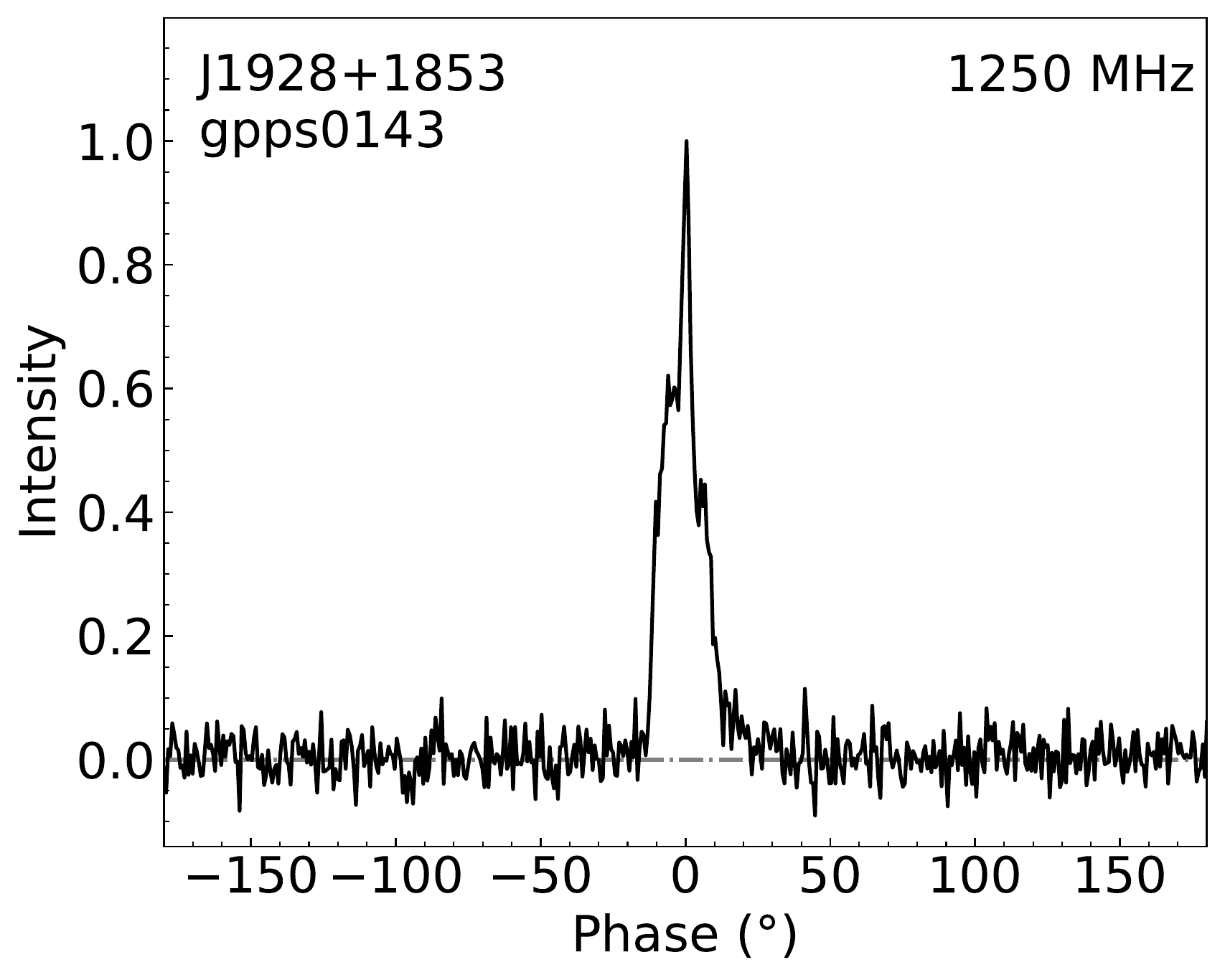}
\includegraphics[width=0.24\textwidth]{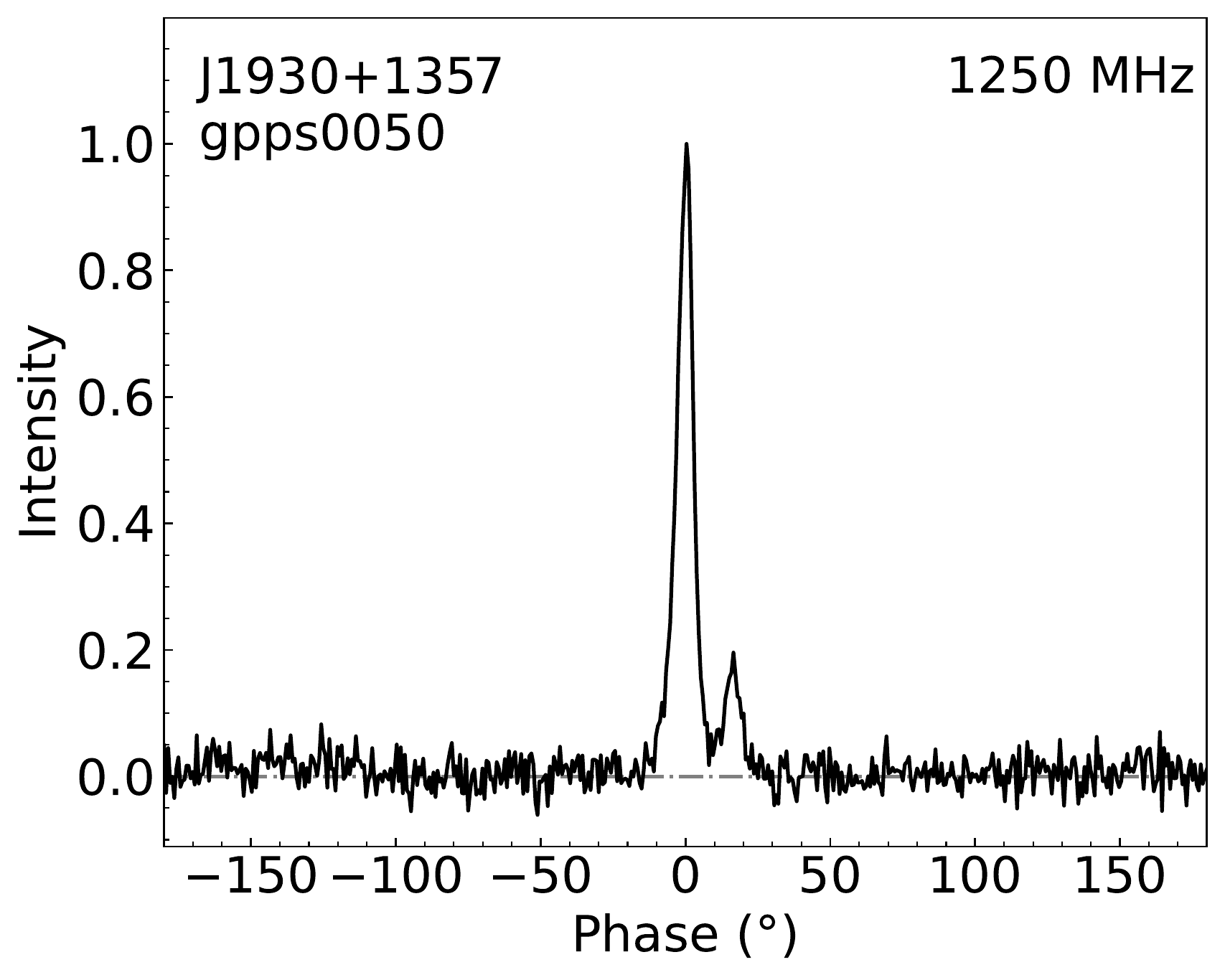}
\caption{Total intensity profiles for 13 previously known pulsars and 11 newly discovered pulsars in the GPPS Survey, obtained by adding data from a number of FAST observations based on the phase-coherent timing solutions. They are the highest-quality profiles ever achieved.  
}
\label{profiles}
\end{figure*}

\begin{figure*}
\includegraphics[width=0.24\textwidth]{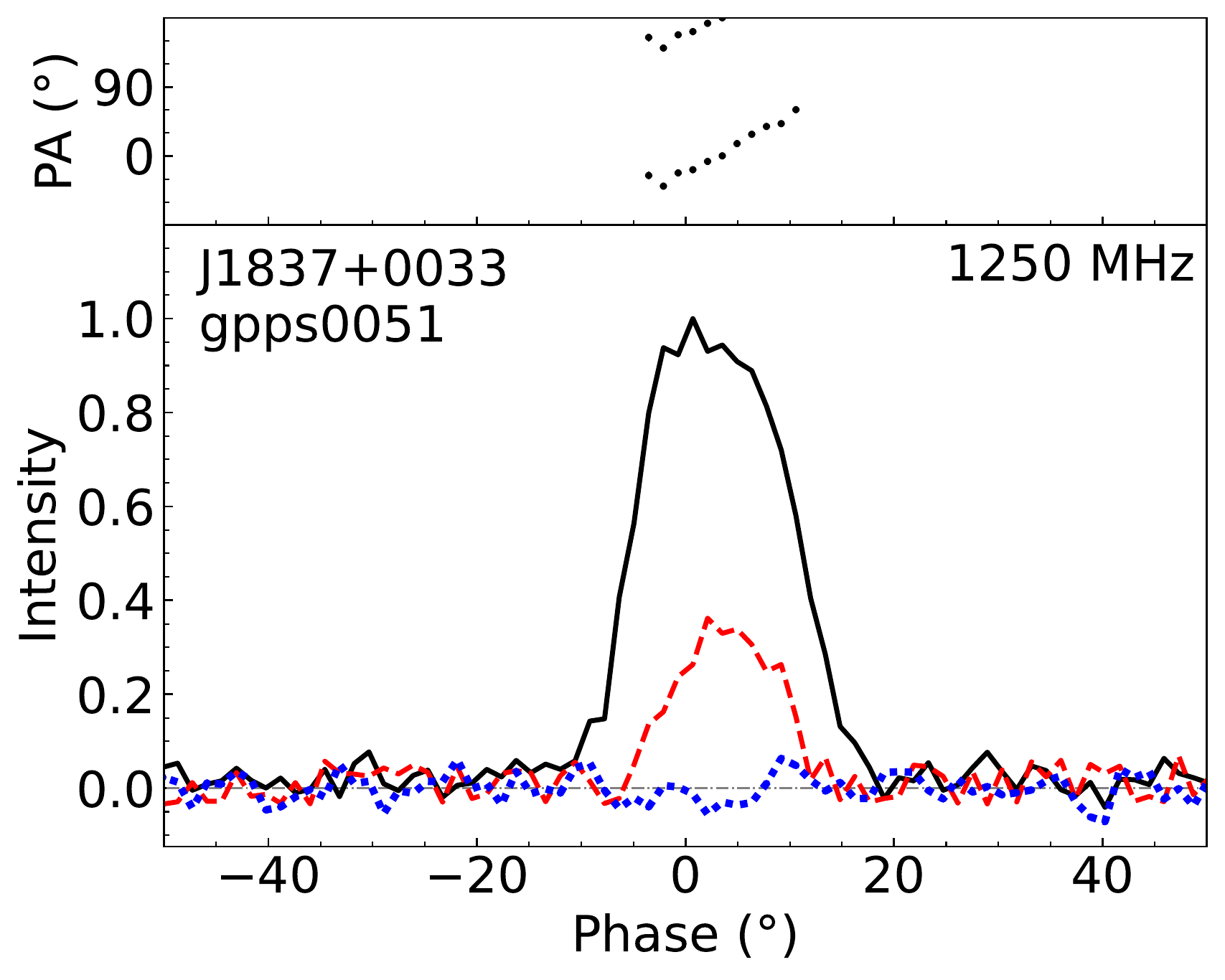}
\includegraphics[width=0.24\textwidth]{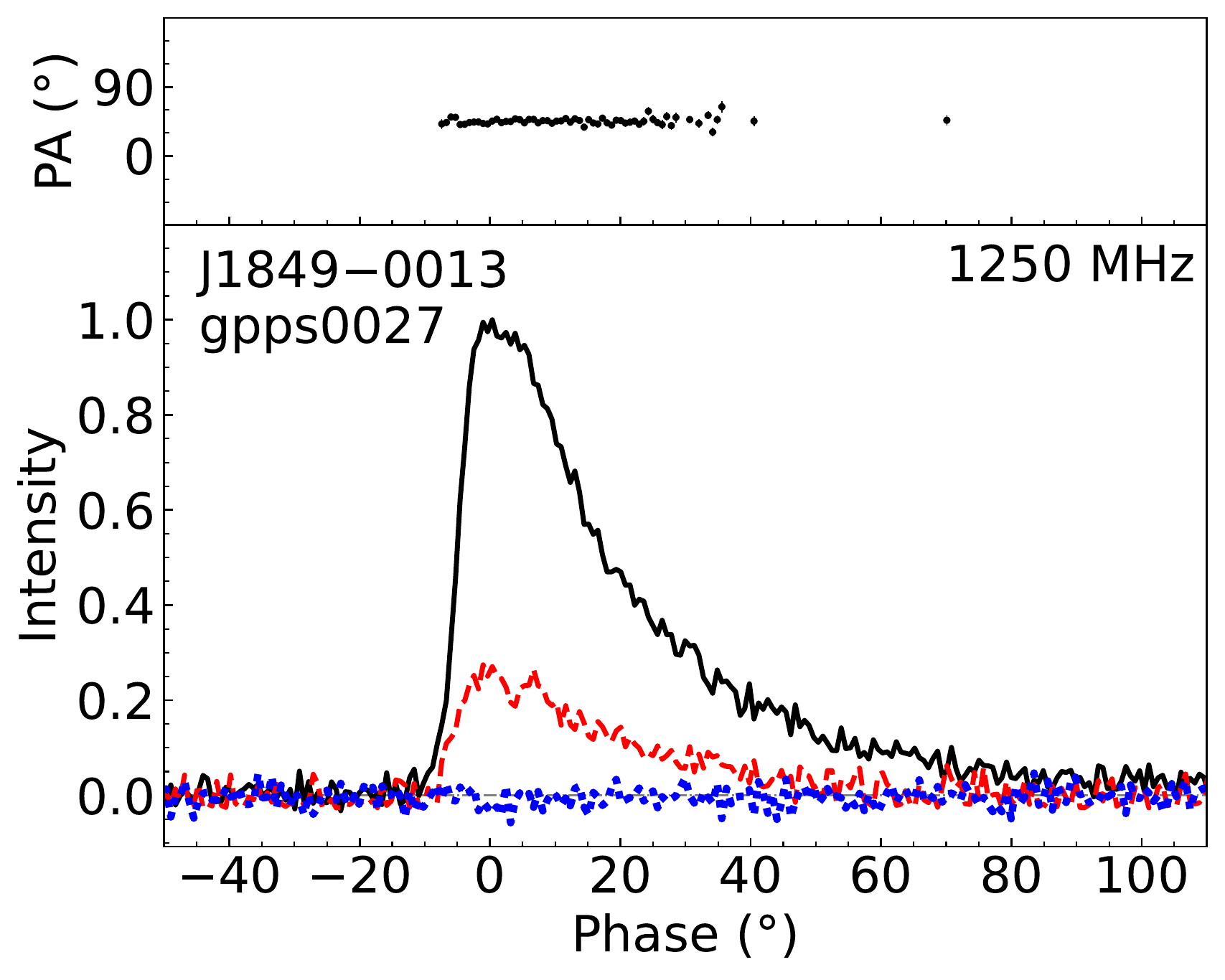}
\includegraphics[width=0.24\textwidth]{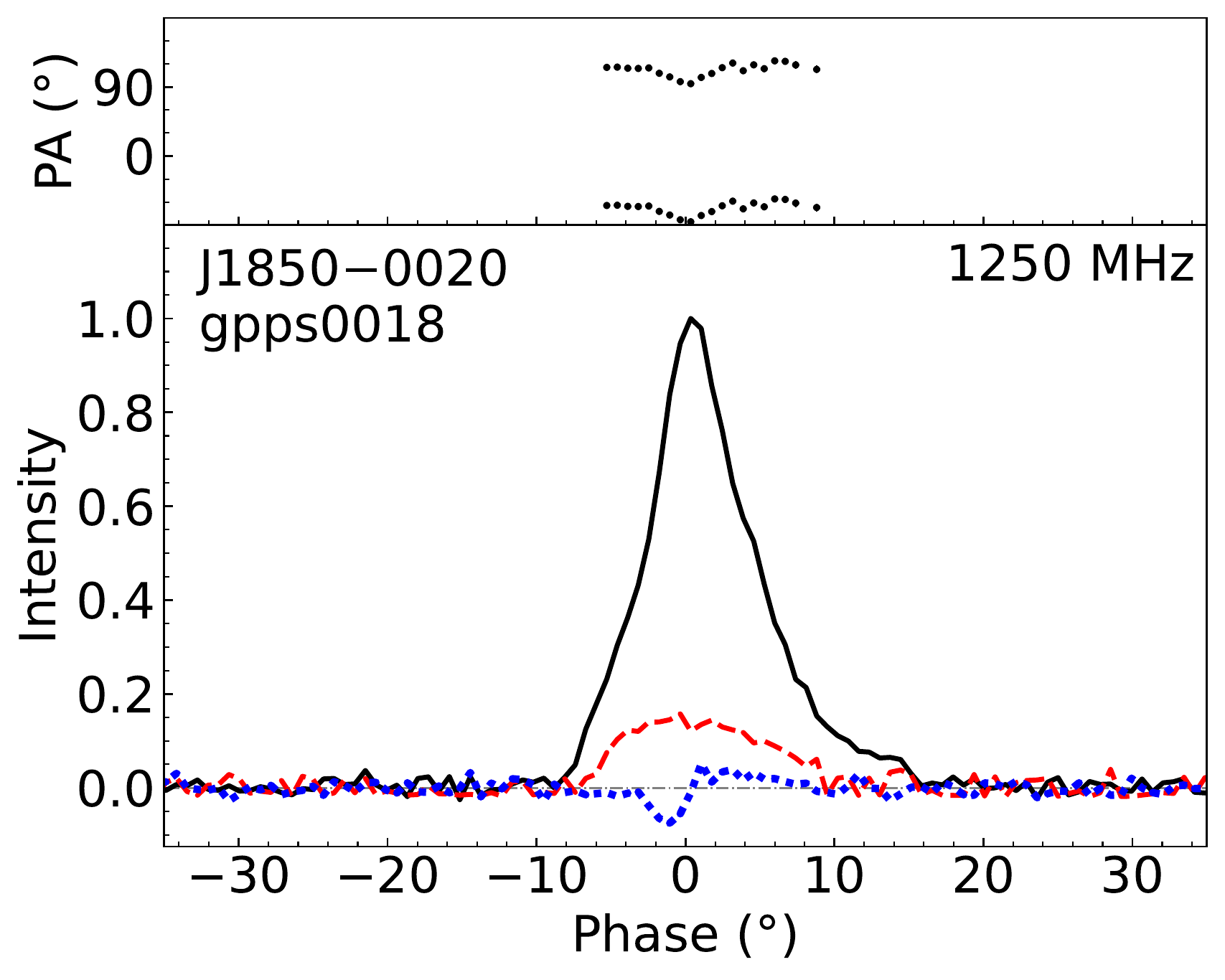}
\includegraphics[width=0.24\textwidth]{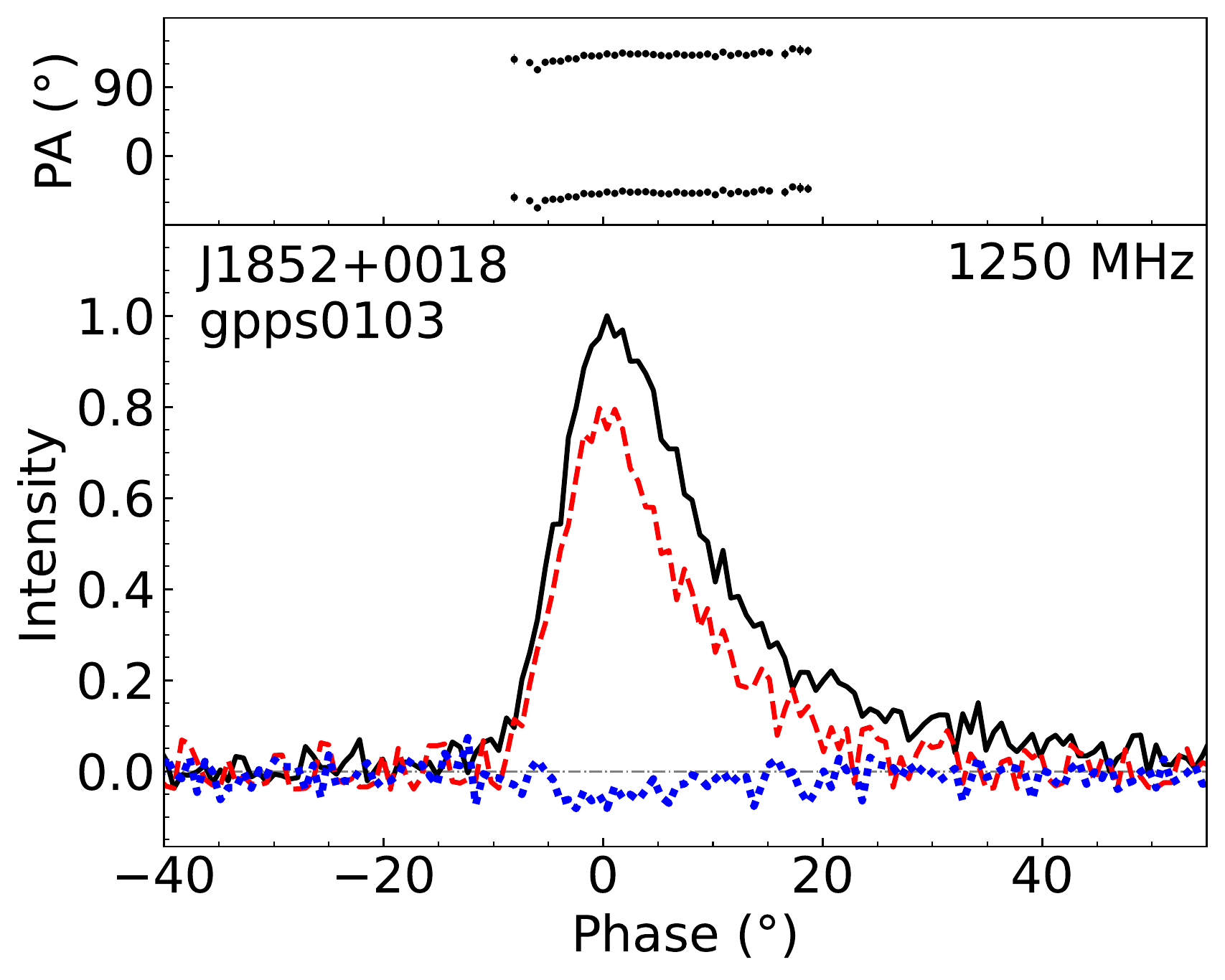}
\includegraphics[width=0.24\textwidth]{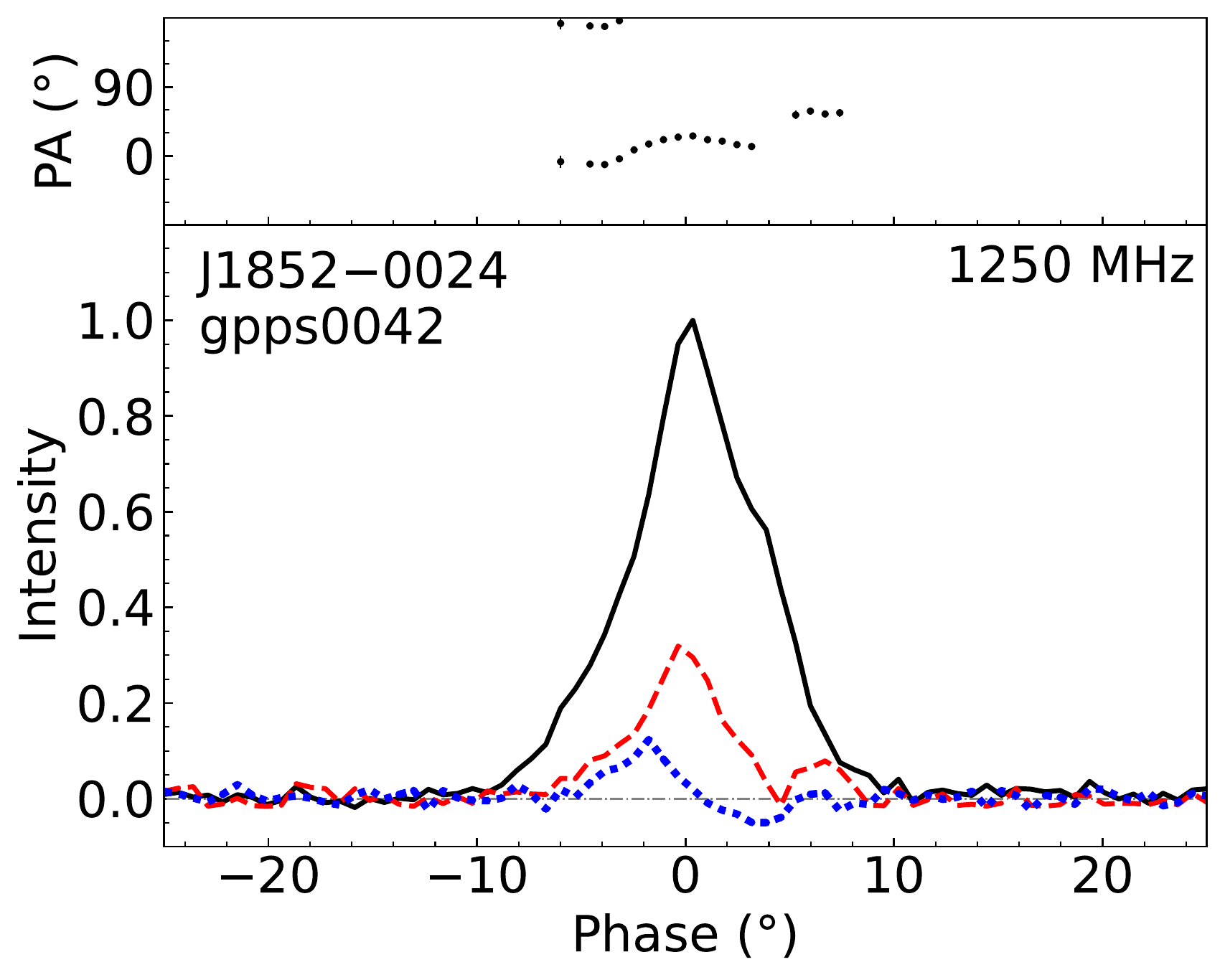}
\includegraphics[width=0.24\textwidth]{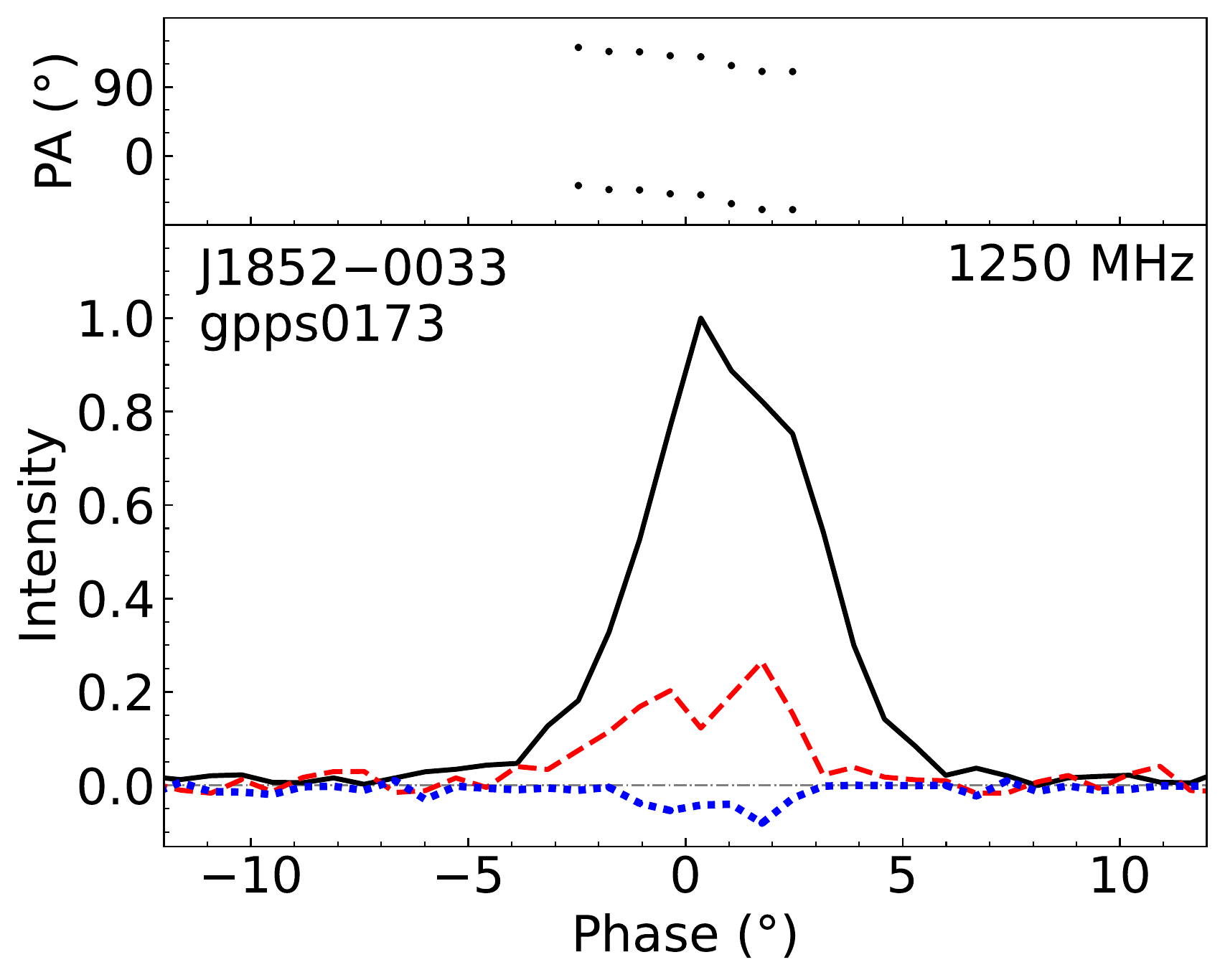}
\includegraphics[width=0.24\textwidth]{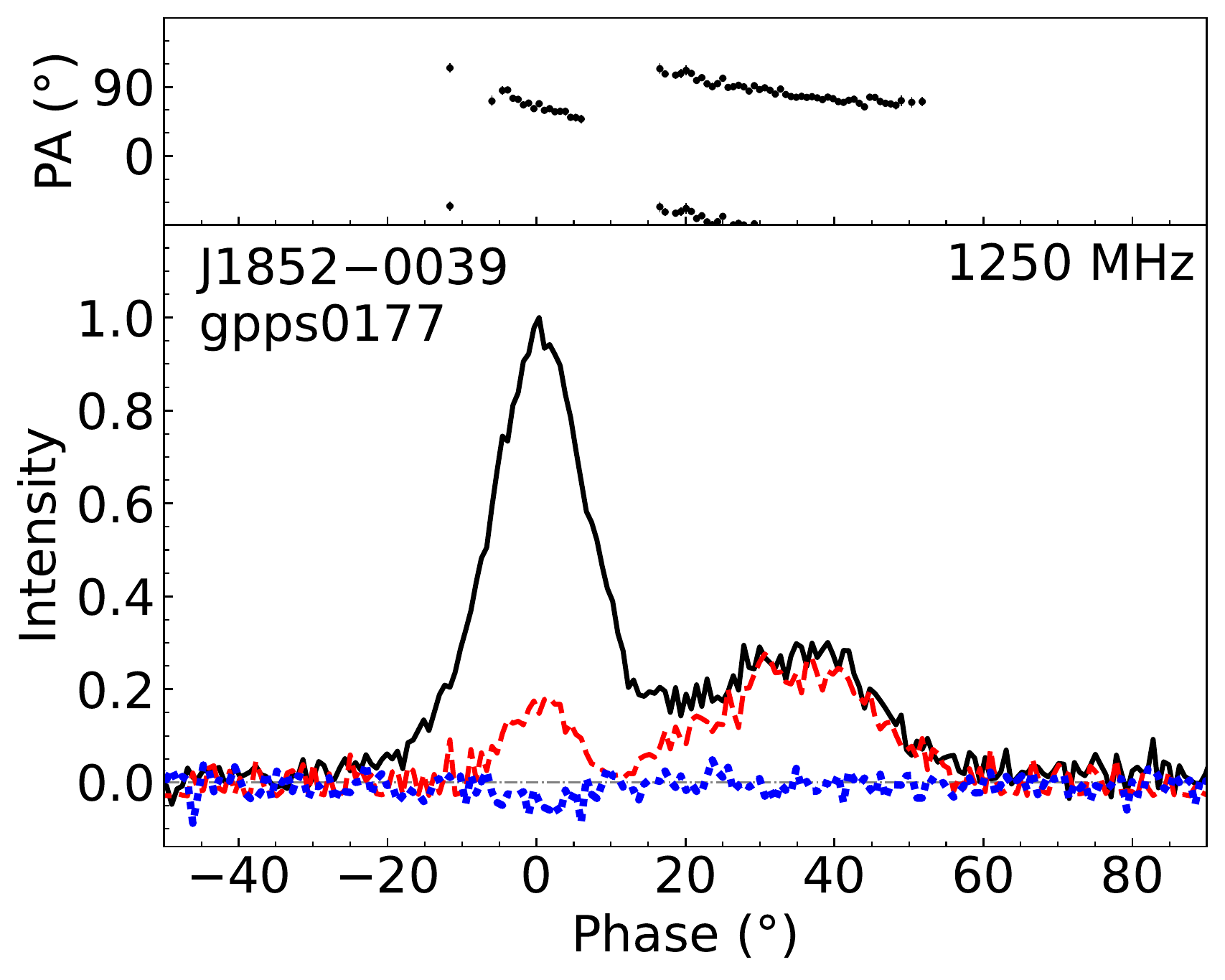}
\includegraphics[width=0.24\textwidth]{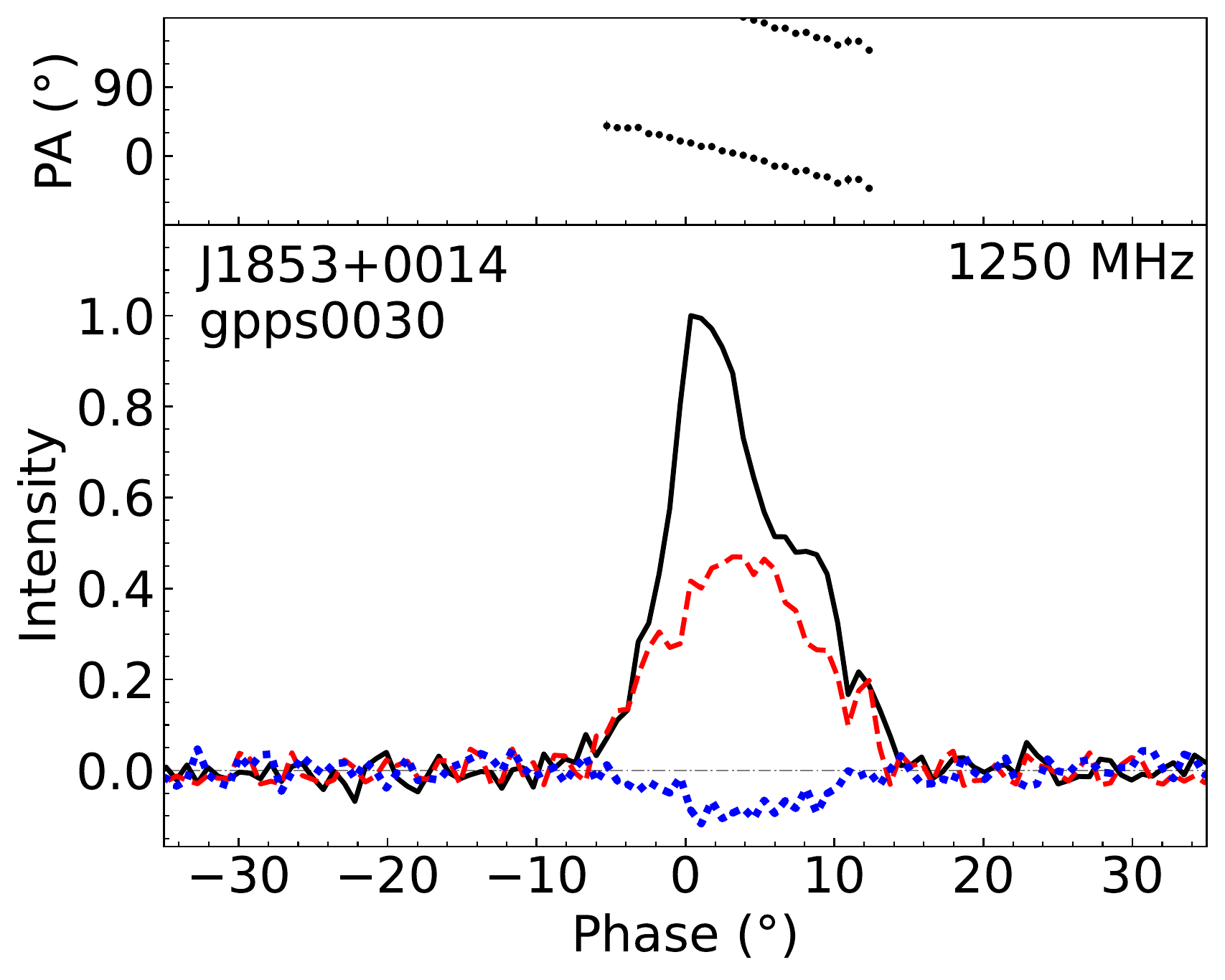}
\includegraphics[width=0.24\textwidth]{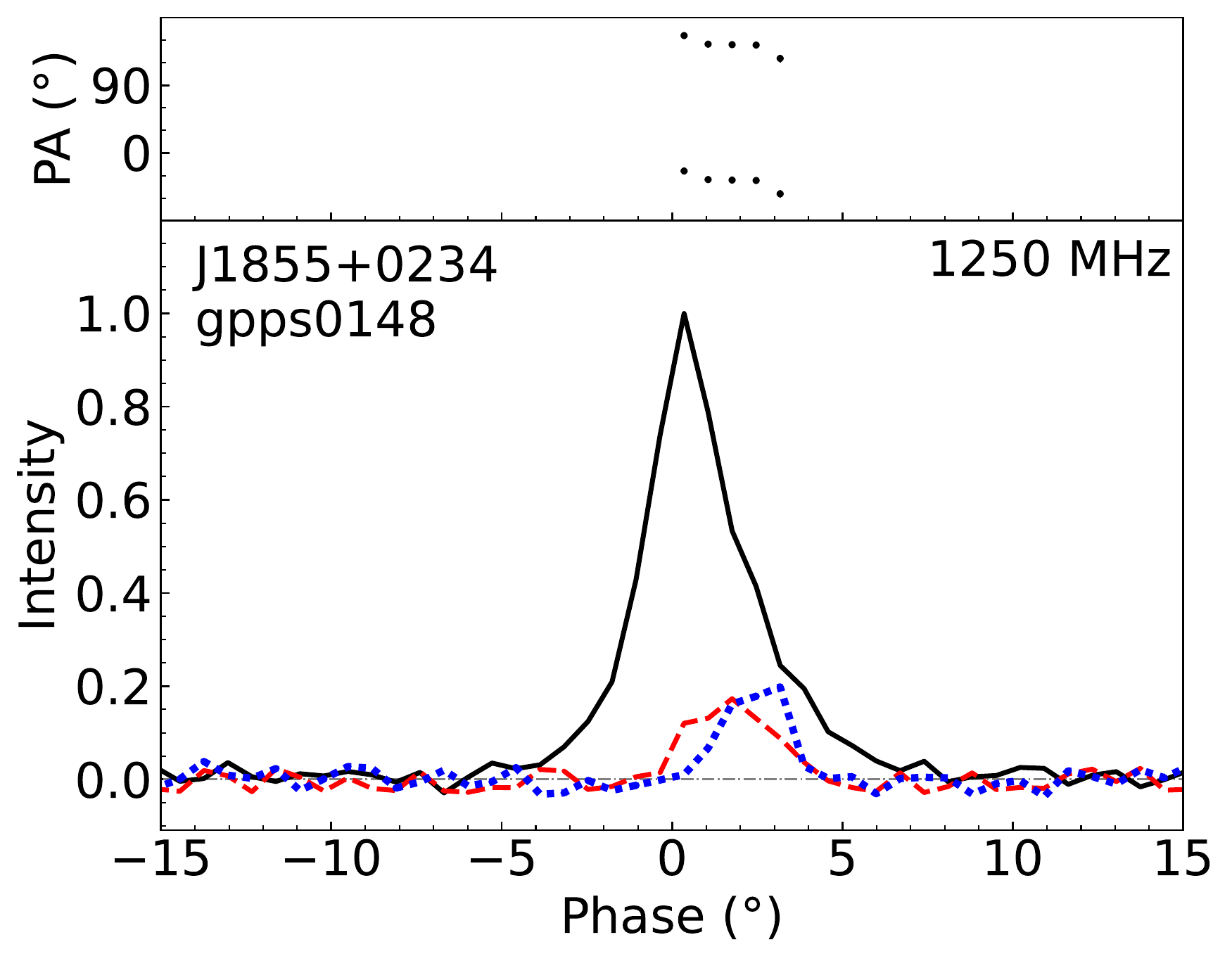}
\includegraphics[width=0.24\textwidth]{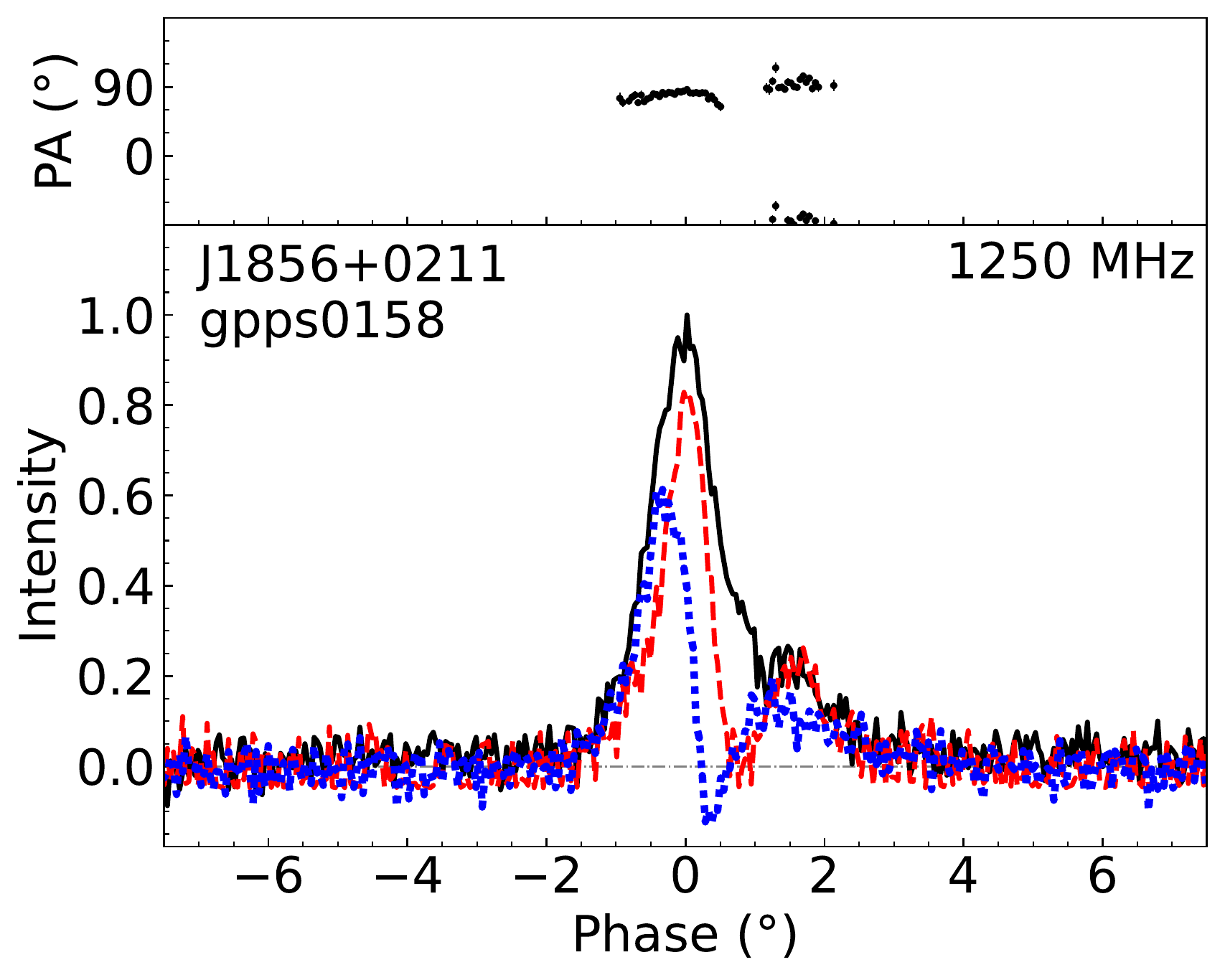}
\includegraphics[width=0.24\textwidth]{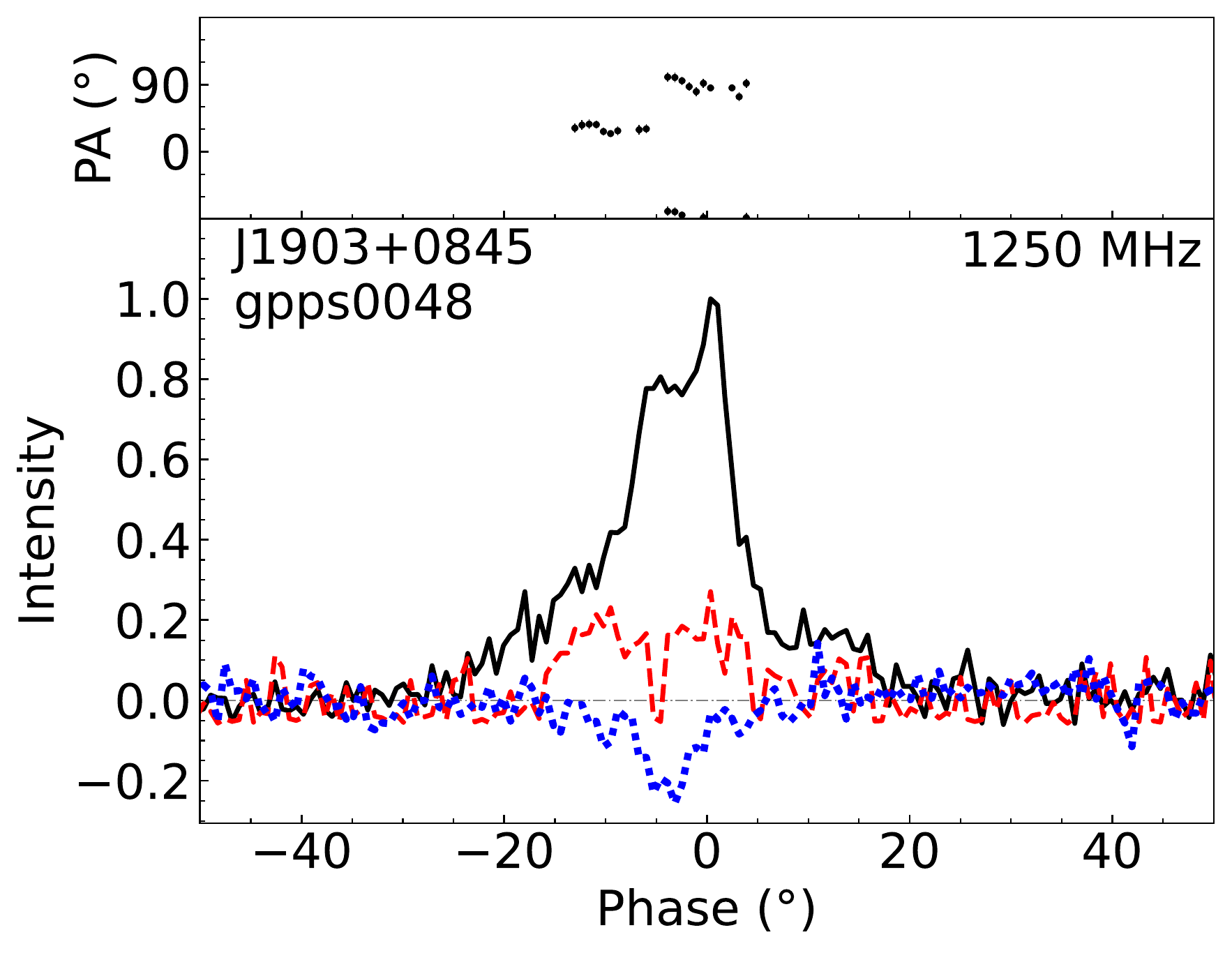}
\includegraphics[width=0.24\textwidth]{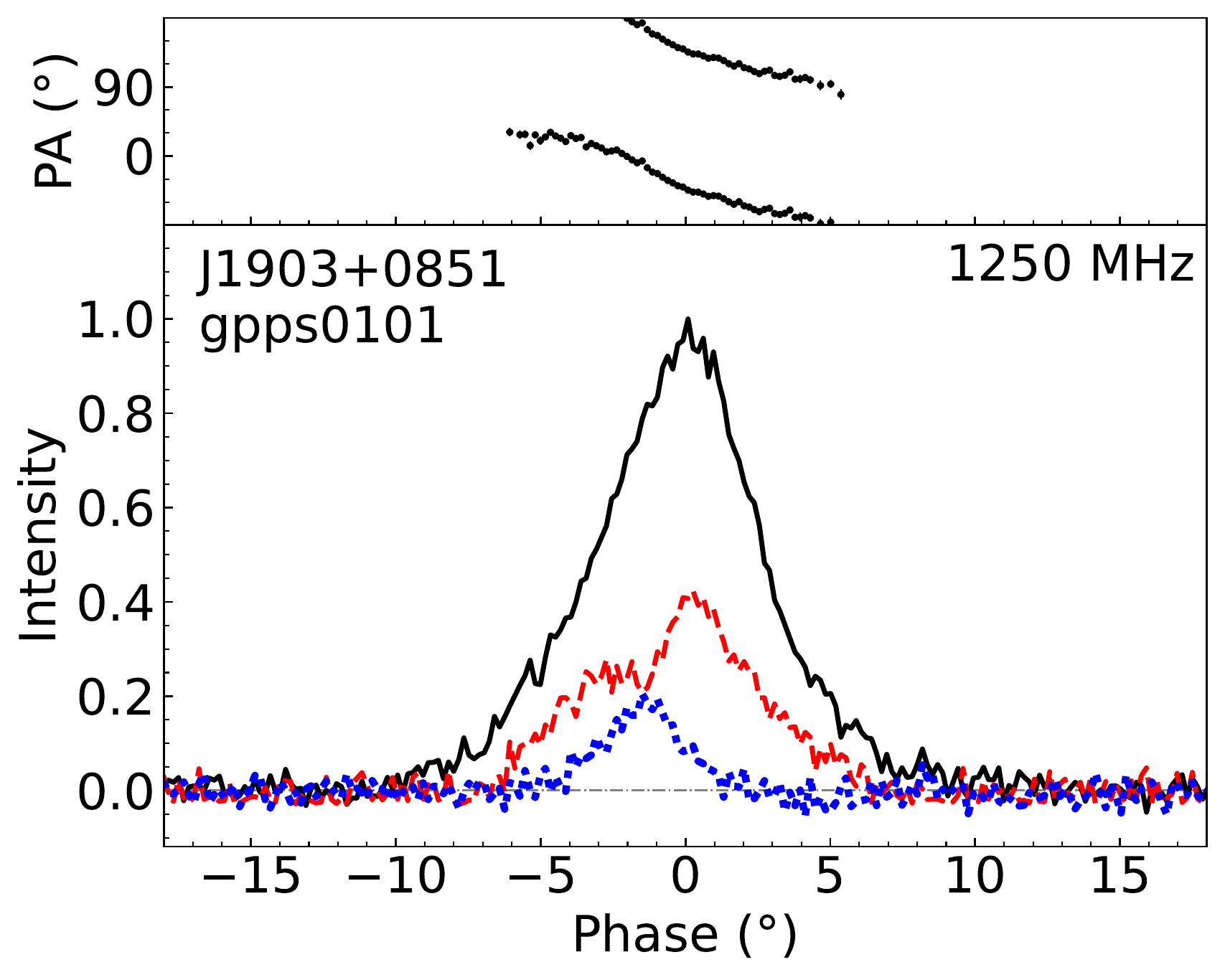}
\includegraphics[width=0.24\textwidth]{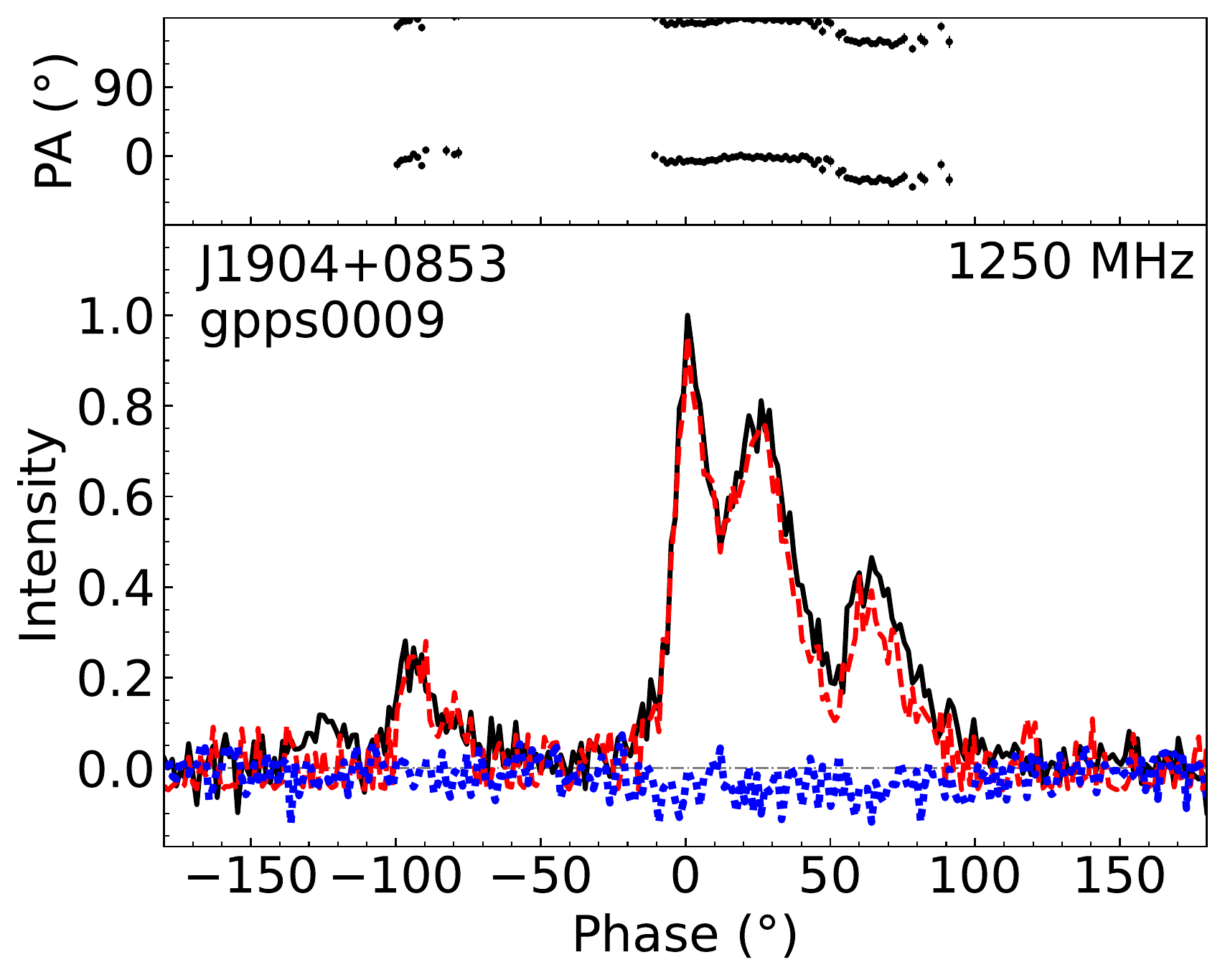}
\includegraphics[width=0.24\textwidth]{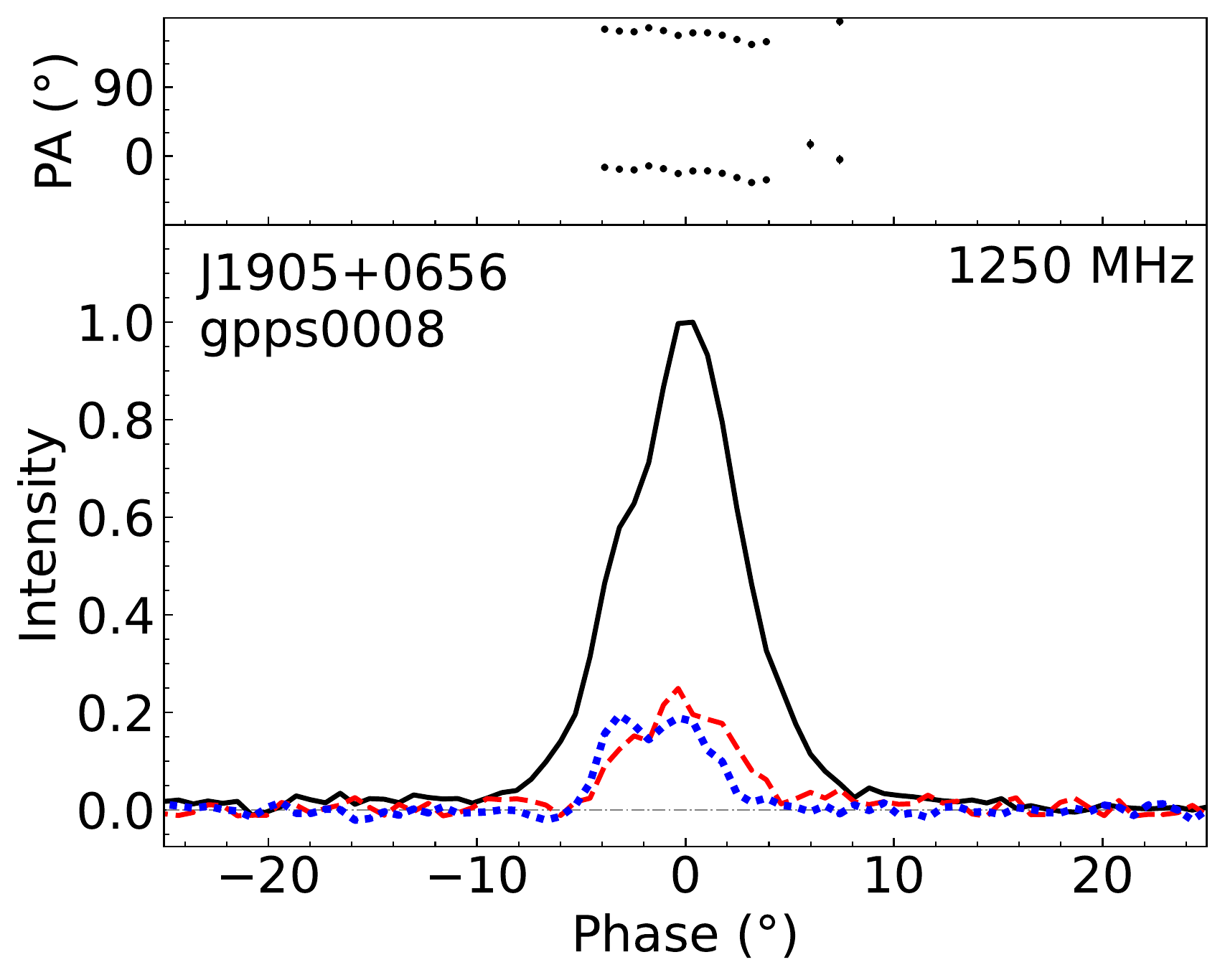}
\includegraphics[width=0.24\textwidth]{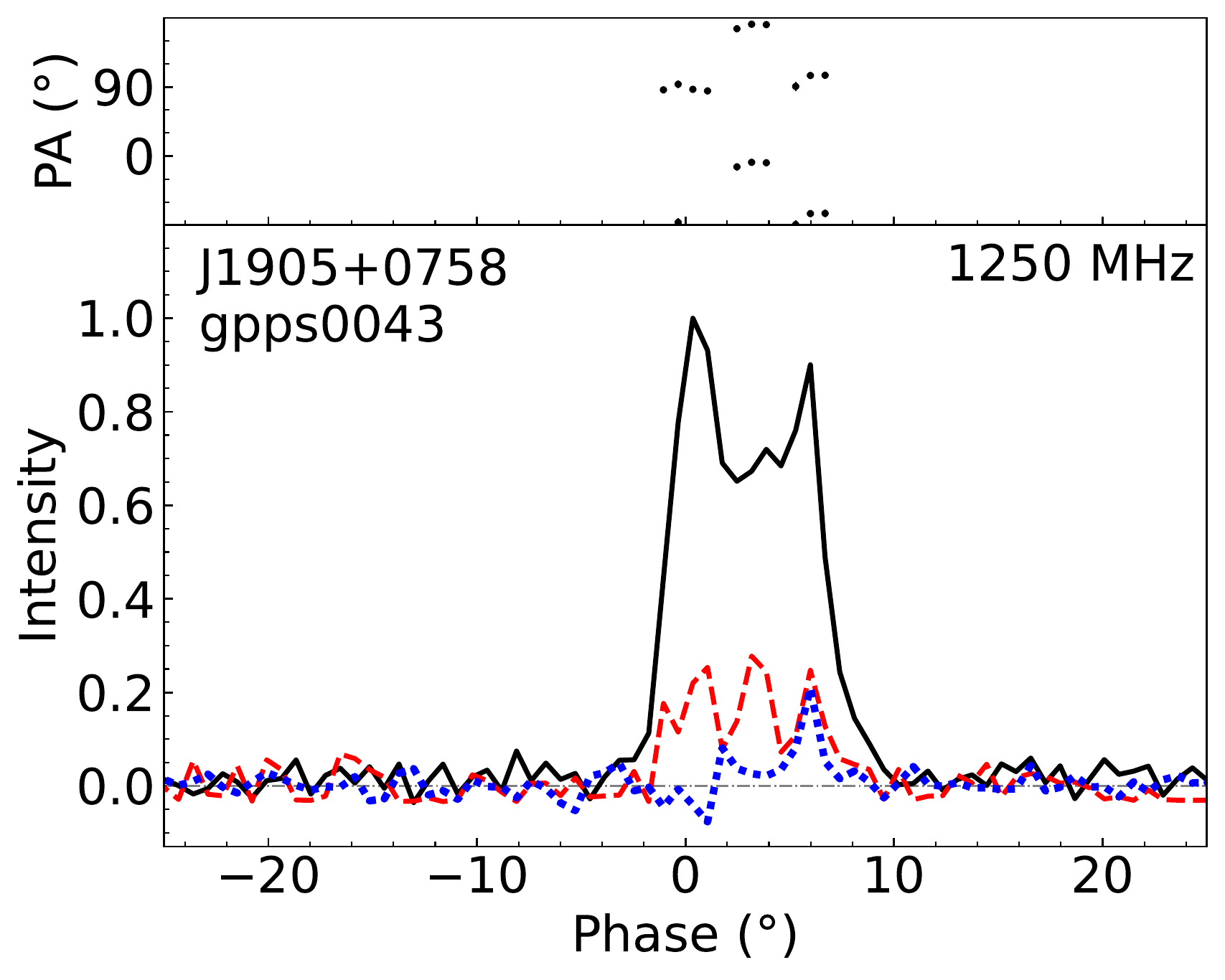}
\includegraphics[width=0.24\textwidth]{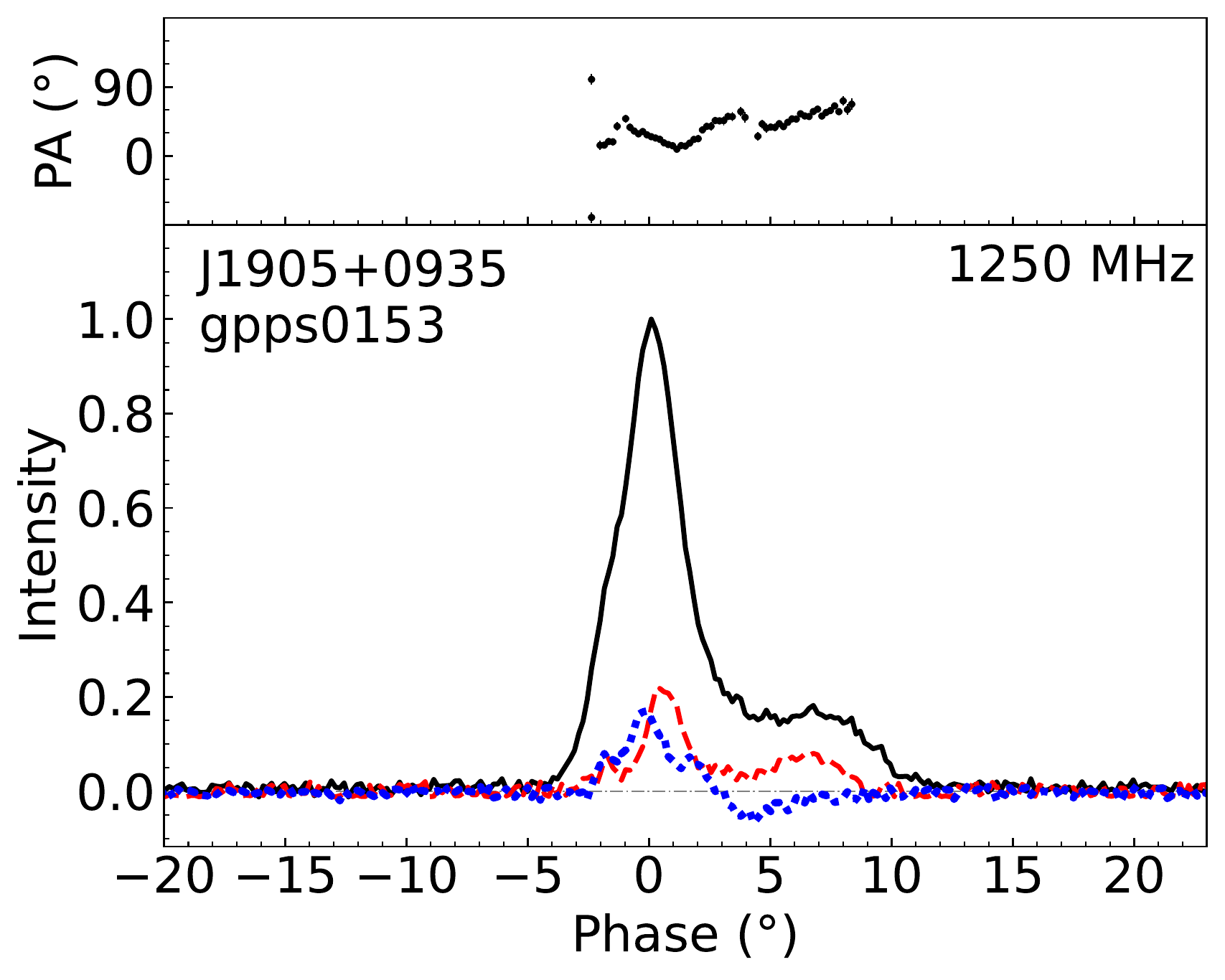}
\includegraphics[width=0.24\textwidth]{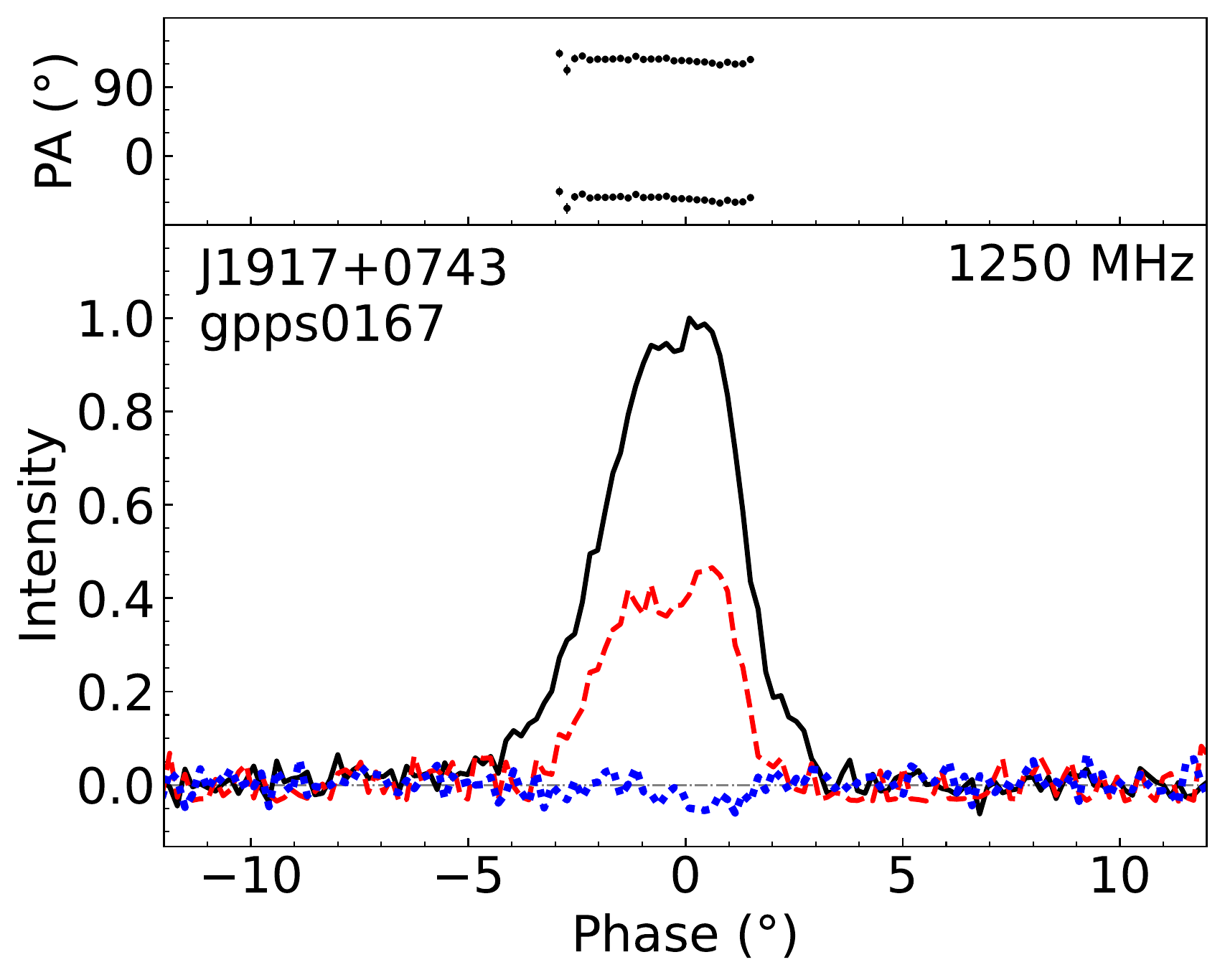}
\includegraphics[width=0.24\textwidth]{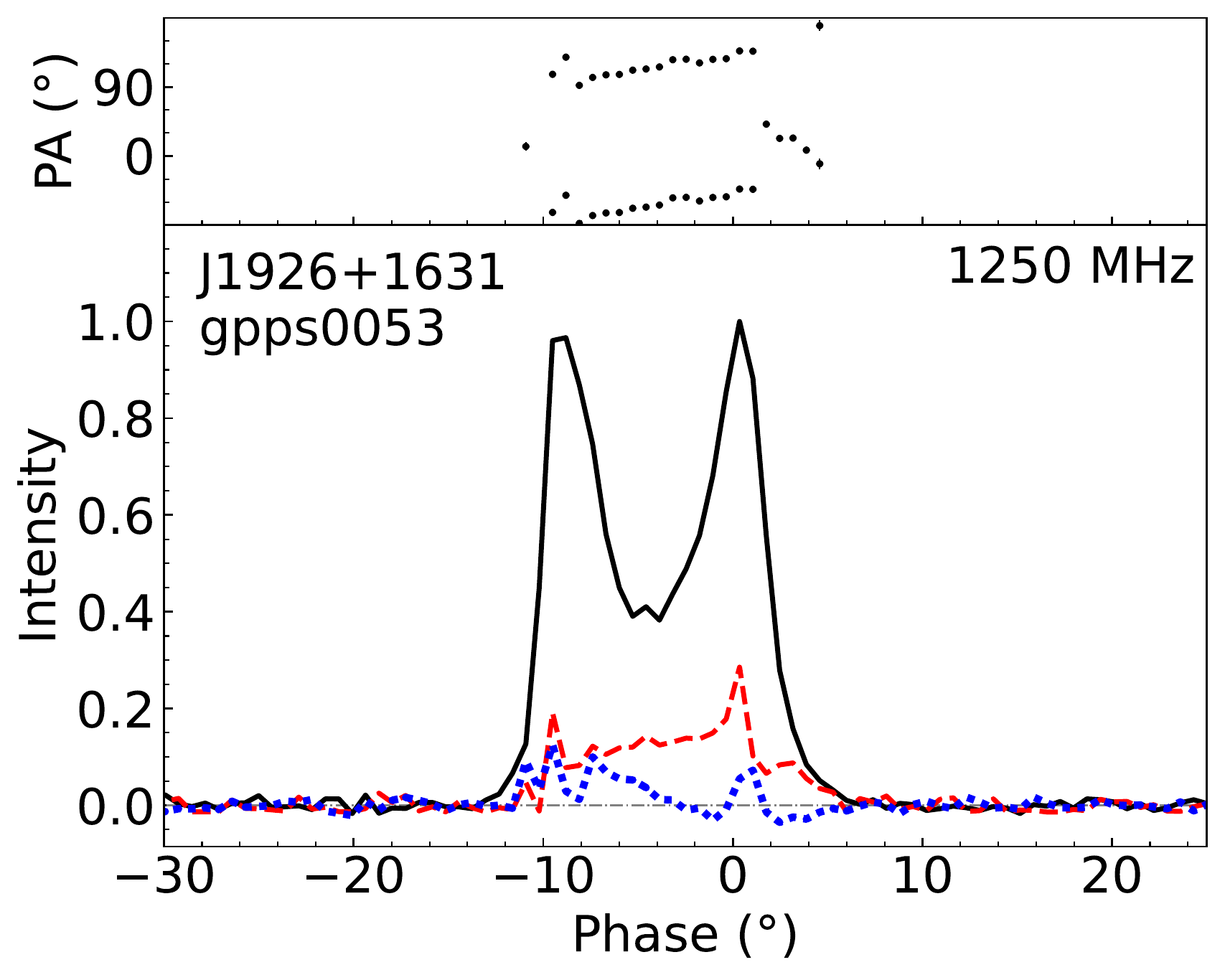}
\includegraphics[width=0.24\textwidth]{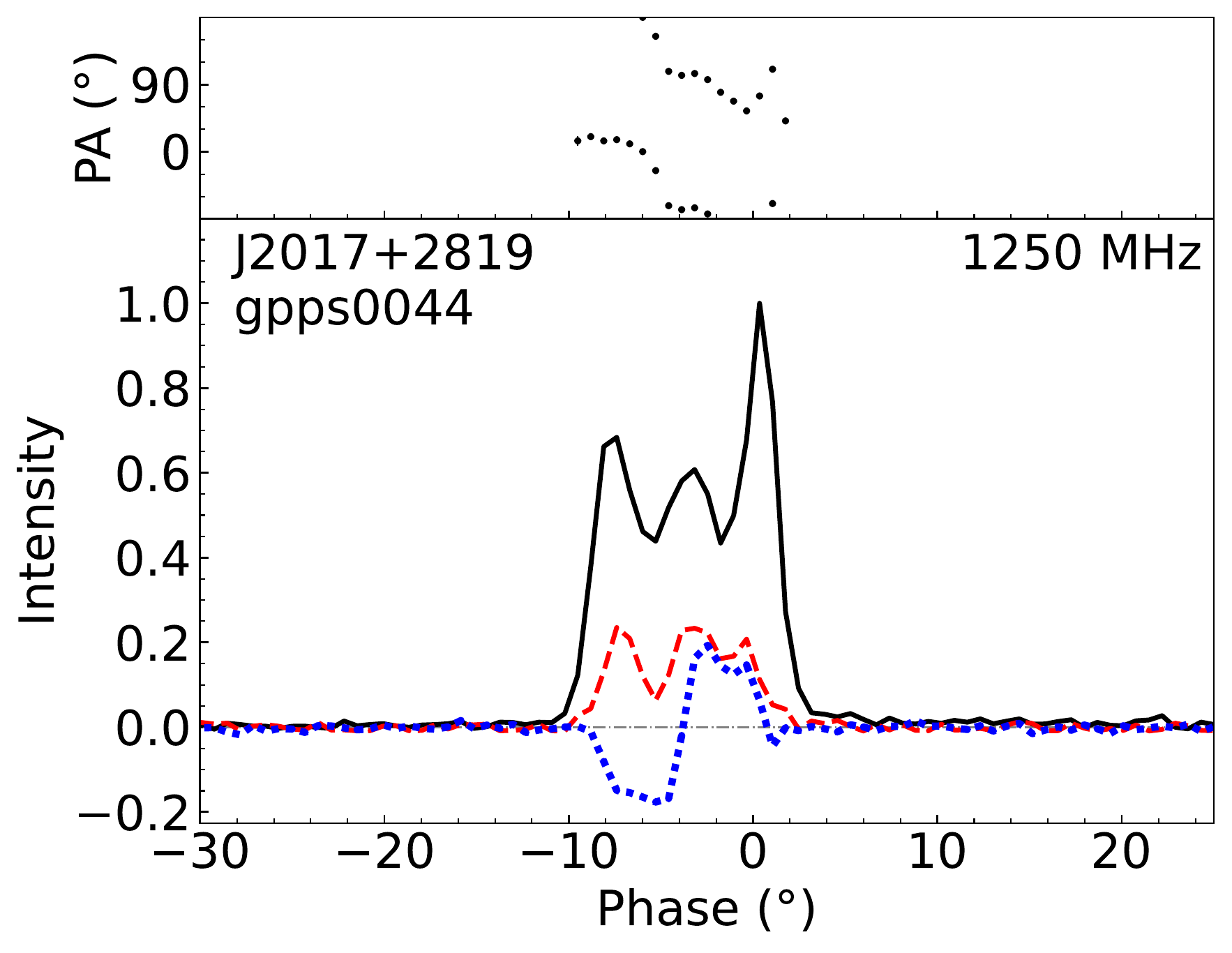}
\caption{Polarization profiles of 19 pulsars discovered in the GPPS Survey. Solid black lines, dashed red lines, and dotted blue lines indicate intensity profiles, linear polarization, and circular polarization, respectively. The polarization position angle curve is displayed in the top panel of each plot. }
\label{polProfiles}
\end{figure*}

\section{Summary and Discussion}

In summary, we collected FAST observations of pulsars and found that pulsar data from any beams of the FAST L-band 19-beam receivers can be used together for timing analysis. We obtained phase-coherent solutions for 13 previously known pulsars and 30 pulsars newly discovered in the GPPS Survey. The basic timing models, including pulsar position, spin period, period derivative, and DM, provide a good fit to the FAST-measured TOAs. Several pulsars, including PSR J1856+0211 with a spin period of 9.89 s, have a very low period derivative, putting them near the death line. 

The pulsar death line is the theoretical limit below which pulsars are unable to produce radio emission owing to insufficient conditions for pair production. The conventional death valley model  defined by \cite{1993ApJ...402..264C} is based on the vacuum gap model \citep{rs75}. If a gap above the polar cap area cannot be formed, for a pure dipole magnetic field or the highly curved field lines in a complicated magnetic structure, then a pulsar cannot have particles for radiation. Recent simulations \citep[e.g.][]{pts20, 2020ApJ...889...69C, bba22} show that the 'gap' can be formed in the form of either a vacuum gap \citep{rs75} or space-charge-limited flow \citep{as79}, or even in the outer magnetosphere beyond the null-charge surface \citep{chr86} or in the annular region \citep{qlw+04} extending from the surface to the outer magnetosphere in the form of a slot gap \citep{mh04}. Pulsar radio emission can be coherently produced by a bunch of particles, as indicated by the extremely high brightness temperature.
Different emission models therefore correspond to different locations of the death line. At least three pulsars in our sample are located near or even below the death lines, and hence can place strong constraints on the death-line models (see Fig.~\ref{PPdot}). 

To date, more than 600 pulsars have been discovered in the GPPS Survey,\footnote{\url{http://zmtt.bao.ac.cn/GPPS/}} but only the 30 pulsars considered here have timing solutions. A vast number of new pulsars are waiting for timing, including 140 MSPs. For 28 pulsars in the period range of 30–100 ms, we are timing them to distinguish among young pulsars, pulsars in binary systems, and disrupted recycled pulsars. The timings of long-period pulsars are needed in order to distinguish old pulsars from magnetars.

As a byproduct, after the phase-coherent timing solution is obtained, all FAST observations can be added together to give the best pulsar profiles. The total intensity profiles of 13 previously known pulsars and 11 newly discovered pulsars in the GPPS Survey are shown in Fig.~\ref{profiles}. Note that PSR J1849+0010 exhibits a weak interpulse component, suggesting that its rotation axis is nearly perpendicular to the magnetic axis. Profiles of PSRs J1850$-$0002 and J1857+0214 have remarkable long tails owing to scatter broadening.

For the other 19 newly discovered pulsars in the GPPS Survey, a number of follow-up observations have been made with the polarization data recorded. Based on the phase-coherent timing solutions, we add all polarization data together to obtain their integrated polarization profiles, as shown in Fig.~\ref{polProfiles}. The polarization properties obtained from these FAST observations are given in Table~\ref{GPPSprofTable}, including the pulse width at 50 per~cent and 10 per~cent peak intensities, the degrees of linear, circular, and absolute circular polarization, and RMs obtained from data of different sessions. Polarization profiles of seven pulsars were obtained for the first time. Polarization profiles of 12 pulsars (PSRs J1849$-$0013, J1850$-$0020, J1852+0018, J1852$-$0024, J1855+0234, J1856+0211, J1903+0851, J1905+0656, J1905+0758, J1905+0935, J1926+1631 and J2017+2819) have been previously reported by \citet{whj+22}, but from only one observation session. Significant improvements of profiles with a much better S/N are achieved here by adding data of multiple sessions together.

\begin{landscape}
\begin{table}
\begin{center}
\caption[]{Polarization properties of 19 newly discovered pulsars in the GPPS Survey, including pulse width at 50 per~cent and 10 per~cent peak intensities, the degrees of linear, circular, and absolute circular polarization, as well as final and session-specific RM values.}
 \tabcolsep 2.5pt  
 \footnotesize
\label{GPPSprofTable}
 \begin{tabular}{lrrrrrrcrcrcrcr}
  \hline\noalign{\smallskip}
Name          & $W_{50}$   & $W_{10}$   & L/I & V/I &|V|/I  & \multicolumn{1}{c}{RM} & Obs. Date1 & \multicolumn{1}{c}{RM1} & Obs. Date2 & \multicolumn{1}{c}{RM2} & Obs. Date3 & \multicolumn{1}{c}{RM3} & Obs. Date4 & \multicolumn{1}{c}{RM4}    \\
              & ($\degr$) & ($\degr$) & (per~cent) & (per~cent) & (per~cent)   & (rad~m$^{-2}$) &  &(rad~m$^{-2}$)  & & (rad~m$^{-2}$) & &(rad~m$^{-2}$) & & (rad~m$^{-2}$) \\
  \hline\noalign{\smallskip}
J1837$+$0033 & 16.9 & $-$ & 24.8 & $-$0.1 & 0.5  & 212.6(36) & 20200821 & 215(6) & 20201107 & 211(5) & 20230411 & 213.1(33)$^*$ & 20230416 & 218(4)$^*$\\
J1849$-$0013 & 22.5 & 71.0 & 24.6 & $-$0.3 & -0.8  & $-$108.0(9) & 20200401 & $-$110.1(35) & 20200815 & $-$108.2(17) & 20200816 & $-$110.4(29) & 20210302 & $-$104.9(17)\\
 & & & & & &  & 20211226 & $-$109.1(15)\\
J1850$-$0020 & 7.7 & 17.6 & 19.4 & $-$0.7 & 3.2  & 95.0(22) & 20200402 & 94(5) & 20200816 & 101.3(29) & 20211226 & 76(6) & 20220629 & 85(7)\\
J1852$+$0018 & 14.8 & $-$ & 68.4 & $-$4.3 & 1.1  & 73.2(5) & 20200417 & 76.4(14) & 20200811 & 75.0(13) & 20200826 & 74.1(14) & 20200915 & 72.1(12)\\
 & & & & & &  & 20200920 & 73.8(30) & 20211208 & 77(5) & 20220306 & 69.8(13) & 20220607 & 71.1(19)\\
J1852$-$0024 & 7.0 & 14.1 & 24.3 & 2.7 & 6.8  & 94.2(8) & 20200905 & 93.9(12) & 20200920 & 94.2(17) & 20220102 & 94.1(17) & 20221111 & 96.8(31)\\
J1852$-$0033 & 4.9 & 8.4 & 23.0 & $-$4.6 & 3.0  & 175.0(13) & 20210317 & 177.2(32) & 20210318 & 178(4) & 20210913 & 179.9(39) & 20211213 & 170.6(32)\\
 & & & & & &  & 20220121 & 173.4(31) & 20220217 & 174.7(39) & 20220414 & 181(4) & 20220509 & 174(5)\\
 & & & & & &  & 20220605 & 172(9) & 20221111 & 151(8)\\
J1852$-$0039 & 15.5 & $-$ & 35.2 & $-$2.6 & -0.1  & 198.0(18) & 20210308 & 185(7) & 20210318 & 201(7) & 20210913 & 205(5) & 20211213 & 195(4)\\
 & & & & & &  & 20220121 & 198(6) & 20220414 & 200.3(34) & 20220509 & 197(5) & 20220605 & 201(6)\\
 & & & & & &  & 20221111 & 193(8) & 20230110 & 201(9)$^*$\\
J1853$+$0014 & 8.4 & $-$ & 58.9 & $-$9.9 & 7.9  & 16.8(8) & 20200401 & 17.6(17) & 20200828 & 18.1(21) & 20200904 & 16.6(18) & 20200906 & 15.8(16)\\
 & & & & & &  & 20200915 & 16.4(15)\\
J1855$+$0234 & 2.8 & $-$ & 15.4 & 11.4 & 10.9  & $-$4(5) & 20210111 & $-$8(6) & 20220306 & 1(7)\\
J1856$+$0211 & 1.0 & $-$ & 62.8 & 45.1 & 47.3  & $-$169.4(20) & 20210113 & $-$169.4(20)  & 20221207 & $-$165.6(10)$^*$ & 20230108 & $-$157.6(12)$^*$ & 20230116 & $-$162.1(14)$^*$\\
J1903$+$0845 & 10.5 & $-$ & 22.2 & $-$13.2 & 9.7  & 250(6) & 20220121 & 254(8) & 20220607 & 246(8)\\
J1903$+$0851 & 5.8 & $-$ & 38.2 & 8.4 & 9.0  & 181.6(6) & 20200905 & 183.5(32) & 20220121 & 178(6) & 20220121 & 181.2(8) & 20220607 & 182.7(12)\\
 & & & & & &  & 20230407 & 186.9(37)$^*$ & 20230412 & 179(5)$^*$\\
J1904$+$0853 & 42.2 & $-$ & 85.2 & $-$8.5 & 3.7  & 373.1(7) & 20211026 & 371.6(17) & 20211127 & 383(5) & 20211230 & 372.0(29) & 20211230 & 373.0(18)\\
 & & & & & &  & 20220116 & 372.1(29) & 20220202 & 370(5) & 20220210 & 370.9(17) & 20220320 & 373.8(11)\\
 & & & & & &  & 20220603 & 378.4(32) & 20221224 & 375.0(18)$^*$\\
J1905$+$0656 & 6.3 & 12.7 & 20.3 & 15.8 & 16.3  & $-$8.9(9) & 20200607 & $-$7.4(17) & 20200826 & $-$8.1(15) & 20201002 & $-$9.5(21) & 20210408 & $-$12.0(22)\\
J1905$+$0758 & 7.0 & $-$ & 22.0 & 3.7 & 6.1  & 424.5(23) & 20200829 & 422(5) & 20200911 & 428.1(35) & 20211226 & 421.9(39)\\
J1905$+$0935 & 3.0 & 11.8 & 18.9 & 5.9 & 12.4  & 696.1(8) & 20210126 & 694.9(28) & 20211114 & 691.8(17) & 20220306 & 697.3(15) & 20220529 & 702.0(32)\\
 & & & & & &  & 20220608 & 697.2(13) & 20221205 & 698.2(18)$^*$ & 20230407 & 697.4(19)$^*$\\
J1917$+$0743 & 3.3 & $-$ & 41.4 & $-$2.4 & 0.4  & 429.9(13) & 20210218 & 429.9(13) & 20221205 & 433.5(16)$^*$ & 20230407 & 433.0(14)$^*$ & 20230412 & 430.1(14)$^*$\\
J1926$+$1631 & 12.0 & 14.8 & 20.8 & 5.0 & 6.8  & 552.0(8) & 20200818 & 552.7(20) & 20210402 & 555.1(19) & 20211226 & 550.0(14) & 20220613 & 551.9(16)\\
J2017$+$2819 & 9.8 & 12.0 & 25.4 & $-$1.2 & 18.7  & $-$170.4(5) & 20200401 & $-$170.7(14) & 20200820 & $-$168.6(16) & 20200914 & $-$171.2(10) & 20201006 & $-$170.5(10)\\
 & & & & & &  & 20210730 & $-$170.1(12)\\
  \noalign{\smallskip}\hline
\end{tabular}
\end{center}
$^*$ RM does not discount the ionospheric contribution owing to the lack of ionospheric data, and it is not included in the mean value calculation. 
\end{table}
\end{landscape}

Polarization profiles of PSRs J1852+0018, J1852$-$0039, J1853+0014, J1856+0211 and J1904+0853 have highly linearly polarized components. The polarization profile of PSR J1856+0211 has an unusually highly circularly polarized component. Orthogonal polarization modes are detected for PSRs J1852$-$0039, J1903+0845 and J1905+0758. The profiles of PSRs J1849$-$0013 and J1852+0018 have long tails caused by interstellar scattering, and their polarization position angle curves are flattened out at the tail part, as discussed by \citet{lh03}. A detailed study of scattering profiles observed by the GPPS Survey is presented in \citet{jhw+23}.

The flux densities of all 30 GPPS pulsars listed in Table~\ref{GPPStab1} are estimated from the total intensity profiles, as in \citet{2021RAA....21..107H}.

\section*{Author contributions}  
The FAST GPPS Survey is a key FAST science project led by JLH. WQS processed all data and drafted this paper under the supervision of JLH and PFW. JLH organized this work and was in charge of the final writing of this paper. PFW developed the processing procedures for the pulsar polarization profile and pulsar timing, which are extensively used in this paper. JPY carried out a fraction of the FAST observations. CW fed all targets for the GPPS observations. DJZ, TW, WCJ, YY, ZLY, NNC, XC, and LX contributed to different aspects of data processing and/or joined many group discussions. PFW and JX made fundamental contributions to the construction and maintenance on the computation platform. Other people jointly proposed or contributed to the FAST key project. All authors contributed to the final version of this paper.


\section*{Data and package availability}
Original FAST observational data will be open source according to the FAST data 1-year protection policy. The folded and calibrated profiles in this paper can be found at the webpage: \url{http://zmtt.bao.ac.cn/psr-fast/}.

\section*{Acknowledgements}
This work has made use of data from the GPPS Survey project, as one of five key projects of FAST, a Chinese national mega-science facility, operated by National Astronomical Observatories, Chinese Academy of Sciences. The authors are supported by the National Natural Science Foundation of China (NSFC, nos 11988101 and 11833009) and the Key Research Program of the Chinese Academy of Sciences (no. QYZDJ-SSW-SLH021).
%
%

\bibliographystyle{mnras}
\bibliography{bibfile}
\appendix

\bsp	
\label{lastpage}
\end{document}